\documentclass[twocolumn, amsmath,amssymb, superscriptaddress]{revtex4}
\setcitestyle{authoryear,semicolon,round} 

\usepackage{dcolumn}
\usepackage{bm}
\usepackage{amsmath}
\usepackage{amsfonts}
\usepackage{graphics} 
\usepackage[dvips]{graphicx}
\usepackage{times}
\usepackage{setspace}
\usepackage{color}   
\usepackage{verbatim}
\usepackage[raggedright]{titlesec}
\usepackage[normalem]{ulem}
\usepackage{framed}

\usepackage{enumerate}
\usepackage[breaklinks=true]{hyperref} 
\usepackage{breakurl} 

\usepackage{balance}

\usepackage[outercaption]{sidecap} 

\usepackage{dcolumn}
\newcolumntype{d}[1]{D{.}{.}{#1}}

\graphicspath{{}}
\hypersetup{colorlinks=true,citecolor=blue, linkcolor=black,urlcolor=blue} 

\newcommand{\beginsupplement}{%
        \setcounter{section}{0} 
        \renewcommand{\thesection}{S\arabic{section}}%
        \setcounter{table}{0}
        \renewcommand{\thetable}{S\arabic{table}}%
        \setcounter{figure}{0}
        \renewcommand{\thefigure}{S\arabic{figure}}%
        \setcounter{equation}{0}
        \renewcommand{\theequation}{S\arabic{equation}}%
     }
 
\begin{document}
\title{Grand challenges and emergent modes of convergence science}

\author{Alexander M. Petersen$^{\ast}$}
\affiliation{Department of Management of Complex Systems, Ernest and Julio Gallo Management Program, School of Engineering, University of California, Merced, CA 95343}
\author{ Mohammed E. Ahmed}
\author{ Ioannis Pavlidis$^{\ast}$}
\affiliation{Computational Physiology Laboratory, University of Houston, Houston, Texas 77204}

\begin{abstract}
\noindent To address complex problems, scholars are increasingly faced with challenges of integrating diverse  knowledge domains. We analyzed the evolution of this convergence paradigm in the broad ecosystem of brain science, which provides a real-time testbed for evaluating two modes of cross-domain integration -- subject area exploration via expansive learning and cross-disciplinary collaboration among domain experts. We show that research involving both modes features a 16\% citation premium relative to a mono-disciplinary baseline. 
Further comparison of research integrating neighboring versus distant research domains shows that the cross-disciplinary mode is essential for integrating across relatively large disciplinary distances.
Yet we find research utilizing cross-domain subject area exploration alone --  a convergence shortcut -- to be growing in prevalence at roughly 3\% per year, significantly faster than the alternative cross-disciplinary mode,
 despite being  less effective at integrating domains and markedly less impactful.  
By measuring  shifts in the prevalence and impact of different convergence modes in the 5-year intervals before and after 2013, our results indicate that these counterproductive patterns may relate to competitive pressures associated with global Human Brain flagship funding initiatives.
Without additional policy guidance, such Grand Challenge flagships may unintentionally incentivize such convergence  shortcuts, thereby  undercutting the advantages of cross-disciplinary teams in tackling challenges calling on convergence. 
\end{abstract}

\maketitle

\footnotetext[1]{$^\ast$To whom correspondence should be addressed;  E-mail:  ipavlidis@uh.edu or apetersen3@ucmerced.edu}

The history of scientific development is characterized by a pattern of convergence-divergence cycles \citep{Roco:2013}.
In convergence, originally distinct disciplines synergistically interact to address complex problems and accelerate breakthrough discovery \citep{NRC:2014}. In divergence,  in addition to   fragmentation  resulting from conflicting social forces  \citep{balietti2015disciplinary}, spin-offs  occur as new techniques, tools and applications spawn. The evolving fusion of multi-domain expertise during the present convergence cycle carries significant intellectual and organizational challenges \citep{committee1900facilitating,fealing_science_2011,bromham2016interdisciplinary,Pavlidis:2014}. The core issue is that contemporary convergence takes place in the context of team science \citep{Wuchty:2007,TeamGrowth2014}.  Accordingly, collaboration across distinct academic cultures and units faces behavioral  \citep{van2011factors} and institutional barriers  \citep{NRC:2014}.

Two early successful examples of convergence are worth mentioning to draw a comparative baseline. First, the Manhattan Project (MP), where physicists, chemists, and engineers successfully worked in the 1940s  to control nuclear fission and produce the first atomic bomb, under a tightly run government program   \citep{Hughes:2003}. A half-century later (1990s-2000s), the Human Genome Project (HGP) forged a  multi-institutional bond integrating  biologists and computer scientists, under an organizational design known as consortium science model  whereby teams of teams organize around a well-posed  central grand challenge \citep{helbing2012accelerating}, with a common goal to share benefits equitably within and beyond institutional boundaries \citep{Petersen:2018}.  In 10 short years, the HGP led to the mapping and identification of the human genetic code, ushering civilization into the genomics era.

 Brain science is presently supported by major funding programs that span the world over \citep{Grillner:2016}. In late 2013, the United States launched the BRAIN Initiative\textsuperscript{\textregistered} (Brain Research through Advancing Innovative Neurotechnologies), a public-private effort aimed at developing new experimental tools that will unlock the inner workings of brain circuits \citep{Jorgenson:2015}. At the same time, the European Union launched the Human Brain Project (HBP), a 10 year funding program based on exascale computing approaches, which aims to build a collaborative infrastructure for advancing knowledge in the fields of neuroscience, brain medicine, and computing \citep{Amunts:2016}. In 2014, Japan launched the Brain Mapping by Integrated Neurotechnologies for Disease Studies (Brain/MINDS), a program to develop innovative technologies for elucidating primate neural circuit functions \citep{Okano:2015}. China followed in 2016 with the China Brain Project (CBP), a 15 year program targeting the neural basis of human cognition \citep{Poo:2016}. Canada \citep{Jabalpurwala:2016}, South Korea \citep{Jeong:2016}, and Australia \citep{Australian:2016} followed suit,  launching their own brain programs in the late 2010s.
 
By nature and  historical precedence, convergence tends to operate on the frontier of science. In the 2010s, brain science was declared the new research frontier  \citep{Quaglio:2017}  promising health and behavioral applications \citep{Eyre:2017}. 
Intensification of brain research has been taking place against a backdrop of an increasingly globalized, interconnected and online scientific commons. This stands in sharp contrast to the nationally unipolar and offline backdrop of the MP and even the HGP. Moreover, the brain funding programs were designed to act as behavioral incentives in an scientific  marketplace, aimed at bringing together diverse scholars and ideas. However, despite being oriented around the  compelling structure-function brain problem, there were few guidelines on how to configure scholarly expertise to address the  brain challenge. 
As such, these characteristics render brain research  a ``live experiment'' in the international evolution of the convergence paradigm. 

 Accordingly,  we apply data-driven methods to reconstruct  the brain science ecosystem as a way to capture the  contemporary ``pulse'' of  convergence, explored through a progressive series of research questions regarding its prevalence, anatomy and scientific impact. Given the pervasive funding  championing the HBS challenge, we further analyze how the trajectory of HBS convergence  has been impacted by the ramp-up of flagship funding initiatives oriented around the world.
While previous work  explored the role of cross-disciplinary collaboration  in the Human Genome Project \citep{Petersen:2018}, here we extend that framework to  differentiate between (a) the disciplinary diversity of the research team and  (b) the  topical diversity of their research -- two alternative means of cross-domain integration.  We refer to the former as disciplinary diversity and to the latter as  topical diversity. We leverage existing taxonomies -- in the case of   disciplines, using the \emph{Classification of Instructional Program} (CIP)  system developed by the U.S. National Center for Education Statistics; and for topics using Medical Subject Heading (MeSH)  ontology developed by the U.S. National Library of Medicine disciplines -- to  distinguish mono-domain versus cross-domain activity. 
Accordingly, we  classify   HBS  research according to four integration types defined by a mono-/cross- \{discipline $\times$ topic\} domain decomposition.

In a highly competitive and open science system with multiple degrees of freedom, our motivating hypothesis is that more than one operational  cross-domain integration mode  is likely to emerge.
With this in mind, we identify five research questions (RQ) addressed in each figure in series.
 The first (RQ1) regards  how to define  convergence,  which we address by developing a  typological framework, one that is  generalizable  to other frontiers of biomedical science, and is relevant to the evaluation of multiple billion-dollar HBS flagship  projects around the world. 
 The second (RQ2) regards the status and impact of brain science convergence: Have HBS  interfaces have developed to the point of  sustaining fruitful cross-disciplinary knowledge exchange? Does the increasing prevalence of teams adopting convergent approaches  correlate with higher scientific  impact research? 
 RQ3 addresses whether convergence is evenly distributed across HBS subdomains? And what  empirical combinations of distinct subject areas (knowledge) and disciplinary expertise (people)  are overrepresented in convergent research? 
RQ4  follows by seeking to identify whether convergence is evenly distributed over time and geographic region? 
And finally, RQ5: does the propensity to pursue convergence science or does the citation impact of convergence science  depend on the convergence mode?  To address this question, we implement hierarchical regression models that differentiate between three convergence modes: research involving cross-disciplinary collaboration, cross-subject-area exploration, or both. 
Given the lucrative nature of flagship funding initiatives, we  hypothesize that the  ramp-up of HBS flagships   correlates with shifts in the prevalence and relative impact of research adopting  these different convergence modes. 

Our results  identify timely and relevant science policy implications. Given contemporary emphasis around accelerating breakthrough discovery  \citep{helbing2012accelerating} by way of strategic research team configurations \citep{borner2010multi},   convergence science originators called for  cross-disciplinary approaches integrating distant disciplines \citep{NRC:2014}.
Instead, our analysis reveals that HBS teams recently tend to integrate diverse topics without necessarily integrating appropriate disciplinary expertise -- an approach we identify as a {\it convergence shortcut}. 

\begin{SCfigure*}
\includegraphics[width=0.7\textwidth]{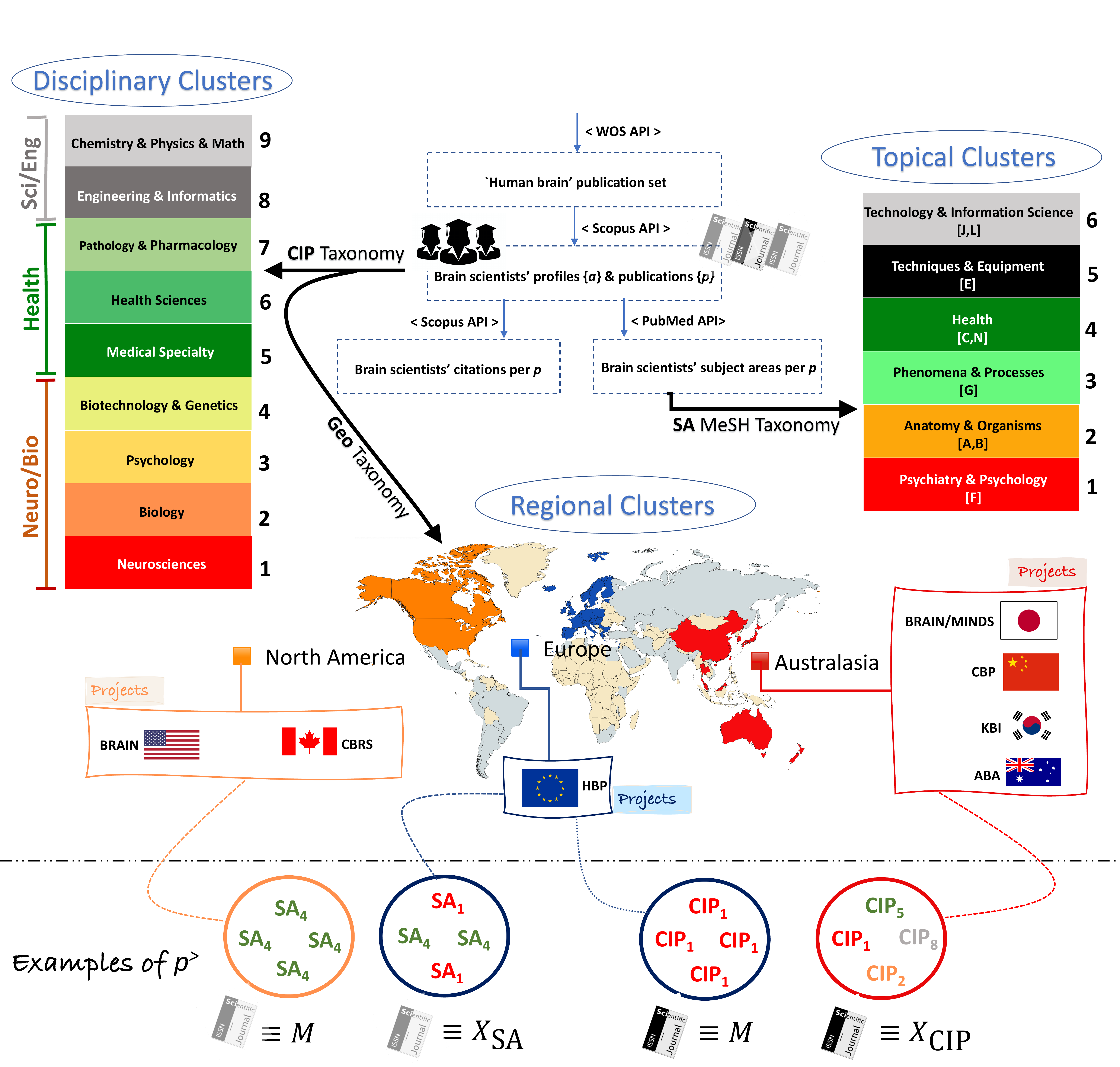}
 \caption{  \label{Figure1.fig} {\bf Data collection and classification schemes.}
 The upper part of the figure shows the data generation mechanism along with the resulting topical (SA) and disciplinary (CIP) clusters. The middle part of the figure shows on the world map regional clusters pertaining to three  large HBS funding initiatives -- North America (NA), Europe (EU), and Australasia (AA).
The lower part of the figure shows an example of how all three categorizations are operationalized for analytic purposes. Circles represent four research articles with authorship from distinct regions. The articles feature different keyword (SA) or disciplinary (CIP) category mixtures assigned one of two diversity measures: mono- $(M)$ and cross-domain $(X)$.}
\end{SCfigure*}

\vspace{-0.25in}
\section*{Theoretical background}
\vspace{-0.2in}
\noindent This work contributes to several literature streams, including the quantitative analysis of recombinant search and innovation  \citep{fleming2001recombinant,fleming2004science,youn2015invention}; cross-domain  integration \citep{fleming2004perfecting,Leahey_Sociological_2014,Petersen:2018}; cross-disciplinarity as a strategic team configuration  \citep{cummings2005collaborative,Petersen:2018} facilitated by division of labor across teams of specialists and generalists \citep{rotolo2013does,melero2015renaissance,teodoridis2018understanding,haeussler2020division}; and science of science \cite {Fortunato_2017_Science} and science policy  \citep{fealing_science_2011} evaluation of  convergence science  \citep{committee1900facilitating,Roco:2013,NRC:2014}.

Efficient long-range exploration facilitated by multi-disciplinary teams   is a defining  value proposition of convergence science  \citep{NRC:2014}, and  provides a testable mechanism underlying the  increased likelihood of  large team science producing  high-impact research \citep{Wuchty:2007}. 
Hence, the emergence and densification  of  cross-domain interfaces   are likely to increase the  potential for breakthrough discovery by catalyzing recombinant  innovation \citep{fleming2001recombinant}, which effectively expands the solution space accessible to problem-solvers.  It then follows that certain   configurations are  likely to amplify the effectiveness of recombinant innovation. Adapting a triple-helix model of medical innovation \citep{petersen2016triple},   recombinant innovation manifests from integrating expertise around the  three   dimensions of supply, demand and technological capabilities: (i) the  fundamental biology domain that supplies a theoretical understanding of the anatomical structure-function relation, (ii) the health domain that identifies demand for  effective science-based solutions, and (iii)  the techno-informatics domain which  develops  scalable products, processes and services to facilitate matching supply from (i) with  demand from (ii) \citep{BMConvergence_2021}.

In order to overcome the challenges of selecting new strategies from the vast number of possible combinations, prior research finds that innovators are more  likely to succeed by way of exploiting their own  local expertise \citep{fleming2001recombinant} rather than individually exploring distant configurations by way of internal expansive learning \citep{Engestrom:2010}. Extending  this argument, exploration at unchartered multidisciplinary interfaces is likely to be more successful when integrating  knowledge across a team of experts   from different domains, thereby  hedging against recombinant uncertainty underlying the exploration process \citep{fleming2004perfecting}. 

A  complementary argument for convergence derives from the advantage of diversity for harnessing collective intelligence and identifying  successful hybrid strategies \citep{page2008difference}. Recent work provides additional empirical support  for the competitive advantage of diversity, using  cross-border mobility  \citep{petersen2018mobility} as an instrument for social capital disruption  to  identify the  positive role of research topic and collaborator diversity.

\vspace{-0.25in}
\section*{Data collection and notation}
\vspace{-0.2in}
{\bf Figure \ref{Figure1.fig}} shows the multiple sources combined in our  study, which integrates publication and author data from Scopus, PubMed, and the Scholar Plot web app \citep{majeti_scholarplot_2020} (see  Supplementary Information (SI) {\it Appendix S1} for detailed  description). In total, our data sample spans 1945-2018 and consists of  655,386 publications derived from 9,121 distinct Scopus Author profiles, to which we apply  the following variable definitions and subscript conventions  to capture both   article- and scholar-level information. At the article level, subscript $p$ indicates publication-level information such as publication year, $y_{p}$; the number of coauthors,  $k_{p}$; and the number of keywords, $w_{p}$.  Regarding the temporal dimension, a superscript $>$ (respectively, $<$) indicates data belonging to the 5-year ``post'' period 2014-2018 (5-year ``pre'' period 2009-2013), while $N(t)$  represents the  total number of articles published in year $t$. Regarding proxies for scientific impact, we obtained  the number of citations  $c_{p,t}$ from Scopus, which are counted through late 2019. Since nominal citation counts suffer from systematic temporal bias, we use a normalized   citation measure, denoted by $z_{p}$ (see  {\it Methods -- Normalization of Citation Impact}). Regarding author-level information, we use the index $a$ -- e.g. we denote the academic age  measured in years since a scholar's first  publication   by $\tau_{a,p}$. 

 To address RQ1 we classified research according to three category systems indicative of  topical,  disciplinary and regional clusters. The first category system captures research topic  clusters grouped into Subject Areas (SA);  counts for each  article are represented by a vector with 6 elements, $\overrightarrow{SA}_{p}$, each corresponding to  top-level \emph{Medical Subject Heading} (MeSH) categories   implemented by PubMed, which are indicated by the letters in brackets: (1) Psychiatry \& Psychology [F],  (2) Anatomy \& Organisms [A,B],  (3) Phenomena \& Processes [G],  (4) Health [C,N], (5) Techniques \& Equipment [E], and (6) Technology \& Information Science [J,L]; notably, regarding the  {\it structure-function} problem that is a fundamental  focus in much of  biomedical science, category (2) represents the domain of {\it structure} while  (3) represents  {\it function}. The variable $N_{SA,p}$  counts the total number of SA categories present in a given article, with min value 1 and max value 6. 
 
The second taxonomy identifies disciplinary clusters determined by author departmental affiliation, which we  categorized according to \emph{Classification of Instructional Program} (CIP) codes. Article-level  CIP category counts are represented by $\overrightarrow{CIP}_{p}$, with 9 elements pertaining to the following categories: (1) Neurosciences,   (2) Biology,   (3) Psychology,   (4) Biotech. \& Genetics,  (5) Medical Specialty,  (6) Health Sciences, (7) Pathology \& Pharmacology,  (8) Engineering \& Informatics, and  (9) Chemistry \& Physics \& Math. The variable $N_{CIP,p}$  counts the total number of CIP categories present in a given article, with min value 1 and max value 9;  {\it Methods} and SI {\it Appendix S1} offer more details.

The third taxonomy captures the broad regional scope of each research article team determined by each Scopus author's affiliation location, and represented by the vector $\overrightarrow{R}_{p}$ which has 4 elements representing North America, Europe, Australasia, and rest of World.  See {\bf Fig. \ref{FigureS2.fig}} for the composition  of SA and CIP clusters, and  SI {\it  Appendix S1} for additional description of how these classification systems are constructed. {\bf Figure \ref{FigureS3.fig} (Fig. \ref{FigureS4.fig})} shows the frequency of each SA (CIP) category and the pairwise  frequency  of all $\{SA,SA\}$ $(\{CIP,CIP\})$ combinations
over the 10-year period centered on 2014, along with their relative changes after 2014;
See  {\it SI Appendix S2-S3} for discussion of the relevant changes in SA and CIP categories after 2014. 

We represent the collection of article features  by $\vec{F}_{p} \equiv \{\overrightarrow{SA}_{p},\overrightarrow{CIP}_{p},\overrightarrow{R}_{p}\}$. As indicated in {\bf Fig. \ref{Figure1.fig}}, based upon the distribution of types tabulated as counts across vector elements,  an article is  either cross-domain, representing a diverse mixture of types  denoted by $X$; or  mono-domain, denoted by $M$.   We use a generic operator notation to specify how articles are classified as $X$ or $M$,  The objective criteria of the feature operator $O$ is specified by its subscript: for example $O_{SA}(\vec{F}_{p})$ yields one of two values: $X_{SA}$ or $M$; similarly, $O_{CIP}(\vec{F}_{p})=X_{CIP}$ or $M$. Note that all scholars map onto a single CIP, hence solo-authored research articles are by definition classified by $O_{CIP}$ as $M$.
 While we acknowledge that is possible for a scholar to have significant expertise in two or more domains, we do not account for this duplicity, as it is likely to occur at the margins; hence, the home department CIP represents the scholar's principle domain of expertise.
We also classify articles  featuring both $X_{SA}$ and $X_{CIP}$  as  $O_{SA\&CIP}(\vec{F}_{p})=X_{SA\&CIP}$ (and otherwise $M$).

To complement these categorical measures, we also developed a scalar  measure of an article's cross-domain diversity  (see {\it Materials \& Methods  -- Measuring cross-domain diversity} for additional details). By way of example, consider the vector $\overrightarrow{SA}_{p}$ (or $\overrightarrow{CIP}_{p}$) which tallies the SA (or CIP counts) for a given article $p$ published in year $t$. We apply the outer tensor product  $\overrightarrow{SA}_{p} \otimes \overrightarrow{SA}_{p}$  (or $\overrightarrow{CIP}_{p} \otimes \overrightarrow{CIP}_{p}$)   to represent all  pairwise co-occurrences  in a  weighted matrix ${\bf D}_{p}(\vec{v}_{p})$ (where $\vec{v}_{p}$ represents a generic category vector; see {\it SI Appendix S4} for examples of the outer tensor product). The sum of elements in this co-occurrence  matrix are normalized to unity so that each ${\bf D}_{p}(\vec{v}_{p})$ contributes equally to  averages computed across all articles from a given year or period. Since the off-diagonal elements represent cross-domain combinations, their relative weight given by $f_{D,p}=1-\mathrm{Tr}({\bf D}_{p}) \in [0,1)$ is a straightforward Blau-like measure of variation and disparity  \citep{harrison2007s}.  

\begin{figure*}[!t]
\centering{\includegraphics[width=0.99\textwidth]{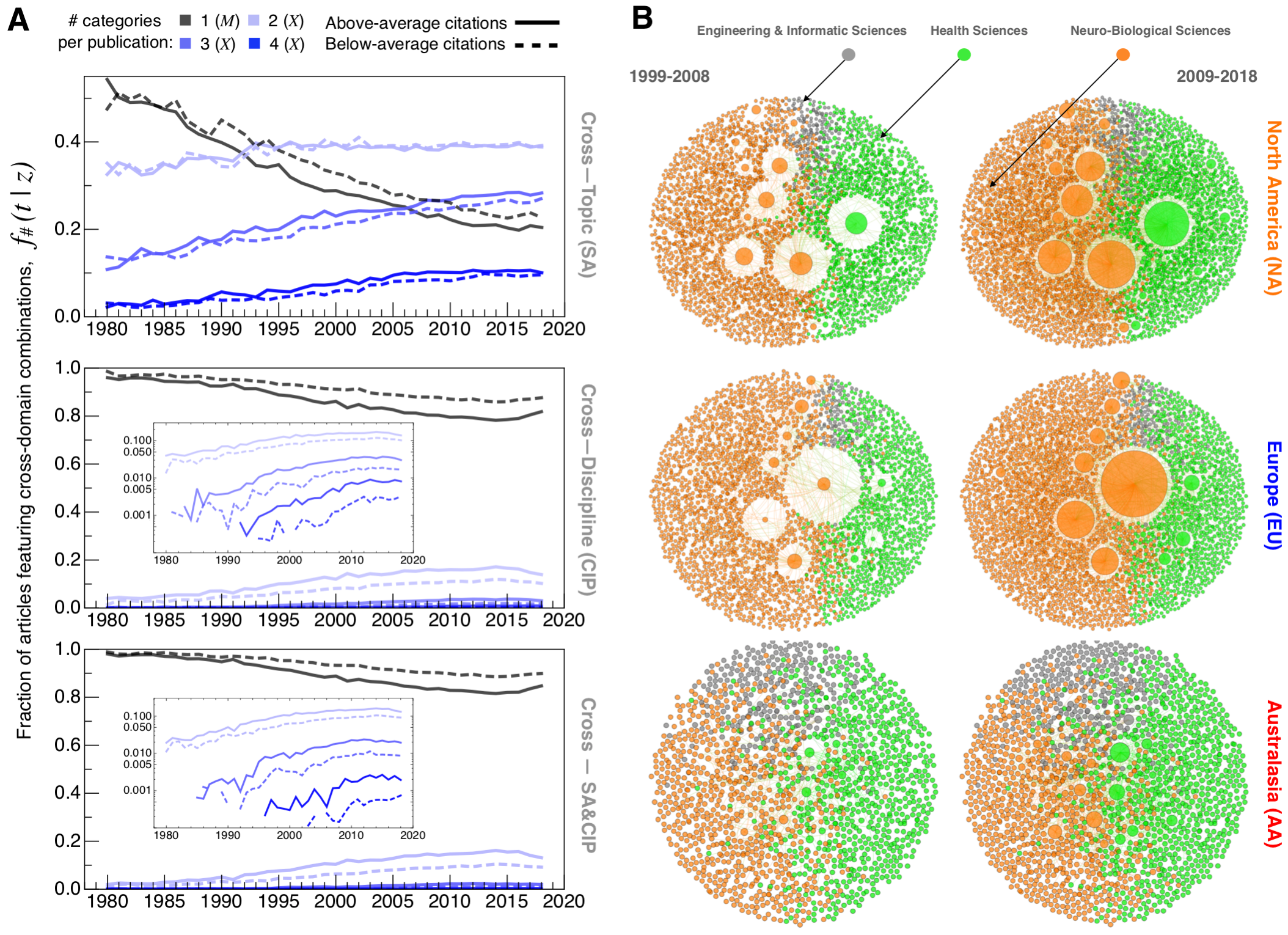}}
 \caption{  \label{Figure2.fig} {\bf Trends in cross-domain scholarship in Human Brain Science.} 
 (A) Fraction $f_{\#}(t \vert z)$ of articles published each year $t$ that feature a particular number ($\#$) of categories. Articles are split into an above-average citation subset ($z_{p}>0$) and below-average citation subset ($z_{p}<0$). Upper panel: Articles categorized by SA.  Middle panel: Articles categorized by CIP; subpanel shows data on logarithmic y-axis; Lower panel: Articles categorized by both SA and CIP.
 Distinguishing frequencies by citation group indicates higher levels of cross-domain combinations among research articles with higher scientific impact --  for both SA and CIP. However, cross-domain activity levels are visibly higher for SA than for CIP, indicating higher barriers to boundary-crossing arising from mixing different scholar expertise. (B) Snapshots of the collaboration network at 10-year intervals indicating researcher population sizes by region, and the densification of convergence science at cross-disciplinary interfaces. Nodes (researchers) are  sized according to the number of  collaborators (link degree) within  each  time window. }
\end{figure*}

\vspace{-0.25in}
\section*{Results}
\vspace{-0.2in}
\subsection*{Descriptive Analysis}
\vspace{-0.15in}
\noindent  \textbf{Increasing prevalence of cross-domain  science.} With the continuing shift towards  large team science \citep{Wuchty:2007,TeamGrowth2014,Pavlidis:2014,petersen2014quantitative}, one might expect a similar shift in the multiplicity of domains spanned by modern research teams -- but to what degree?  {\bf Figure \ref{Figure2.fig}}(A) addresses RQ2 by showing the frequencies of mono-domain ($M$) research articles versus cross-domain articles ($X$) in our HBS sample. Articles were separated into above- and below-average citation impact ($z$) for each publication-year cohort ($t$), and within each of these two subsets we calculated the fraction $f_{\#}(t \vert z)$ of articles containing combinations across $\# = 1, 2, 3 \ \text{and} \ 4 \ \text{categories}$. 
The fraction of mono-domain articles is trending downward, which we observe  for both research  topics (SA) and authors' disciplinary affiliations (CIP). The decline is much more steep for SA   than for CIP. Correspondingly, cross-domain articles have become increasingly prevalent, in particular for SA. For both SA and CIP  the two-category mixtures dominate the three- and four-category mixtures in frequency, in sequence. Accordingly, in the sections that follow  we do not distinguish between cross-domain articles with different  $\#$. 

As a first indication of the comparative advantage associated with $X$,  we observe a robust inequality $f_{\#}(t \vert z >0) > f_{\#}(t \vert z <0)$ for cross-domain research ($\# \geq 2$), meaning a higher frequency of cross-domain combinations is observed among articles with higher impact. Contrariwise, in the case of mono-domain research the opposite phenomenon occurs, $f_{1}(t \vert z >0) < f_{1}(t \vert z <0)$. Taking into consideration temporal trends, these robust patterns indicate a  faster depletion of impactful mono-domain articles, coincident with an increased prevalence of impactful research drawing upon  integrative recombinant innovation.\\

\begin{figure*}[!t]
\centering{\includegraphics[width=0.89\textwidth]{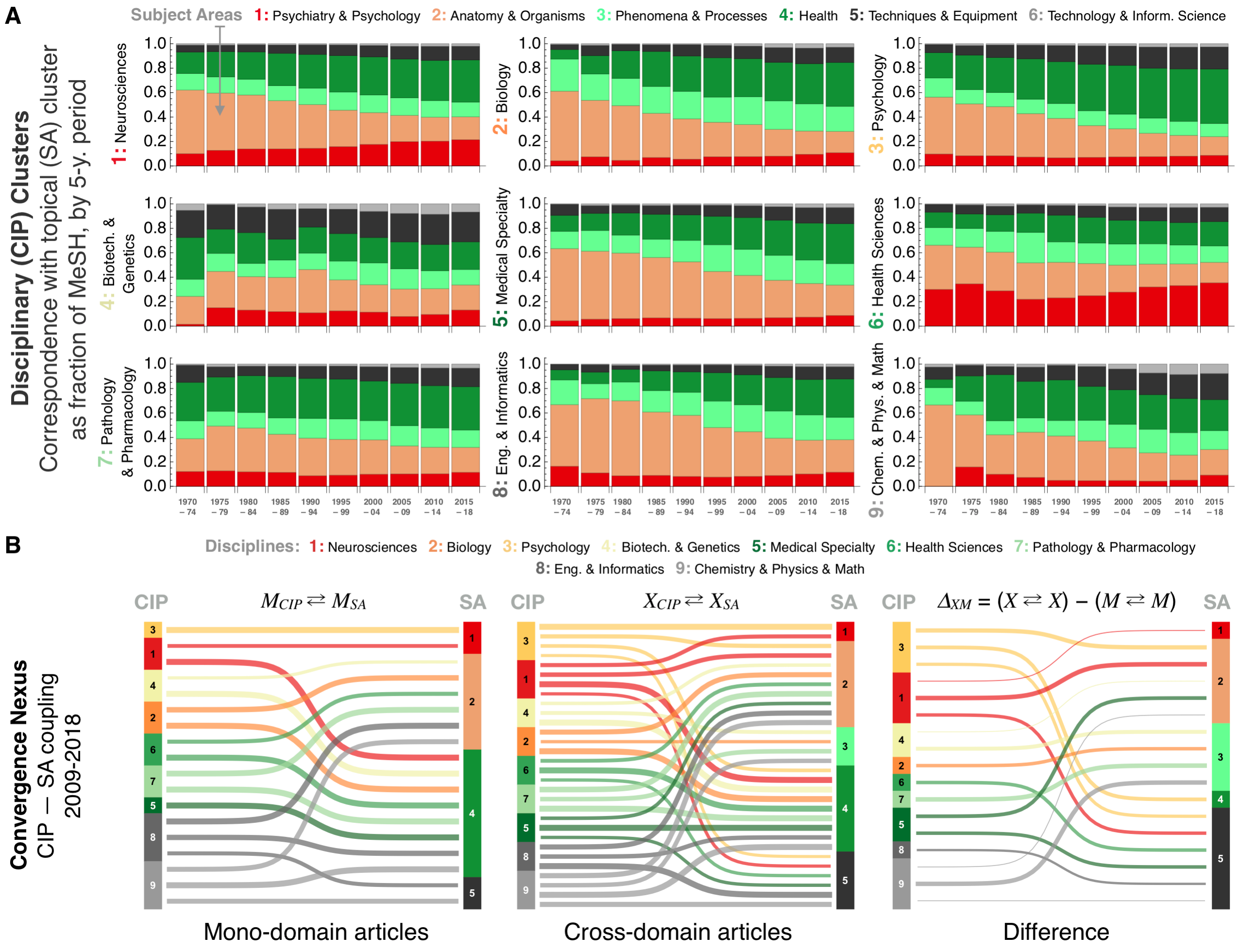}}
 \caption{  \label{Figure3.fig} {\bf Evolution of SA boundary-crossing within and across disciplinary clusters.} 
 (A) SA composition of HBS research within disciplinary (CIP) clusters. Each subpanel represents articles published by researchers  from a given CIP cluster, showing the fraction of article-level MeSH belonging to each SA, shown over 5-year intervals across the period 1970-2018. The increasing prominence of SA 5 \& 6 in nearly all domains, in particular CIP 4 (Biotech. \& Genetics) indicates the critical role of  informatic capabilities in facilitating  biomedical convergence science \citep{BMConvergence_2021}.
   (B) Empirical  CIP-SA association networks calculated for non-overlapping sets of  mono-domain $(M_{CIP}\rightleftarrows M_{SA})$  and cross-domain ($X_{CIP}\rightleftarrows X_{SA}$) articles, and based upon the {\it Broad} configuration. The difference  between these two bi-partite networks ($\Delta_{XM}$) indicates the emergent research channels that are facilitated by simultaneous $X_{CIP}$ and  $X_{SA}$ boundary crossing -- in particular integrating SA 2 with 3 (i.e. the structure-function nexus) facilitated by teams combining disciplines  1, 2, 4 and 9.} 
\end{figure*}

\noindent \textbf{Recombinant innovation at the convergence nexus.} Comprehensive analysis of biomedical science indicates that convergence has largely been mediated around the integration of modern techno-informatics capabilities \citep{BMConvergence_2021}. 
Yet within any domain, in particular HBS, the questions remains as to the development of a functional  nexus that sustains and possibly even accelerates high-impact discovery by both expanding the number of possible functional expertise configurations and supporting rich cross-disciplinary exchange of new knowledge and best practices.
The robust inequality $f_{\#}(t \vert z >0) > f_{\#}(t \vert z <0)$ provides support at the aggregate level, but does not lend any structural evidence.

 To further address  RQ2, {\bf Fig. \ref{Figure2.fig}}(B) illustrates the composition of the HBS convergence nexus, showing  integration of cross-disciplinary expertise across three broad yet distinct biomedical domains. Shown are the populations of HBS researchers by region, represented as collaboration networks compared  over two non-overlapping 10-year intervals to indicate dynamics. Each node represents a researcher, colored according to three disciplinary CIP superclusters: (i) neuro-biological sciences (corresponding to CIP 1-4), (ii) health sciences (CIP 5-7), and (iii) engineering \& information sciences (CIP 8-9).
 Node locations are fixed  to facilitate  visual representation of network densification. Inter- and cross-regional comparison alludes to the  emergence and densification  of  cross-domain interfaces   (see also {\bf Fig. \ref{FigureS11.fig}}).
 Because the network layout  is determined by the underlying   structure, there is a high degree of clustering by node color,  emphasizing both the relative sizes of the subpopulations that are well-balanced across region and time, and also the convergent interfaces where cross-disciplinary collaboration and knowledge exchange are likely to catalyze.
  As such, these communities of expertise conjure the image of a  P\'{o}lya urn, whereby successful configurations reinforce the adoption of similar configurations.  
 
The links that span disciplinary boundaries are fundamental conduits across which scientists' strategic affinity for  exploration  \citep{rotolo2013does,foster2015tradition} is effected via cross-disciplinary collaboration that  brings ``together distinctive components of two or more disciplines'' \citep{nissani1995fruits,Petersen:2018}.
Our analysis of cross-disciplinary collaboration indicates that the fraction of articles featuring convergent collaboration have continued to grow over  the last two decades (see  {\bf Fig. \ref{FigureS11.fig}}).
In what follows we further distinguish between integration across neighboring \citep{Leahey_Sociological_2014} and  distant domains, with the latter   appropriately representing  convergence   \citep{committee1900facilitating,Roco:2013,NRC:2014}. \\

 \begin{figure*}[!t]
\centering{\includegraphics[width=0.99\textwidth]{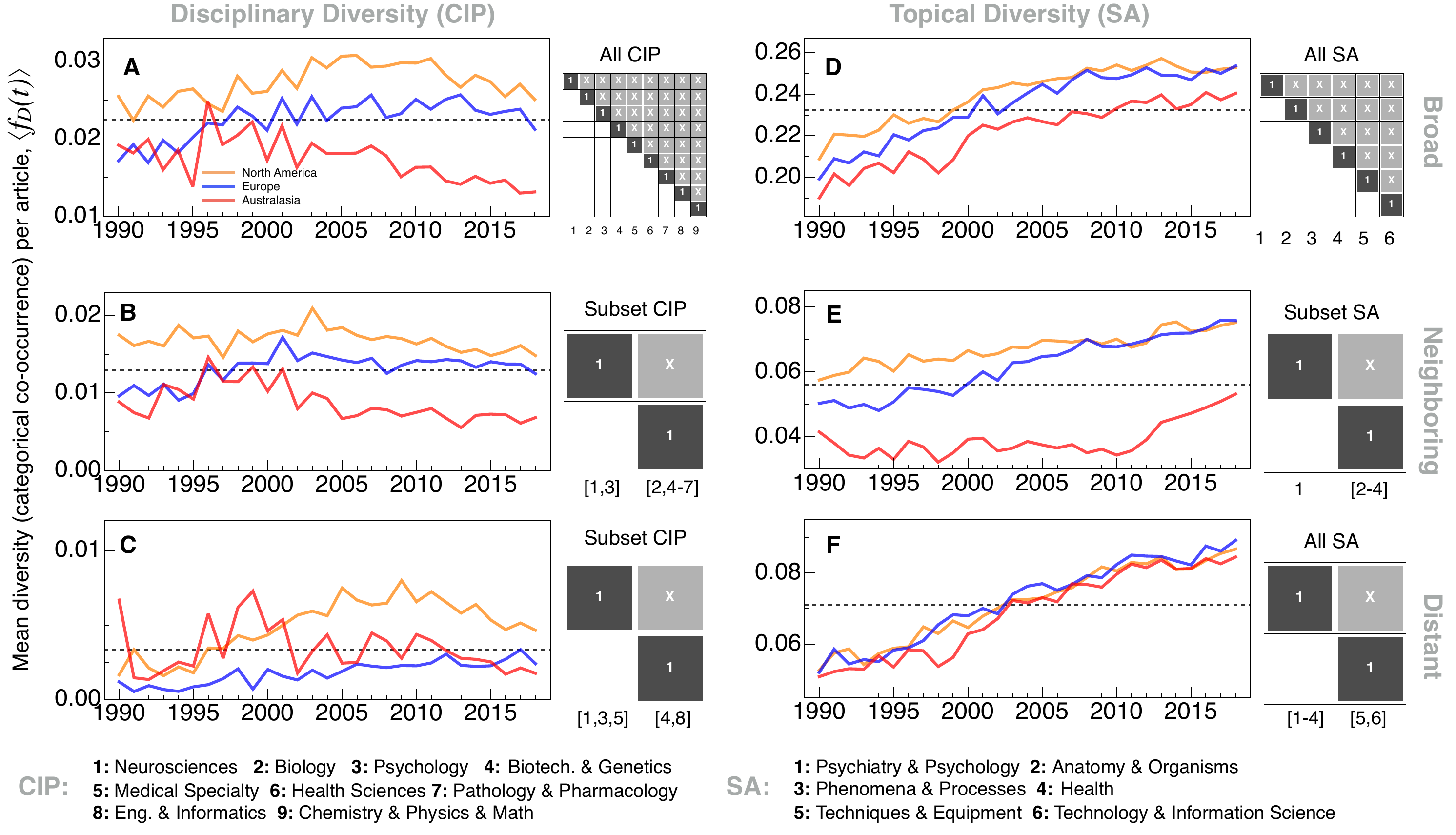}}
 \caption{    \label{Figure4.fig} {\bf Evolution of CIP and SA diversity in Human Brain Science research.} (A-F) Each $\langle f_{D}(t) \rangle$  represents the average article diversity measured as categorical co-occurrence, by geographic region: North America (orange), Europe (blue), and Australasia (red). Each matrix motif indicates the set of CIP or SA categories used to compute ${\bf D}_{p}$ in Eq. (\ref{Dmatrix});  categories included in brackets are considered in union. For example, panel (A) calculates $\langle f_{D,CIP}(t) \rangle$ across all 9 CIP categories; instead, panel (B) is based upon counts for two super-groups, the first consisting of the union of CIP counts for categories 1 and 3, and the second comprised of categories 2, 4, 5, 6 and 7. 
 (A,D) {\it Broad} diversity is calculated using all categories considered as separate domains; 
  (B,E)  {\it Neighboring}  represents the shorter-distance boundary across the neuro-psychological $\leftrightarrow$ bio-medical interface; 
  (C,F)  {\it Distant}  represents longer-distance convergence across the neuro-psycho-medical $\leftrightarrow$ techno-informatic interface.} 
\end{figure*}

\noindent \textbf{Cross-domain convergence of expertise (CIP)  and knowledge (SA).} In the context of the bureaucratic structure-function problem,   team assembly should be optimized   by strategically matching scholarly expertise and research topics to address the particular demands of a particular challenge. 
Hence, with 9 different disciplinary (CIP) domains historically faced with a variety of challenges, RQ3 addresses to what degree these domains differ in terms of their composition of targeted SA.
 {\bf  Fig. \ref{Figure3.fig}}(A) illustrates the evolution of  topical  diversity within and across each CIP cluster, revealing several common patterns. First, nearly all domains show a reduction in  research pertaining to  structure (SA 2), with the exception of Biotechnology \& Genetics, which was oriented around the structure-function problem from the outset. As such,  this domain features a steady balance between SA 2-5, while being an early adopter of techno-informatics concepts and methods (SA 6). Early balance around the innovation triple-helix \citep{petersen2016triple} may explain to some degree the longstanding success of the  genomics revolution, as the core disciplines of biology and computing were primed for a fruitful union  \citep{Petersen:2018}. 
Other HBS disciplinary clusters are also integrating techno-informatic capabilities, reflecting a widespread pattern observed across all of biomedical science  \citep{BMConvergence_2021}.

Which CIP-SA combinations are are overrepresented in boundary-crossing HBS research?
Inasmuch as mono-domain articles identify the topical boundary closely associated with individual disciplines,  cross-domain articles are useful for identifying otherwise obscured boundaries  that call for both $X_{CIP}$ and  $X_{SA}$ in combination.
We identified  these novel CIP-SA relations  by collecting   articles   that are purely mono-domain for both CIP and SA (i.e., those with $O_{CIP}(\vec{F}_{p})= O_{SA}(\vec{F}_{p})= M$) and a complementary non-overlapping subset of articles  that are simultaneously  cross-domain for both CIP and SA (i.e.,  $O_{SA\&CIP}(\vec{F}_{p})=X_{SA\&CIP}$). 

Starting with  mono-domain articles,   we identified the SA that are most frequently associated with each CIP category. Formally, this amounts to calculating the bi-partite network between CIP and SA, denoted by  $M_{CIP}\rightleftarrows M_{SA}$. These CIP-SA associations are calculated by averaging the $\overrightarrow{SA}_{p}$ for mono-domain articles from each CIP category, given by $\langle \overrightarrow{SA} \rangle_{CIP}$.  {\bf  Figure \ref{Figure3.fig}}(B) highlights  the most prominent  CIP-SA links  (see  SI {\it Appendix S5} for more details).
Likewise, we also calculated the bi-partite network $X_{CIP}\rightleftarrows X_{SA}$  using the subset of $X_{SA\&CIP}$ articles.

To identify the cross-domain frontier, we calculated the network difference  $\Delta_{XM} \equiv X_{CIP}\rightleftarrows X_{SA} -  M_{CIP}\rightleftarrows M_{SA}$, and plot  the links with positive values -- i.e. CIP-SA links that are over-represented in  $X_{CIP}\rightleftarrows X_{SA}$ relative to $M_{CIP}\rightleftarrows M_{SA}$.  
Results  identify   SA that are reached by way of cross-disciplinary teams. SA 2 (Anatomy and Organisms) and 3 (Phenomena \& Processes) representing  the structure-function problem, stand out as a potent convergence nexus accessible by teams combining  disciplines 1, 2, 4 and 9. 

A related key insight concerns the relative increase in SA integration   achieved by increased CIP diversity.
{\bf Figure \ref{FigureS12.fig}} compares the average number of SA integrated by teams with varying number of distinct CIP, $N_{CIP,p}$.  On average, mono-disciplinary teams ($N_{CIP,p}=1$) span 2.2 SA, whereas teams with $N_{CIP,p}=3$ span 19\% more SA, confirming that cross-disciplinary configurations are functional in achieving research breadth.

\vspace{-0.2in}
\subsection*{Quantitative Model}
\vspace{-0.15in}
\noindent   \textbf{Trends in cross-domain activity}.  To address the temporal and geographic parity associated with RQ4,  we define three types of cross-domain  configurations  -- {\it Broad}, {\it Neighboring}, and {\it Distant} --   defined according to a particular combination of SA and CIP categories featured by a given article.

{\it Broad} is the most generic  cross-domain  configuration, based upon combinations of any two or more SA  (or CIP) categories, and represented by our operator notation as $O_{SA}(\vec{F}_{p})=X_{SA}$ (or  $O_{CIP}(\vec{F}_{p})=X_{CIP}$, respectively).  {\it Neighboring} is the $X$ configuration that captures the neuro-psychological $\leftrightarrow$ bio-medical interface representing  articles that contain MeSH from SA (1) and also from  SA  (2, 3 or 4), represented summarily as $[1] \times [2-4]$); and for CIP, combinations containing CIP (1 or 3) and (2, 4, 5, 6 or 7), represented as $[1,3] \times [2,4-7]$. 
Articles featuring these configurations are represented using our operator notation as  $O_{{\text{Neighboring},SA}}(\vec{F}_{p}) = X_{\text{Neighboring},SA}$, $O_{{\text{Neighboring},CIP}}(\vec{F}_{p}) =X_{\text{Neighboring},CIP}$, or $O_{{\text{Neighboring},SA\&CIP}}(\vec{F}_{p}) =X_{\text{Neighboring},SA\&CIP}$; alternatively, articles not containing the specific category combinations are represented by  $M$.
  
{\it Distant} is the $X$ configuration that captures the neuro-psycho-medical $\leftrightarrow$ techno-informatic interface.
 The specific set of category combinations representing this configuration are  SA [1-4] $\times$ [5,6]; and for CIP, [1,3,5] $\times$  [4,8]; as above, articles  featuring (or not featuring) categories spanning these categories are represented by $X_{\text{Distant},SA}$ (belong to a counterfactual set indicated by $M$), $X_{\text{Distant},CIP}$ (resp., $M$), $X_{\text{Distant},SA\&CIP}$ (resp., $M$). 
By way of example, the bottom of {\bf Figure \ref{Figure1.fig}}  illustrates an article  combining SA 1 and 4, which is thereby classified as both $X_{SA}$ and $X_{\text{Neighboring},SA}$; and, an article featuring CIP 1,3,5,8, which is thereby both $X_{CIP}$ and $X_{\text{Distant},CIP}$.

To complement these categorical variables, we also developed a Blau-like measure of  cross-domain  diversity, given by $f_{D,p}$ (see Methods {\it Measuring cross-domain diversity}).
{\bf Figure \ref{Figure4.fig}} shows the trends in mean diversity $\langle f_{D}(t) \rangle$ for the  {\it Broad}, {\it Neighboring}, and {\it Distant} configurations. For each  configuration we provide a schematic motif illustrating the  combinations measured by ${\bf D}_{p}(\vec{v}_{p})$,  with diagonal components representing mono-domain articles (indicated by 1 on the matrix diagonal) and upper-diagonal elements capturing cross-domain combinations  (indicated by $X$). 
Comparing SA and CIP, there are higher diversity levels  for SA, in addition to a  prominent upward trend.
In terms of CIP,  {\bf Fig. \ref{Figure4.fig}}(A) indicates a decline in {\it Broad} diversity in recent years, with North America (NA) showing higher levels than Europe (EU) and Australasia (AA); these general patterns are also evident for {\it Neighboring}  diversity, see {\bf Fig. \ref{Figure4.fig}}(B).
 {\it Distant} CIP diversity shown in  {\bf Fig. \ref{Figure4.fig}}(C) indicates a recent decline for AA and NA, with NA peaking around 2009; contrariwise, EU shows a steady increases consistent with the computational framing of the Human Brain Project.

In contradistinction,  all three regions show steady increase irrespective of configuration in the case of SA diversity, consistent with scholars  integrating  topics without integrating scholarly expertise, possibly owing to differential  costs associated with each.
For both {\it Broad} and {\it Neighboring} configurations, NA and EU show remarkably similar levels of SA diversity above AA; however, in the case of {\it Neighboring},  AA appears to be catching up quickly since 2010, see {\bf Fig. \ref{Figure4.fig}}(D,E). In the case of {\it Distant}, all regions show steady increase that appears to be in lockstep for the entire period. 
See {\bf Figs. \ref{FigureS6.fig}-\ref{FigureS7.fig}} and SI  {\it  Appendix Text S6} for trends in SA and CIP diversity across additional   configurations.\\

\noindent   \textbf{Regression model --  propensity for and impact of $X$.} To address RQ5, we constructed   article-level and author-level panel data to facilitate measuring factors relating to SA and CIP diversity and shifts  related to the  ramp-up of   HBS flagship projects circa 2013 around the globe. 
To address these two outcomes, we modeled two dependent variables separately: In the first model the dependent variable is the propensity for  cross-domain research  (indicated   by $X$; depending on the focus around topics, disciplines or both, then $X$  is specified by  $X_{SA}$, $X_{CIP}$ or $X_{SA\&CIP}$). We use a Logit specification to model  the likelihood $P{(X})$.
In the second model the dependent variable is the article's scientific impact,  proxied by $c_{p}$. Building on previous efforts \citep{Petersen:2018,petersen2018mobility}, we  apply a  logarithmic transform to $c_{p}$ that facilitates removing the time-dependent trend in the location and scale of the underlying  log-normal citation distribution \citep{Radicchi:2008} (see {\it Methods -- Normalization of Citation Impact}).  {\bf Figure \ref{FigureS9.fig}} shows the covariation matrix between the principal variables of interest. \\

\noindent   \textbf{Model A: Quantifying the propensity for $X$ and the role of funding.} As defined, $O(\vec{F}_{p}) = X$ or $M$ is a two-state outcome variable with complementary likelihoods,    $P(X)+P(M)=1$. Thus,  we apply logistic regression to model the odds $Q\equiv \frac{P(X)}{P(M)}$, measuring the  propensity to adopt cross-domain configurations. 
We then estimate the annual growth in $P(X)$ by modeling the odds as $\log (Q_{p}) = \beta_{0} + \beta_{y} y_{p}+ \vec{\beta} \cdot \vec{x}$, where $\vec{x}$ represents the additional controls for  confounding sources of variation, in particular increasing $k_{p}$ associated with the growth of team science \citep{Wuchty:2007,TeamGrowth2014}.
See SI  {\it  Appendix Text S7}, in particular Eqns. (\ref{M1aSA})-(\ref{M1aSACIP}), for the full model specification; and,  {\bf  Tables \ref{TableS1.tab}-\ref{TableS3.tab}} for parameter estimates.

\begin{figure*}[!t]
\centering{\includegraphics[width=0.99\textwidth]{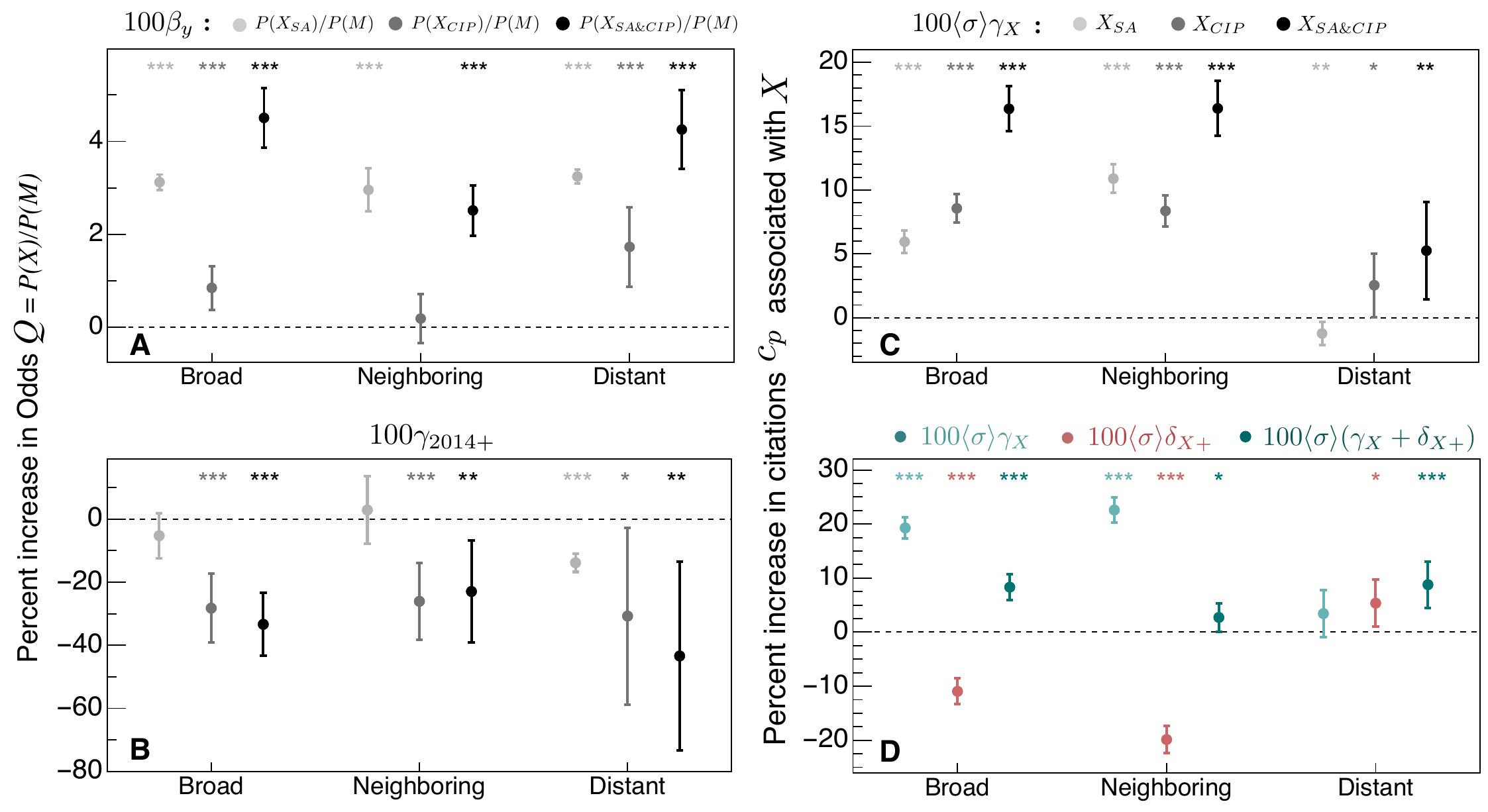}}
 \caption{ \label{Figure5.fig} {\bf Propensity for $X$ and citation impact attributable to cross-domain activity at the article level.}  
 (A) Annual growth rate  in the likelihood $P(X)$ of  research having cross-domain attributes represented generically by $X$.
 (B) Decreased likelihood $P(X)$ after 2014.
(C) Citation premium estimated as the percent increase in  $c_{p}$ attributable to cross-domain mixture $X$, measured relative to mono-domain (M) research articles representing the counterfactual baseline. Calculated using a researcher fixed-effect model specification which accounts for time independent individual-specific factors; see Tables \ref{TableS4.tab}-\ref{TableS5.tab} for full model estimates. 
Note that ``Broad'' corresponds to $X_{SA}$,$X_{CIP}$,$X_{SA\&CIP}$; ``Neighboring''  corresponds to $X_{\text{Neighboring},SA}$,$X_{\text{Neighboring},CIP}$,$X_{\text{Neighboring},SA\&CIP}$; and ``Distant''  corresponds to $X_{\text{Distant},SA}$,$X_{\text{Distant},CIP}$,$X_{\text{Distant},SA\&CIP}$.
 (D)  Difference-in-Difference ($\delta_{X+}$) estimate of the ``Flagship project effect'' on the citation impact of cross-domain research. 
 Shown are point estimates with 95\% confidence interval. Asterisks above each estimate indicate the associated $p-$value level:  $^{*} \ p<0.05$, $^{**} \ p<0.01$, $^{***} \ p<0.001$.
}
\end{figure*}

 Summary results shown in {\bf Fig. \ref{Figure5.fig}}(A) indicate a roughly 3\% annual growth in  $P(X_{SA})$,  consistent with descriptive trends shown in {\bf Fig. \ref{Figure2.fig}}. In contradistinction,  growth rates  for $P(X_{CIP})$  are generally smaller, indicative of the additional barriers to integrating individual expertise as opposed to  just combining different research topics.
 In the case of $P(X_{SA\&CIP})$, the growth rate is  higher for {\it Distant}, where the need for cross-disciplinary expertise cannot be short-circuited as easily as in {\it Neighboring}.

A relevant dimension of RQ5 is how HBS projects have altered the propensity for $X$. Hence, we added an  indicator variable $I_{2014+}$ which takes the value 1 for articles with $y_{p}\geq 2014$ and 0 otherwise. 
{\bf Figure  \ref{Figure5.fig}}(B) indicates significant decline  in $P(X)$ for $X_{CIP}$ and $X_{SA\&CIP}$ for each configuration on the order of -30\%; this result is consistent with the recent increase in $f_{1}(t \vert z)$ visible in  {\bf Fig. \ref{Figure2.fig}}(B). \\

\noindent \textbf{Model B: Quantifying the citation premium associated with $X$ and  funding.}  We model the normalized citation impact  $z_{p} = \alpha_{a} + \gamma_{X_{SA}} I_{X_{SA,p}}+ \gamma_{X_{CIP}} I_{X_{CIP,p}}+ \vec{\beta} \cdot \vec{x}$, where $\vec{x}$ represents the additional control variables and $ \alpha_{a}$ represents an author fixed-effect to account for unobserved time-invariant  factors specific to each researcher. The primary test variables are $I_{X_{SA,p}}$ and $I_{X_{CIP,p}}$,  two binary factor variables with $I_{X_{SA,p}}$= 1 if $O_{SA}(\vec{F}_{p})=X_{SA}$ and 0 if $O_{SA}(\vec{F}_{p})=M$,  defined similarly  for CIP. To  distinguish estimates by  configuration, for {\it Neighboring} 
we specify  $I_{X_{\text{Neighboring},SA}}$ and $I_{X_{\text{Neighboring},CIP}}$, with similar notation for {\it Distant}. Full model estimates are shown  
in {\bf  Tables \ref{TableS4.tab}} - {\bf  \ref{TableS5.tab}}.

{\bf  Figure \ref{Figure5.fig}}(C) summarizes the model estimates -- $\gamma_{X_{SA}}$, $\gamma_{X_{CIP}}$ and $\gamma_{X_{SA\&CIP}}$ --  quantifying the citation premium attributable to $X$. To translate the effect on  $z_{p}$ into the associated citation premium in $c_{p}$, we calculate the  percent change $100 \Delta c_{p}/c_{p}$  associated with a shift in $I_{X,p}$  from 0 to 1.  Observing that $\sigma_{t} \approx \langle \sigma \rangle =  1.24$ is approximately constant over the period 1970-2018 and due to the property of logs, the citation percent change is given by $100 \Delta c_{p}/c_{p}  \approx 100 \langle \sigma \rangle  \gamma_{X}$, (see SI  {\it  Appendix S7B}).

Our results indicate a robust statistically significant positive relationship between cross-disciplinarity ($X_{CIP}$) and citation impact, consistent with the effect size in a different case study of the genomics revolution \citep{Petersen:2018}, which supports the generalizability of our findings to other convergence frontiers. To be specific, we calculate a 8.6\% citation premium for the {\it Broad} configuration  ($\gamma_{X_{CIP}} = 0.07$; $p<0.001$), meaning that the average cross-disciplinary publication is more highly cited than the average mono-disciplinary publication. We calculate a smaller 5.9\% citation premium associated with $X_{SA}$ ($\gamma_{X_{SA}} = 0.05$; $p<0.001$). Yet  the effect associated with articles featuring $X_{CIP}$ and $X_{SA}$ simultaneously is considerably larger  (16\% citation premium; $\gamma_{X_{SA\&CIP}} = 0.13$; $p<0.001$), suggesting an additive effect.

Comparing results for the {\it Neighboring} configuration to the baseline estimates for  {\it Broad},  the citation premium is relatively larger for $X_{SA}$  (11\% citation premium; $\gamma_{X_{\text{Neighboring},SA}} = 0.088$; $p<0.001$)  and roughly the same for $X_{CIP}$ and $X_{SA\&CIP}$. This result reinforces our findings regarding the convergence ``short-cut'' (when $X_{CIP}$ is absent), indicating that this approach is more successful when integrating domain knowledge  across shorter distances, consistent with innovation theory \citep{fleming2001recombinant}.

The configuration most representative of convergence is {\it Distant}, which compared to {\it Broad} and {\it Neighboring} features smaller effect size for $X_{SA\&CIP}$ (5.2\% citation premium; $\gamma_{X_{\text{Distant},SA\&CIP}} = 0.04$; $p<0.001$). 
 The reduction in $\gamma_{X_{\text{Distant},SA\&CIP}}$ relative to values for {\it Broad} and {\it Neighboring} configurations likely reflects the challenges bridging communication, methodological and theoretical gaps across the {\it Distant}  neuro-psycho-medical $\leftrightarrow$ techno-informatic interface. 
 More interestingly, this configuration is distinguished by a negative $X_{SA}$  estimate, indicating that the convergence shortcut yields less-impactful research than mono-domain research.
Nevertheless, it is notable that for this convergent configuration, there is a clear hierarchy indicating the superiority of cross-disciplinary  collaboration approaches to integrating research across distant domains.

As in the Article-level model, we also tested for shifts in the citation premium attributable to the advent of Flagship HBS project funding using a similar DiD approach. {\bf  Figure \ref{Figure5.fig}}(D) shows the citation premium $\gamma_{X_{SA\&CIP}}$ for articles published prior to 2014, and the difference  $\delta_{X+}$ corresponding to the added effect for articles published after 2013. 
For {\it Broad}  and {\it Distant} we observe $\delta_{X+}<0$, indicating a reduced citation premium for  post-2013 research. By way of example for the {\it Broad} configuration: whereas cross-domain articles published prior to 2014 show a 19\% citation premium ($\gamma_{X_{SA\&CIP}} = 0.15$; $p<0.001$), those published after 2013 have just a 19\%-11\% = 8\% citation premium ($\delta_{X_{SA\&CIP}+} = -0.09$; $p<0.001$). The reduction of the citation premium is even larger for  {\it Neighboring} ($\delta_{\text{Neighboring},X_{SA\&CIP}+} = -0.16$; $p<0.001$). Yet for  {\it Distant},  we observe a different trend -- research combining both $X_{SA}$ and $X_{CIP}$ simultaneously  has advantage over those with just $X_{CIP}$ or  $X_{SA}$, in that order ($\delta_{\text{Dist.},X_{SA\&CIP}+} =  0.04$; $p = 0.016$; 95\% CI = [.01,  .08]).

We briefly summarize coefficient estimates for the other control variables.
Consistent with prior research on   cross-disciplinarity  \citep{Petersen:2018}, we observe a positive relationship between team-size and citation impact ($\beta_{k} = 0.415$; $p<0.001$), which translates to a $\langle \sigma \rangle \beta_{k}  \approx $  0.5\% increase in citations associated with a 1\% increase in team size (since $k_{p}$ enters in log in our specification).
We also observe a positive relationship for topical breadth ($\beta_{w} = 0.03$; $p<0.001$), which translates to a much smaller $\langle \sigma \rangle \beta_{w}  \approx $  0.04\% increase in citations associated with a 1\% increase in the number of major MeSH  headings.
 And finally, regarding the career life-cycle, we observe  a negative  relationship with increasing career age ($\beta_{\tau} = -0.011$; $p<0.001$)  consistent with prior studies \citep{Petersen:2018}, translating to a $100 \langle \sigma \rangle \beta_{\tau}  \approx  $  -1.3\% decrease in $c_{p}$ associated with every additional career year. 
 See {\bf Tables \ref{TableS4.tab}-\ref{TableS5.tab}} for the full set of model parameter estimates.

\vspace{-0.2in}
\subsection*{Behind the Numbers}
\vspace{-0.15in}
Further qualitative inspection of prominent research articles in this category identifies four key convergence themes associated with past or developing breakthroughs:  

{\it Magnetic Resonance Imaging (MRI).} MRI  technology has been instrumental in identifying structure-function relations in brain networks, and has reshaped brain research since the 1990s.  As a method that involves both sophisticated technology and core brain expertise,  MRI has been a focal point for $X_{\text{Distant},SA\&CIP}$ scholarship. For example, ref. \citep{Van:2012}  addresses the  problem of motion, a pernicious confounding factor that can invalidate  MR brain results. Hence, this research article exemplifies how a fundamental problem threatening an entire line of research acts as an attractor of distant cross-disciplinary collaborations with an all-encompassing theme,  including authors from CIP 5 (medical specialists) and CIP 8 (engineers and computer scientists), while thematically spans four topical domains: SA 2 (Anatomy \& Organisms), SA 3 (Phenomena \& Processes), SA 5 (Techniques \& Equipment), and SA 6 (Technology \& Information Science).

{\it Genomics.} Following the completion of the Human Genome Project (HGP) in the early 2000s, genomics and biotechnology methods  have established a foothold in brain research. This convergent  frontier  made headway in solving long-standing morbidity riddles and formulating novel therapies, e.g. providing a deeper understanding of the genetic basis of developmental delay  \citep{Cooper:2011} and developing treatment for glioblastoma  using  a recombinant poliovirus \citep{Desjardins:2018}.
Both these articles include authors from CIP 4  and  5; thematically, these articles cast a wide net, with the former  spanning SA 1,  3, 4  and 5, while the latter 
covers SA 2, SA 4 and SA 5.

{\it Robotics.} In the early 2010s neurally controlled robotic prosthesis reached fruition by way of collaboration between neuroscientists (CIP 1) and biotechnologists (CIP 4). A prime example of this emerging bio-mechatronics frontier is  research on robotic arms for tetraplegics \citep{Hochberg:2012}, which thematically covers all  SA 1-6.  

{\it Artificial Intelligence (AI) and Big Data.} Following  developments in machine learning capabilities (ML), deep AI methods were brought to bear on MR data, pushing  brain imaging towards more quantitative, accurate, and automated diagnostic methods. Research on brain legion segmentation using Convolutional Neural Networks (CNN) \citep{Kamnitsas:2017}  is an apt example produced by collaboration between medical specialists (CIP 5) and engineers (CIP 8), and spanning SA 2-4 and SA 6. Simultaneously, massive brain datasets combined with powerful AI engines made their appearance along with methods to control noise and ensure their validity, as exemplified by ref. \citep{Alfaro:2018} produced by neuroscientists (CIP 1), health scientists (CIP 6), and engineers (CIP 8), and also featuring  a nearly exhaustive topical scope (SA 2-6).

All together, case analysis indicates $X_{\text{Distant},SA\&CIP}$ products are typically characterized by significant SA integration, typically including 3-4 non-technical SA plus 1-2 technical SA. This thematic coverage exceeds the disciplinary bounds implied by the CIP set of the authors, which typically includes one non-technical CIP plus one technical CIP.

\vspace{-0.25in}
\section*{Discussion}
\vspace{-0.1in}
 In a highly competitive and open science system with multiple degrees of freedom, more than one operational mode is likely to emerge. To assess the different configurations that exist, we developed an \{author discipline $\times$ research topic\}  classification that enables examination of  several operational modes and their relative scientific impact. 
 
 {\it Competing Convergence Modes:} Our key result regards the identification and assessment of a prevalent  convergence shortcut characterized by research combining different SA ($X_{SA}$) but not integrating cross-disciplinary expertise ($M_{CIP}$).
Assuming  the HBS ecosystem to be  representative of other competitive science frontiers, our results suggest that the two operational modes of convergence  evolve as substitutes rather than complements. Trends from the last five years indicate an increasing tendency for scholars to shortcut  cross-disciplinary approaches, and instead integrate by way of expansive  learning.  
This  appears to be in tension with the intended mission of flagship HBS programs.  Instead, our analysis provides strong evidence that the rise of expedient convergence tactics  may be an unintended consequence of the race among teams to secure  funding.

In order to provide timely assessment of  convergence science, we addressed our fundamental RQ1 -- how to measure convergence?-- by developing a generalizable framework that differentiates between diversity in team expertise and research topics.
While it is true that a widespread paradigm shift towards increasing team size has transformed the scientific landscape \citep{Wuchty:2007,TeamGrowth2014}, this work challenges the prevalent view that larger  teams are innately more adept at prosecuting cross-domain research. Indeed,  convergence  does not only depend on team size but also on its composition.
In reality, however, research teams targeting the class of hard problems calling for convergent approaches are faced with coordination costs and other constraints  associated with crossing disciplinary and organizational boundaries \citep{cummings2005collaborative,cummings2008collaborates,van2011factors}.
Consequently, teams are likely to economize in disciplinary expertise,  and instead integrate cross-domain knowledge in part (or in whole) by way of polymathic generalists  comfortable with the expansive learning approach. 
As a result, a team's composite disciplinary pedigree tend to be a subset of the topical dimensions of the problem under investigation.  

As a consistency check, we also find  this convergence shortcut  to be more widespread in research  involving topics that are  epistemically close, as represented by the {\it Neighboring} configuration we analyzed. Contrariwise, in the neuro-psycho-medical $\leftrightarrow$ techno-informatic interface, belonging to the {\it Distant} configuration, convergent cross-disciplinary collaboration runs strong. Perhaps not by serendipity, mixed analysis further indicates that this is exactly the configuration where transformative science has long been occurring.

Arguably, a certain degree of  expansive learning is needed for multidisciplinary teams to operate in harmony. For example, in the case of a psychologist collaborating with a medical specialist, it would be ideal if each one knew a little bit about the other's  field, so that they establish an effective knowledge bridge. After all, this is what transforms a multidisciplinary team to a cross-disciplinary team, such that convergence becomes operative. However, this approach is not the dominant trend in HBS (see the {\it Article level Model}), and is possibly a response to the broad and longstanding  paradigm promoting interdisciplinarity \citep{nissani1995fruits} with less emphasis on cross-domain collaboration. Again using our simple example, it may be that the medical specialist  prefers not to partner at all with psychologists in the prosecution of bi-domain research, i.e., opting for the streamlined substitutive strategy of total replacement over the   strategy of partial redundancy, which comes with the risks associated with cross-disciplinary coordination.

A  limitation to our framework is that  we do not specify what task (e.g. analysis, conceptualization, writing) a given domain expert performed, and hence do not account for division of labor in the teams here analyzed. Indeed, recent work provides evidence that larger teams tend to have higher levels of task specialization \citep{haeussler2020division}, which thereby provides a promising avenue for future investigation, i.e., to provide additional clarity on how bureaucratization \citep{walsh2015bureaucratization} offsets the recombinant uncertainty \citep{fleming2001recombinant} associated with cross-disciplinary exploration.
Another limitation regards the nuances of HBS programs that we do not account for, e.g. different grand objectives, funding levels and disciplinary framing which varies across flagships. 
Yet as a truly multidisciplinary ecosystem, we believe HBS provides an ideal testbed for evaluating the prominence, interactions, and impact of the constitutional aspects of convergence   \citep{Quaglio:2017,Eyre:2017,Grillner:2016,Jorgenson:2015}.

Our results also provide clarity regarding recent efforts to evaluate the role of cross-disciplinarity in the domain of genomics \citep{Petersen:2018}, where we used a similar scholar-oriented  framework that did not incorporate the SA  dimensions.  One could argue that the cross-disciplinary citation premium reported in the genomics revolution arises simply from the genomics domain being primed for success. Indeed, {\bf Fig. \ref{Figure3.fig}}(A)  shows that HBS scholars in the domain of Biotech. \& Genetics discipline maintained high levels of SA diversity extending back to the 1970s.
We do not observe similar patterns for other HBS sub-disciplines. Yet, our measurement of a $\sim$16\% citation premium for research featuring both modes  $(X_{SA\&CIP})$ are remarkably similar in magnitude to the analog measurement of a $\sim 20$\% citation premium reported in \citep{Petersen:2018}.

{\it Econometric Analysis:}  In order to accurately measure shifts in the prevalence and impact of  cross-domain integration, in addition to how they depend on the convergence mode, we employed an econometric regression specification that leverages author fixed-effects and accounts for research team size, in addition to a battery of other CIP and SA controls.
Regarding the growth rate of HBS  convergence science, {\bf Fig. \ref{Figure5.fig}}(A) indicates that research integrating topics and disciplinary expertise is growing between $\sim$2-4\% annually, relatively to the mono-disciplinary baseline; however, this upward trend reversed after the ramp-up of HBS flagships, as indicated in {\bf Fig. \ref{Figure5.fig}}(B).
Our results also indicate that the citation impact of publications from polymathic teams ($X_{\text{Neighboring},SA}$ and $X_{\text{Distant},SA}$) is significantly lower than the  impact of publications from more balanced cross-disciplinary teams ($X_{\text{Neighboring},SA\&CIP}$ and $X_{\text{Distant},SA\&CIP}$), see {\bf Fig. \ref{Figure5.fig}}(C). 
On a positive note, a difference-in-difference strategy provides support that HBS research featuring the $X_{\text{Distant},SA\&CIP}$ configuration has increased in citation impact following the ramp-up of HBS flagships, see {\bf Fig. \ref{Figure5.fig}}(D). 
There are various possible explanations to consider, most prominent of which is that the cognitive and resource demands required to address grand scientific challenges have outgrown the capacity of even mono-disciplinary teams, let alone solo genius \citep{simonton_after_2013}. 

Reflecting upon these results together, it is somewhat troubling that the polymathic trend proliferates and competes with the gold standard, that is, configurations featuring a balance of cross-disciplinary teams and diverse topics ($X_{SA\&CIP}$). Counterproductively, flagship HBS projects appear to have incentivized expansive research strategies  manifest in a relative shift towards $X_{SA}$ since the ramp-up of flagship projects  in  2014. This trend may depend upon the particular flagship's objective framing. Take for instance the US BRAIN Initiative, with the expressed aim to support multi-disciplinary mapping and  investigation of dynamic brain networks. As such, its corresponding research goals promote the integration of {\it Neighboring} topics, where scientists with polymathic tendencies may feel more emboldened to short-circuit expertise. In addition, there are practical  pressures associated with proposal calls. Another possible explanation regarding team formation, is that  it may be easier and faster for researchers to find collaborators from their own discipline  when faced with the pressure  to meet proposal deadlines. Additionally,  funding levels are not unlimited and bringing additional reputable specialists into the team comes with great financial consideration. Hence, a natural avenue for future investigation is to test whether other convergence-oriented  funding initiatives also unwittingly amplify such suboptimal teaming strategy. 

{\it Theoretical insights -- expansive learning:}  Indeed, the polymathic trends described here pre-existed the flagship HBS projects, and so must have deeper roots. One hypothesis is that this trend represents an emergent scholarly behavior owing to efficient  21st century means to pursue new topics by way of expansive learning \citep{Engestrom:2010}, since the  learning costs associated with certain tasks characterized by explicit knowledge  have markedly decreased with the advent of the internet and other means of  rapid  high-fidelity  communication. Indeed, many of the activity signals brought to the fore by this study bear the hallmarks of expansive learning. Perhaps the most telling  signal is the propensity towards topically diverse  publications -- {\bf Fig. \ref{Figure4.fig}}(D-F), which largely stems from horizontal movements in the research focus of individual scientists rather than vertical integration among experts from different disciplines -- {\bf Fig. \ref{Figure4.fig}}(A-C). The scientific system is increasingly interconnected, as  evident from the densification of collaboration networks and emergent cross-disciplinary interfaces -- {\bf Fig. \ref{Figure2.fig}}(B). 
 These interfaces satisfy the conditions that are conducive to boundary crossing, especially with respect to research topics, which  can act as structures facilitating ``minimum energy'' expansion \citep{Toiviainen:2007}. 
To this point, we also assessed wether the relationship between CIP diversity and SA integration depends on wether the configuration represents neighboring or distant domains.
Analyzing  the set of $X_{SA\&CIP}$ articles, we find that  expansive integration is consistently most effective in {\it Distant} configurations, e.g. teams with $N_{CIP,p}=3$ span roughly 32\% more SA than their mono-disciplinary counterparts -- {\bf Fig. \ref{FigureS12.fig}}(B).

{\it Policy Implications:} Consistent also with other studies in expansive learning, actions taken by participants do not necessarily correspond to the intentions by the interventionists \citep{Rasmussen:2009}. The participants are brain scientists in this case, and the interventionists are the funding agencies and the scientific establishment at large. While the latter aim to promote research powered by true multidisciplinary teams, the former appear to prefer to shortcut around this ideal.

 Policy makers and  other decision-makers within the scientific commons are faced with the persistent challenge of efficient resource allocation, especially in the case of grand scientific challenges that foster aggressive timelines \citep{stephan_how_2012}.
The implicit uncertainty  and risk associated with such endeavors is bound to affect  reactive scholar strategies, and this interplay between incentives and behavior is just one source of complexity among many that underly the scientific system \citep{fealing_science_2011}. 

To begin to address this issue, policies addressing the challenges of historical  fragmentation in Europe offer guidance.  European Research Council (ERC) funding programs have been powerful vehicles for integrating national innovation systems by way of supporting cross-border collaboration, brain-circulation and knowledge diffusion -- yet with unintended outcomes that increase the burden of the challenge \citep{Petersen_braindrain_2017}. To address this fragmentation,  many major ERC collaborative programs require multi-national partnerships as an explicit funding criteria.
Motivated by the effectiveness of this straightforward integration strategy,  convergence programs can can include analog  cross-disciplinary criteria or review assessment to address the convergence shortcut. Such guidelines could help to align polymathic vs. cross-disciplinary pathways towards more effective cross-domain integration. Much like the vision for brain science -- towards a more complete understanding of the emergent structure-function relation in an adaptive complex system --  a better understanding of cross-disciplinary team assembly, among other team science considerations \citep{borner2010multi}, will be essential in other challenging frontiers calling on convergence. 

\vspace{-0.25in}
\section*{Methods}
\vspace{-0.2in}
\noindent \textbf{Normalization of citation impact.} 
We  normalized each Scopus citation count, $c_{p,t}$, by leveraging the well-known log-normal properties of citation distributions  \citep{Radicchi:2008}.
To be specific, we grouped articles by publication year $y_{p}$, and  removed the time-dependent trend in the location and scale of the underlying  log-normal citation distribution. The  normalized citation value  is given by
\begin{equation}
z_{p}=(\ln ({c}_{p,t}+1) - \mu_{t})/\sigma_{t} \ ,
\label{Zp}
\end{equation}
 where $\mu_{t} \equiv \langle \ln ({c}_{t}+1) \rangle$ is the mean and $\sigma_{t} \equiv  \sigma[ \ln ({c}_{t}+1)]$ is the standard deviation of the citation distribution for a given $t$; we add 1 to ${c}_{p,t}$ to  avoid the  divergence of $\ln 0$ associated with uncited publications -- a common method which does not alter the interpretation of results.   
 
{\bf Figure  \ref{FigureS8.fig}}(G) shows the probability distribution $P(z_{p})$ calculated across  all $p$  within  five-year  non-overlapping time periods. The resulting normalized citation measure is well-fit by the Normal $N(0,1)$  distribution, independent of $t$, and thus is a stationary measure across time. Publications with $z_{p} > 0$ are thus above the average log citation impact $\mu_{t}$, and since they are measured in units of standard deviation $\sigma_{t}$, standard intuition and statistics of $z$-scores apply. The annual  $\sigma_{t}$ value is rather stable across time, with average and standard deviation  $\langle \sigma \rangle \pm \text{SD} = 1.24 \pm 0.09$ over the 49-year period 1970-2018.\\

\noindent \textbf{Subject Area classification using MeSH.} 
Each MeSH descriptor has a tree number that identifies its location within one of  \href{https://meshb.nlm.nih.gov/treeView}{16 broad categorical branches}. We merged   9 of the science-oriented MeSH branches (A,B,C,E,F,G,J,L,N)  into 6  Subject Area ($SA$) clusters (see {\bf  Fig. \ref{Figure1.fig}}). {\bf  Figure \ref{FigureS2.fig}} shows  the 50 most prominent MeSH descriptors for each SA cluster.  Hence, we take the set of MeSH for each $p$ denoted by $\vec{W}_{p}$, and map these MeSH  to the corresponding MeSH branch (represented by the operator $O_{SA}$), yielding a count vector with six elements: $O_{SA}(\vec{W}_{p})=\overrightarrow{SA}_{p}$.  {\bf  Figure \ref{FigureS8.fig}}(D) shows the distribution $P(N_{SA})$ of the number of $SA$ per publication: 72\% of articles have two or more $SA$; the mean (median) $SA_{p}$ is 2.1 (2), with standard deviation 0.97, and maximum 6.\\

\noindent \textbf{Disciplinary classification using CIP.} 
We obtained host department information from each scholar's Scopus Profile. Based upon this information provided in the profile description, and in some cases using additional web search and data contained in the Scholar Plot web app \citep{majeti_scholarplot_2020}, we  manually annotated each scholar's home department name according to National Center for Education Statistics
\href{https://nces.ed.gov/ipeds/cipcode/browse.aspx?y=56}{Classification of Instructional Program} (CIP) codes. We then merged these CIP codes into  9 broad clusters and three super-clusters (Neuro/Biology, Health, and Science 
\& Engineering, as indicated in {\bf  Fig. \ref{Figure1.fig}}); for a list of constituent CIP codes for each cluster see {\bf  Fig. \ref{FigureS2.fig}}(C).  Analogous to the notation for assigning $\overrightarrow{SA}_{p}$, we take the set of authors for each $p$ denoted by $\vec{A}_{p}$, and map their individual departmental affiliations  to the corresponding CIP cluster (represented by the operator $O_{CIP}$), yielding a count vector with nine elements: $O_{CIP}(\vec{A}_{p})=\overrightarrow{CIP}_{p}$.\\

\noindent \textbf{Measuring cross-domain diversity.} 
We developed a measure of cross-domain diversity defined according to  categorical co-occurrence within individual research articles. 
  Each article $p$ has a count vector $\vec{v}_{p}$: for discipline categories  $\vec{v}_{p} \equiv \overrightarrow{CIP}_{p}$ and for topic categories $\vec{v}_{p} \equiv \overrightarrow{SA}_{p}$.  We then measure article co-occurrence levels by way of the normalized outer-product 
\begin{eqnarray}
{\bf D}_{p}(\vec{v}_{p}) \equiv \frac{U(\vec{v}_{p} \otimes  \vec{v}_{p})}{\vert\vert U(\vec{v}_{p} \otimes  \vec{v}_{p}) \vert\vert} \ , 
\label{Dmatrix}
\end{eqnarray}
where $\otimes$ is the outer tensor product, $U({\bf G})$ is an operator yielding the upper-diagonal elements of the matrix ${\bf G}$ (i.e. representing the undirected co-occurrence network among the categorical elements). In essence, ${\bf D}_{p}(\vec{v}_{p})$ captures a weighted combination of all category pairs.  The resulting matrix represents dyadic combinations of categories as opposed to permutations (i.e., capturing the subtle difference between an  undirected  and directed network). While we did not explore it further, this matrix formulation may also give rise to higher-order measures of diversity associated with the eigenvalues of the outer-product matrix.  The notation  $\vert\vert ... \vert\vert$ indicates the matrix normalization implemented by summing all matrix elements. 
The objective of this normalization scheme is to control for the variation in $\vec{v}_{p}$ in a systematic way. 
As such, this  co-occurrence is a article-level measure of diversity which controls for variations in the total number of categories and different count statistics for elements belonging to $\overrightarrow{CIP}_{p}$ and $\overrightarrow{SA}_{p}$. Consequently,  totaling ${\bf D}_{p}(\vec{v}_{p})$ across articles from a given publication year yields the total number of articles published in a given year, $\sum_{p \vert y_{p} \in t} \vert \vert {\bf D}_{p,t} \vert \vert  = N(t)$. 

We also define a categorical  diversity measure for each article given by $f_{D,p}=1-\mathrm{Tr}({\bf D}_{p}) \in [0,1)$, which corresponds to  the sum of the off-diagonal elements in ${\bf D}$.  The average article diversity by publication year is denoted by $\langle f_{D}(t) \rangle$. In simple terms, articles featuring a single category have $f_{D,p} = 0$ whereas articles featuring multiple categories have $f_{D,p} > 0$. While the result of this approach is nearly identical to the Blau index (corresponding to 1- $\overrightarrow{SA}_{p}\cdot \overrightarrow{SA}_{p}/\vert SA_{p}\vert ^{2}$, also referred to as the Gini-Simpson index), $f_{D,p}$ is motivated by way of dyadic co-occurrence rather than the standard formulation  motivated around repeated  sampling. \\

\noindent \textbf{Data accessibility:} All data analyzed here are openly available from Scopus and PubMed APIs. \\
\noindent \textbf{Competing Interests} The authors declare that they have no competing financial interests.\\
\noindent \textbf{Author Contributions} AMP performed the research, participated in the writing of the manuscript, collected, analyzed, and visualized the data; MA developed software to collect, analyze, and visualize the data; and IP designed the research, performed the research, and participated in the writing of the manuscript.\\
\noindent \textbf{Funding:} AMP and IP acknowledge funding from NSF grant 1738163 entitled `From Genomics to Brain Science'.\\ 
\noindent \textbf{Acknowledgements:} The authors acknowledge support from the Eckhard-Pfeiffer Distinguished Professorship Fund. AMP acknowledges financial support from a Hellman Fellow award that was critical to this project. Any opinions, findings, and conclusions or recommendations expressed in this paper are those of the authors and do not necessarily reflect the views of the funding agencies.







\newpage
\clearpage

\begin{widetext}

\beginsupplement

\begin{center}
{\Large \bf Supplementary Information:\\ \bigskip Appendices S1-S7, Figures S1-S12, and Tables S1-S5}
\end{center}

\bigskip

\setcounter{page}{1}

\bigskip
\begin{center}
{\large \bf Grand challenges and emergent modes of convergence science} \\
\bigskip
Alexander M. Petersen,$^{1}$ Mohammed E. Ahmed,$^{2}$ Ioannis Pavlidis$^{2}$\\
\bigskip
$^{1}$Department of Management of Complex Systems, Ernest and Julio Gallo Management Program, School of Engineering, University of California, Merced, California 95343\\
$^{2}$Computational Physiology Laboratory, University of Houston, Houston, Texas 77204 \\
\bigskip
\end{center}
\bigskip

\setlength{\tabcolsep}{5mm}

\vspace{-0.25in}
\section{Data Collection}
\vspace{-0.10in}
 In the effort to capture a comprehensive representation of the  HBS ecosystem, this work contributes to  Science of Science \citep{Fortunato_2017_Science} efforts, in particular where the entry point  is a scholar-oriented dataset spanning a broad research domain, such as sustainability science \citep{bettencourt2011evolution}, genomics \citep{Petersen:2018}, computer science \citep{morgan2018automatically}, climate change \citep{petersen2019CCC}, physics \citep{sinatra2015century,battiston2019taking}  and artificial intelligence \citep{frank2019evolution}.
Here we construct  a comprehensive scholar-oriented HBS  dataset  to facilitate measuring factors related to cross-domain activity and corresponding strategic shifts associated with HBS flagship projects. These projects are considered within their continental framework, and since they started ramping up in late 2013, they naturally divide the timeline under examination into ``post'' and ``pre'' periods.

All together, we constructed our comprehensive scholar-centric representation of the HBS  ecosystem  by merging  data from three  publication indices -- Web of Science (WOS), Scopus, and PubMed, as illustrated in {\bf  Fig. \ref{Figure1.fig}}.  \\

\noindent \textbf{Author keystone via Web of Science (WOS).} 
In building a scholar-centric database, one needs a keystone for developing a list of  authors who have published on a particular topic. For that, we chose WOS, using the topic field query ``Human Brain''  (HB) to search its ``Core Collection'' over the period 1955-2016. This search resulted in 224,201  records with distinct WOS article identifiers. From these records, we extracted the full first and last names of all authors with $\geq 5$ publications, along with their affiliations.\\  

\noindent \textbf{Author name disambiguation.} 
An important challenge in constructing a scholar-centric dataset is name disambiguation of authors. Here we overcame this challenge by using curated publication sets for each scholar obtained from Scopus via their profile-oriented API, which requires an author's full name and affiliation in order to identify their Scopus profile.\\

\noindent \textbf{Brain Science data via Scopus and PubMed.} 
Having amassed a comprehensive set of brain science researchers, the next step was to build a database that represents the totality of their work. Since  brain science is multidisciplinary, these researchers come from different domains, publishing in diverse areas that feed brain science, thus creating an ecosystem prime for cross-domain analysis. We used the full name and affiliation-location of each WOS author to query the \href{https://www.scopus.com/freelookup/form/author.uri}{Scopus Author Profile} repository.  Among these profiles, 9,121 contained  geographic and departmental affiliation information. In order to identify HB research articles, as opposed to other content such as comments and editorials and also non-biomedical research in the physical sciences, we  searched for each article DOI in MEDLINE/PubMed. We only analyzed articles annotated with Medical Subject Heading (MeSH) keywords, which are  indicators that this  research-oriented content is  biomedical in nature
 -- resulting in a HB dataset  with 655,386 articles over the period 1945-2018; see {\bf  Fig. \ref{FigureS8.fig}}(A) for $N(t)$.  
For each research article $p$ published in year $t$, we also obtained the number of Scopus citations $c_{p,t}$ tallied through the API download  (census) date in November 2019.  \\

\noindent \textbf{Topical keyword classification using MeSH.} 
Medical Subject Headings (MeSH) are  a quasi-hierarchical biomedical ontology developed by the National Library of Medicine and implemented across articles indexed by PubMed by expert annotators to classify articles according to their topical and methodological contributions. With on average 12 MeSH per article, this ontology facilitates topic mapping and topic co-occurrence analysis at multiple levels of specificity \citep{petersen2016triple}. We restrict our analysis to only the  ``Major Topic Headings'' MeSH, which are indicated in PubMed by an asterisk, and account for roughly 1 in 3 MeSH descriptors.  As such, we use these publication-level MeSH  to determine the topical subject area of the research reported in each article. In total, we encountered 14,212 distinct Major Topic MeSH. \\ 

\noindent \textbf{Geographic Regions.} 
We obtained geographic location data from each scholar's Scopus Profile, associating  each individual  with one of 77 countries; the top five countries represented are the United States with 5030 scholars, Germany with 1192, UK with 1074, China with 1049, and Japan with 894. These coauthors associate each $p$ with a set of countries, which we cluster into four localized regions indexed by $R$: North America, corresponding to $R=1$ (United States and Canada); Europe, $R=2$ (33 European Union and non-European Union countries including Norway, Switzerland, Israel, Iceland, and Serbia); Australasia, $R=3$ (Peoples Republic of China, Japan, South Korea, Australia, Taiwan, New Zealand, Singapore, Malaysia, and Thailand); and  World, $R=4$ (remaining countries including Brazil, India, Turkey, and South Africa, among others). 88\% of the publications are covered by regional clusters $R=1,2,3$.  

\vspace{-0.25in}
\section{Shifts in SA and CIP portfolios  in  the decade of multi-national Human Brain flagship projects}
\vspace{-0.10in}
 {\bf Figure \ref{FigureS3.fig}}(A) shows the relative frequency  $f^{<}_{R,CIP}$ ($f^{>}_{R,CIP}$) by region, calculated in the 5-year period before ($<$) and after ($>$) the HB flagship project ramp-up year 2014. Each $f_{R,CIP}$ value represents the average $\overrightarrow{CIP}_{p}$ vector calculated across all articles belonging to a particular region, and normalized to unity to facilitate comparison, i.e. $\sum_{CIP=1}^{9} f_{R,CIP} = 1$. 
 
In both the pre-2014 period [2009-2013]  and post-2014 period [2014-2018], the most prominent disciplines are Neurosciences [CIP 1] and Medical Specialty [5] in the North American (NA) and European (EU) regions. The Australasian  (AA) region shows higher levels of scholars from disciplines in Engineering \& Informatics [8] and Chemistry \& Physics \& Math [9] than their NA and EU counterparts in the pre-2014 period. However, after 2014 we observe a realignment of AA with the remarkably similar NA and EU profiles. This realignment is achieved by decreases in  Engineering \& Informatics [8] and Chemistry \& Physics \& Math [9], and increases in Neurosciences [1] \& Medical Specialty [5].  {\bf Fig. \ref{FigureS3.fig}}(B) shows these relative shifts  calculated as the difference $\Delta f_{R,CIP} = f^{>}_{R,CIP}-f^{<}_{R,CIP}$. Overall, there appears to be a remarkable  synchrony in the direction and magnitude of $\Delta f_{R,CIP}$ for the NA and EU regions, primarily associated with decreases in  Neurosciences [1] and Pathology \& Pharmacology [7] and increases in Psychology [3] and Medical Specialty [5]. NA is the only region showing increase in both Science  \& Engineering domains [CIP 8\&9].

Similarly, {\bf Fig. \ref{FigureS4.fig}}(A) shows the  analog  frequencies  $f^{<}_{R,SA}$ ($f^{>}_{R,SA}$) for each SA by region. In the pre-2014 period, the most prominent SA categories are Anatomy \& Organisms [SA 2] and Health [4], with all regions showing similar distribution profiles.  The most prominent distinction in AA is a reduced prominence  of Psychiatry \& Psychology [1]. By and large, the profiles remain consistent in the post-2014 period, with AA  and NA experiencing prominent increases in Health [4], and AA showing a modest increase in Psychiatry \& Psychology [1], which nevertheless does not fully compensate for the initial deficit in this category with respect to both NA and EU. 

{\bf Figure \ref{FigureS4.fig}}(B)  indicates that all regions experienced a consistent decline in research involving the {\it structure}-oriented  topics associated with Anatomy \& Organisms [2], as well as the {\it function}-oriented topics associated with Phenomena \& Processes [3]. The most prominent distinction between regions is for NA and EU, which both feature increases in  Technology \& Information Science [6] that are relatively larger than observed for AA and World, likely reflecting the technological capacity related to the tech. hubs in these regions; another distinction relates to the Psychiatry \& Psychology [SA 1] which increases in EU and AA more than for NA and World; and also for  Health [4] which increases in NA and AA more than for EU and World.

 \vspace{-0.25in}
\section{Levels and changes in SA and CIP co-occurrence before and after 2014}
\vspace{-0.10in}
We also seek to identify which category pairs are frequently combined in articles, and to assess their frequency shifts after 2014. To this end, we introduce a tensor-product method to readily measure SA an CIP co-occurrence statistics for the purpose of identifying particular cross-domain orientations observed in cross-domain HB  science. 

In order to juxtapose the relative frequencies of mono-category articles separately from multi-category articles, we define a modified outer-product matrix designed purely for visualization purposes:
\begin{eqnarray}
{\bf \tilde{D}}_{p}(\vec{v}_{p}) \equiv 
\begin{cases}
    \frac{U(\Upsilon)}{\vert\vert U(\Upsilon) \vert\vert}  , & \text{if $\vec{v}_{p}$ contains 2 or more categories}.\\
    
   \\
      
     \frac{U(\vec{v}_{p} \otimes  \vec{v}_{p})}{\vert\vert U(\vec{v}_{p} \otimes  \vec{v}_{p}) \vert\vert} = \text{DiagonalMatrix(Sign[} \vec{v}_{p}\text{])}   , & \text{otherwise}.
  \end{cases}
\label{DTildematrixSI}
\end{eqnarray}
where $\otimes$ is the outer tensor product and  $\circ$ indicates the element-wise or Hadamard product. Note that this definition is slightly different than  ${\bf D}_{p}(\vec{v}_{p})$ defined in Eq.(\ref{Dmatrix}).
The difference occurs in the first case, for which the matrix ${\bf \Upsilon }\equiv \vec{v}_{p} \otimes  \vec{v}_{p}- \vec{v}_{p} \circ \mathbb{1} \circ  \vec{v}_{p}$
for which the  diagonal elements are eliminated via subtraction, i.e. $\mathrm{Tr}({\bf \Upsilon})=0$. 

Simply stated, when $\vec{v}_{p}$ (representing $\overrightarrow{SA}_{p}$  or $\overrightarrow{CIP}_{p}$) has  2 or more non-zero elements then we primarily count the off-diagonal elements of the outer-product matrix and are not concerned with the relative frequencies of the on-diagonal elements. Contrariwise, in the case that there is just one category present -- e.g. $\vec{v}_{p} = \{0,3,0,0,0,0\}$ -- then we track only the diagonal element, which counts the occurrence of the single category. In this second case, the resulting  matrix ${\bf \tilde{D}}_{p}(\vec{v}_{p}) = \text{DiagonalMatrix(Sign[} \vec{v}_{p}\text{])}$ has only one non-zero element, which occurs for the diagonal value $\tilde{D}_{22}=1$;  and all other matrix elements = 0. Note that in either case the total sum of all elements are normalized to unity, $ \vert\vert {\bf \tilde{D}}_{p}(\vec{v}_{p})\vert\vert  = 1$. This normalization implies that  totaling ${\bf \tilde{D}}_{p}(\vec{v}_{p})$ across articles from a given publication year yields the total number of articles, $N(t)$. 

We then calculated the aggregate co-occurrence matrix, denoted by ${\bf C}^{<} = \sum_{y_{p} \in [2009-2013]}  {\bf \tilde{D}}_{p}$, using all articles published in the pre-period.
It then follows from our normalization procedure that the total across all matrix elements is proportional to the total number of articles published in a given period, i.e. $\vert\vert {\bf C}^{<} \vert\vert = \sum_{t =2009}^{2013} N(t) = N_{ [2009-2013]}$. {\bf  Figures \ref{FigureS3.fig}}(C) and {\bf  \ref{FigureS4.fig}}(C)   show ${\bf C}^{<}_{CIP}$ and ${\bf C}^{<}_{SA}$, respectively. 

To measure relative changes, we then calculated the percent difference in each matrix element $\Delta C_{ij} = 200(C^{>}_{ij} - \theta C^{<}_{ij})/(C^{>}_{ij}+\theta  C^{<}_{ij})$, where $\theta = N_{[2014-2018]}/N_{[2009-2013]}$ corrects for bias associated with differences in the number of  articles published each of the pre- and post-periods. To illustrate why this correction is important, we randomized  the counts contained in $\vec{v}_{p} = \overrightarrow{SA}_{p}$ and plot the resulting  ${\bf C}^{<}_{\text{rand.},SA}$ and $\Delta {\bf C}_{\text{rand.},SA}$ matrices in {\bf Fig. \ref{FigureS5.fig}}. As anticipated, this randomization scheme eliminates the variation among on-diagonal elements and off-diagonal elements in panel (A); Moreover,  in panel (B) the off-diagonal elements all show percent change values that are in the range of $\pm 3\%$, thereby indicative of the threshold for distinguishing statistically significant percent changes in the  real data.

Returning to the real data and the calculation of  ${\bf C}^{<}_{CIP}$, the most notable results of this visualization are the consistently strong couplings between CIP category [1] and all other categories [2,3,4,5,6];  between categories [1,2,3]  and [5];  and also between categories [1,2,5] and [6]. Also of note is the  higher-order clique among [1,5,6] where each CIP is strongly coupled to each other. Contrariwise, we observe relatively weak coupling between  [7,8,9] and most all other CIP. 

Other prominent  CIP that couple by region: NA shows relatively higher coupling between [1,4] and [4,5] and [5,8] compared to other regions; and EU shows relatively higher coupling between [1,9] and [2,9].
Regarding the shifts from the pre- to post-2014 captured by  $\Delta {\bf C}_{CIP}$, NA and EU regions show consistent increase in CIP pairs [4,7] and [4,9], [3,8] and [2,7]; and consistent decrease between [1,8] and [2,9] and [6,9] and all combinations between 5 and [7,8,9]. Notably, AA exhibits higher \% change levels, following from the fact that several elements in ${\bf C}^{<}_{CIP}$ that are nearly 0.

{\bf  Figure \ref{FigureS4.fig}}(C) shows  ${\bf C}^{<}_{SA}$ calculated by region. 
For all regions, the matrix elements corresponding to SA pairs [2,3] and [2,4] and [3,4] are relatively strong, thus forming another  clique among these SA representing traditional branches of biology. Other strongly coupings are SA [4] with both [1,5]; and [5] with  [2,4,6].  As  with  ${\bf C}^{<}_{CIP}$, the technologically-oriented SA are the most weakly coupled categories. 

Regarding the shifts from the pre- to post-2014 captured by  $\Delta {\bf C}_{SA}$, the  most consistent increases are between  SA [1] and each of [2,4,6]; and between 4 and both [5,6]. 
Contrariwise, the most consistent decreasing coupling is between  [3] and both [2,6], and between [5,6]. The matrices for  NA and EU are rather similar, with the most prominent distinction  between [2,6] -- showing  a -12\%  change for NA and a +5\% change for EU; and also between [3,6] -- showing also a -12\% decrease for NA but  no significant change for EU. 
This latter disparity is an example of where EU may be taking the lead in in-silico-oriented approaches to HB science, consistent with the framing of the  Human Brain Project. 

The most notable distinction for AA relative to NA and EU is in the larger magnitude of shifts, representing a period of international convergence  for all couplings involving SA [1], and  in particular between [1,2] and between [1,6]; contrariwise for AA, there is a   prominent decoupling between SA  [5,6] which is consistent with the relative shifts away from these two SA to compensate for the prominent redirection towards [1]  and [4], as also indicated by {\bf  Fig. \ref{FigureS4.fig}}(B).

\vspace{-0.25in}
\section{Calculation of cross-domain co-occurrence: an illustrative example of the Tensor Product}
\vspace{-0.10in}
Calculating $f_{D,p}$ begins with the outer-product between a categorical count vector, e.g. $\vec{SA}_{p} \otimes  \vec{SA}_{p}$, where $\otimes$ is the outer tensor product. The resulting matrix represents dyadic combinations of categories as opposed to permutations (i.e., capturing the subtle difference between an  undirected  and directed network). The subtle difference between the Blau index and $f_{D,p}$ arises from $U({\bf G})$, which is imposed to capture the difference between combinations rather than permutations  (or directed versus undirected network). Hence, this perspective offers a new pathway to the formulation of the common Blau diversity index by way of co-occurrence rather than repeated sampling.

Take for example an article $p$ with 4 metadata entities  belonging to 3 categories,  $\vec{v}_{p}=\{1,2,0,0,1,0\}$. Calculation of  the co-occurrence matrix ${\bf D}_{p}(\vec{v}_{p}) $ using the normalized outer-product defined in Eq.(\ref{Dmatrix}) yields    

\bigskip
${\bf D}_{p}(\vec{v}_{p}) =$ 
\(\left(
\begin{array}{cccccc}
 \frac{1}{11} & \frac{2}{11} & 0 & 0 & \frac{1}{11} & 0 \\
 \\
 0 & \frac{4}{11} & 0 & 0 & \frac{2}{11} & 0 \\
  \\
 0 & 0 & 0 & 0 & 0 & 0 \\
  \\
 0 & 0 & 0 & 0 & 0 & 0 \\
  \\
 0 & 0 & 0 & 0 & \frac{1}{11} & 0 \\
  \\
 0 & 0 & 0 & 0 & 0 & 0 \\
\end{array}
\right)\)
\bigskip

\noindent with $\vert\vert U(\vec{v}_{p} \otimes  \vec{v}_{p}) \vert\vert=11$. The categorical  diversity is calculated as the  total across off-diagonal elements,  $f_{D,p}=1-\mathrm{Tr}({\bf D}_{p}) = 5/11$.

For completeness, consider the representation of a mono-disciplinary article with the same number of metadata entities that all fall into the second category,  $\vec{v}_{p}=\{0,4,0,0,0,0\}$. Then

\bigskip
${\bf D}_{p}(\vec{v}_{p}) =$ 
\(\left(
\begin{array}{cccccc}
 0 & 0 & 0 & 0 &0 & 0 \\
 \\
 0 &1 & 0 & 0 & 0 & 0 \\
  \\
 0 & 0 & 0 & 0 & 0 & 0 \\
  \\
 0 & 0 & 0 & 0 & 0 & 0 \\
  \\
 0 & 0 & 0 & 0 & 0& 0 \\
  \\
 0 & 0 & 0 & 0 & 0 & 0 \\
\end{array}
\right)\)
\bigskip
\noindent  with corresponding $f_{D,p}=1-\mathrm{Tr}({\bf D}_{p}) = 1 - 1 = 0$.

 What does this measure measure? Notably, $f_{D,p}$ accounts for both  categorical differences (Shannon-like) and concentration disparity (Gini-like)  \citep{harrison2007s}.  One the first hand, articles with more variation in SA categories will correspond to larger $f_{D,p}$ values, as the number of non-zero off-diagonal elements is proportional to ${{M_{p}}\choose{2}}\sim M_{p}^{2}$, where $M_{p}$ is the number of distinct SA present, which contributes to larger $f_{D,p}$; and on the second hand, the off-diagonal elements will be relatively larger in combination if the count values contained in $SA_{2}$ are more evenly distributed, i.e., are not  highly concentrated in just one category.

\vspace{-0.25in}
\section{Bi-partite network between CIP and SA}   
\vspace{-0.10in}
We quantify the empirical association between CIP and SA categories by aggregating the information contained in $\overrightarrow{CIP}_{p}$ and $\overrightarrow{SA}_{p}$. We first applied this method to the subset of  mono-domain  articles comprised of $p$  with $O_{CIP}(\vec{F}_{p})= O_{SA}(\vec{F}_{p})= M$.
By definition,  each of these article features just a single CIP,  making it possible to identify the SA that are most frequently associated with mono-domain researchers from that CIP category. Formally, this amounts to calculating the bi-partite network between CIP and SA, operationalized by averaging the $\overrightarrow{SA}_{p}$ for mono-domain articles from each CIP category, given by $\langle \overrightarrow{SA} \rangle_{CIP} = \sum_{p \in CIP} (\overrightarrow{SA}_{p} / N_{SA,p})$; importantly,  this definition  accounts for variability in $N_{SA}$ by normalizing the sum of the SA counts contained in $\overrightarrow{SA}_{p}$ by $N_{SA,p}$ so that each article contributes equally to the average. 
Less prominent CIP-SA links are pruned from our Sankey chart  visualization in order to emphasize the most meaningful CIP-SA relations. 
To this end, we remove the weakest links contained in $\langle \overrightarrow{SA} \rangle_{CIP}$, excluding  those with value $ < 0.5 \ {\text Max}[\langle \overrightarrow{SA} \rangle_{CIP}]$. 
Hence, the chart labelled  $M_{CIP}\rightleftarrows M_{SA}$ in  {\bf  Figure \ref{Figure4.fig}}(G) shows only the most prominent CIP-SA links.

For juxtaposition, we also calculated the bi-partite network  using the non-overlapping subset of articles with $O_{SA\&CIP}(\vec{F}_{p})=X_{SA\&CIP}$. Since these articles by construction have $ N_{CIP,p}\geq 2$,  we  define the average association between CIP and SA as $\langle \overrightarrow{SA} \rangle_{CIP} = \sum_{p \in CIP}( \overrightarrow{SA}_{p} / N_{SA,p}) /  N_{CIP,p}$, where  the vector $\overrightarrow{SA}_{p} / N_{SA,p}$ contributes to the average for all CIP present in $\overrightarrow{CIP}_{p}$. The  bi-partite network labeled $X_{CIP}\rightleftarrows X_{SA}$ in  {\bf  Figure \ref{Figure4.fig}}(G) also shows just  the most prominent CIP-SA links, applying the same threshold that excludes links that have weight less than half of the most prominent  weighted CIP-SA link for a given CIP. 

Let $\bf{A}$ ($\bf{B}$) represent the matrix representation of  $X_{CIP}\rightleftarrows X_{SA}$ ($M_{CIP}\rightleftarrows M_{SA}$) -- after pruning less prominent CIP-SA links. We then compute the difference between the matrices, ${\bf \Delta_{XM}} \equiv  \bf{C} = \bf{A}-\bf{B}$, such that positive (negative) elements of $\bf{C}$ indicate prominent links that are relatively over-represented  in cross-domain (mono-domain) articles. The Sankey chart labeled $\Delta_{XM}$ in  {\bf  Figure \ref{Figure4.fig}}(G) shows just the positive elements, which tend to be larger in magnitude than the (relatively few) negative elements. 

In {\bf  Fig. \ref{Figure3.fig}}(B) we presented the bi-partite network of prominent CIP-SA relations for the {\it Broad} configuration.
 {\bf Figure  \ref{FigureSX.fig}} complements those results showing the bipartite network for both the {\it Neighboring} and  {\it Distant} configurations, which provide cross-validation for the choice of CIP and SA categories they represent. 

\vspace{-0.25in}
\section{Historical trends in SA \& CIP diversity: 2000-2018}
\vspace{-0.10in}
We investigate historical trends in SA  \& CIP diversity using  the matrix ${\bf D}_{p}$ defined in Eq. (\ref{Dmatrix}), which simultaneously measures  mono-dimensional and multi-dimensional features of each article. More specifically, we define   $f_{D,p}=1-\mathrm{Tr}({\bf D}_{p})$ as the fraction of the article's co-occurrence matrix capturing combinatorial diversity.  Hence, in the limiting case that the article features just a single category, then $f_{D,p} =0$; and when all categories are present in equal quantities then $f_{D,p}=(d-1)/(d+1)$, where  $d$ is the dimension of the categorical vector $\vec{v}_{p}$.  As $d$ increases then $f_{D,p}$ approaches 1. Hence, for sufficiently large $d 
$ then $0 \leq f_{D,p} \lesssim1$. {\bf  Figures \ref{FigureS8.fig}}(D,E) show the unconditional distributions, $P(N_{SA})$ and $P(N_{CIP})$, with  observed values spanning  across the full range  $d=6$ and $d=9$, respectively.

As a bounded quantity, the average article-level diversity $\overline{f}_{D}(t)=N(t)^{-1} \sum_{p \in N(t)}  f_{D,p}$ is an appropriate measure of a characteristic article, where $N(t)$ is the number of articles being considered from year $t$. However,    $\overline{f}_{D}(t)$ is nevertheless sensitive to bias associated with a systematic increase over time in  $\langle N_{CIP,t} \rangle$ and $\langle N_{SA,t} \rangle$, the average number of categories present per article per year. We address this  issue by applying a temporal deflator which adjusts the annual averages to account for systematic shifts in the underlying data generating process. To be specific, we define $\langle f_{D,SA}(t) \rangle = \overline{f}_{D,SA}(t) \times [\langle \tilde{\mu}_{SA} \rangle / \tilde{\mu}_{SA}(t)]$, where $\tilde{\mu}_{SA}(t) = \langle N_{SA,t} \rangle / \sigma_{N_{SA,t}} $ is the inverse coefficient of variation (also called the signal-to-noise ratio) with respect to   the number of SA per article, represented by $N_{SA,p}$; and $\langle \tilde{\mu}_{SA} \rangle$ is the average value calculated across the roughly 3 decades of analysis. {\bf  Figure \ref{FigureS5.fig}}(C) shows that $\tilde{\mu}_{SA}(t)$ is increasing steadily with time. Hence,   adjusting for this secular growth is essential  so that observed increases are not simply artifacts of the underlying growth in $N_{SA,p}$ or $N_{CIP,p}$. 
We apply the same method to adjust for systematic shifts in $\langle N_{CIP,t} \rangle$. 

To illustrate the  utility of this deflator method, we randomized the SA for all articles (by randomly shuffling the counts in each $\overrightarrow{CIP}_{p}$ or $\overrightarrow{SA}_{p}$). {\bf  Figure \ref{FigureS5.fig}}(D) demonstrates that there is no trend in the corresponding $\langle f_{D,SA}(t) \rangle$, indicating that this method removes the underlying bias.

Returning to the empirical data, {\bf Fig. \ref{FigureS6.fig}}   shows the evolution of disciplinary diversity captured by coauthors'  departmental affiliations.  Each panel shows  $\langle f_{D,CIP}(t) \rangle$ calculated for a specified combination of categories contained in each $\overrightarrow{CIP}_{p}$ vector, as indicated by the schematic motif provided alongside each panel. For example, {\bf Fig. \ref{FigureS6.fig}}(A) calculates $\langle f_{D,CIP}(t) \rangle$ from all 9 CIP categories considered independently, whereas {\bf Fig. \ref{FigureS6.fig}}(B) collects the counts associated with the combined categories [1-4] and [5-9] and calculates the diversity based upon the  fraction of ${\bf D}_{p}$  belonging to the single off-diagonal element  $D_{12,p}$, which records the disciplinary mixing  between these two supergroups. 

 {\bf Figure \ref{FigureS6.fig}}(A) is calculated using  the {\it Broad} configuration, and exhibits a slow  increase in CIP diversity from 1990 to the mid 2000s in North America (NA) and European (EU) regions, which stalled thereafter, and even declined in the last decade for NA and AA, but not for EU. {\bf Figure \ref{FigureS6.fig}}(B) shows relatively lower levels and trends in the diversity at the intersection of super-categories [1-4]  (representing traditional neuro/biology departments) and  [5-9] (representing all other CIP jointly). By way of comparison, this trend indicates that the decline in   panel (A) is not derived from the intersection explored in panel (B). Instead,  {\bf Figs. \ref{FigureS6.fig}}(C,D) indicate that the decline in (A) is attributable to declines at the individual intersections between all permutations of CIP categories 1-7, and to a lesser extent between  the three disciplinary subdomains: neuro/biology [1-5], health [5-7] and science and engineering [8-9]. Overall, we also observe higher levels of CIP diversity in NA, followed by EU, and then followed by AA.

Likewise, {\bf Figure \ref{FigureS7.fig}} shows the evolution of research topic diversity captured by  SA counts in each $\overrightarrow{SA}_{p}$. We observe much stronger trends for SA, suggesting that scholars tend to also cross disciplines as mono-disciplinary teams rather than via cross-disciplinary collaboration. {\bf Fig. \ref{FigureS7.fig}}(A) shows  $\langle f_{D,SA}(t) \rangle$ calculated for the {\it Broad} configuration which includes all SA categories. The diversity trend is increasing since 1990 for all regions, but with reduced pace since the early 2010s. Similar to our findings for CIP, we observe AA lagging the other two regions; however, in this case of SA we do observe more similar levels of diversity between EU and NA. {\bf Figure \ref{FigureS7.fig}}(B) indicates that much of the increase in SA diversity is attributable to research combining Health [SA 4] and the other categories -- in other words,  the domain of health science appears to be a persistent driving force behind  convergence trends. Supporting evidence for this observation is also captured in the  hierarchical clustering of SA represented by the minimum spanning tree (MST) representation of the aggregate SA co-occurrence matrix ${\bf \tilde{D}}_{SA,p}$  -- see {\bf Fig. \ref{FigureS2.fig}}(B). By way of comparison, the analog  MST representation of  ${\bf \tilde{D}}_{CIP,p}$ in  {\bf Fig. \ref{FigureS2.fig}}(D) features a less prominent hierarchy across the CIP categories.

We analyzed several additional SA category subsets and super-category combinations to more deeply explore the anatomy of research topic diversity. {\bf Figure \ref{FigureS7.fig}}(C,D) show that increasing diversity associated with Health  [4] is largely captured via the incorporation of technology- and informatics-oriented capabilities [5,6] -- as opposed to integrating  more traditional biological SA representing research domains associated with questions relating to how Anatomy \& Organisms (structure) [2] and Phenomena \& Processes (function) [3] relate to complex human behavior addressed by Psychiatry \& Psychology [1] -- as illustrated in {\bf Fig. \ref{FigureS7.fig}}(E). 

Similarly, a significant component of the increasing diversity captured between SA [4,5,6] derives from the increase between research that is centered around Techniques \& Equipment [5] and Technology \& Information Science [6]; although this contribution shown in {\bf Fig. \ref{FigureS7.fig}}(F) only contributes to increases in diversity until 2010, after which there is a prominent  decline. Interestingly, this is a configuration which emphasizes the leading role of AA since 2010 in combining these two areas.
To further emphasize the role of Health, we exclude this category [4] from the diversity measures shown in  {\bf Fig. \ref{FigureS7.fig}}(G), indicating that combinations of SA across the traditional domains of biology and the technology-oriented domains have also saturated around 2010, and their contribution to SA diversity primarily appears when considered the biology [1-3] and technology-oriented [5,6] as super-clusters  illustrated in {\bf Fig. \ref{FigureS7.fig}}(H).
 
\vspace{-0.25in}
\section{Panel regression: model specification}
\vspace{-0.10in}
We constructed   article-level and author-level panel data to facilitate measuring factors relating to SA and CIP diversity and shifts  related to the  ramp-up of  three regional  HB flagship projects circa 2013, and several others thereafter.  {\bf Figure \ref{FigureS8.fig}} shows the distribution of various article-level features; and  {\bf Figure \ref{FigureS9.fig}} shows the covariation matrix between the principle variables of interest.

We use the following operator notation to specify how we classify articles as being cross-domain ($X$) or mono-domain ($M$). Starting with the feature vector $\vec{F}_{p}\equiv \{\overrightarrow{SA}_{p},\overrightarrow{CIP}_{p},\overrightarrow{R}_{p}\}$, we obtain a binary diversity classification for each article denoted by $X$ and $M$. We specify the objective criteria of the  feature operator $O$ by its subscript. For example, $O_{SA}(\vec{F}_{p})=X_{SA}$ if two or more SA categories are present, otherwise  the value is $M$; and by analogy, $O_{CIP}(\vec{F}_{p})=X_{CIP}$ if two or more categories are present, and otherwise  $O_{CIP}(\vec{F}_{p})=M$. In the case of models oriented around articles featuring $X_{SA}$ and $X_{CIP}$ simultaneously   (represented by $O_{SA\&CIP}(\vec{F}_{p})=X_{SA\&CIP}$), we exclude the set of articles classified as $X_{SA}$ but not $X_{CIP}$ and those  classified as $X_{CIP}$ but not $X_{SA}$. Hence, in what follows, the counterfactual baseline group for $X_{SA\&CIP}$ articles are also the subset of  mono-domain  articles, which facilitates comparison of effect sizes across models oriented around $X_{CIP}$, $X_{SA}$ and $X_{SA\&CIP}$. 

\vspace{-0.15in}
\subsection{Article-level Model A}
\vspace{-0.15in}
\subsubsection*{Quantifying factors associated with propensity for CIP and SA diversity}
\vspace{-0.15in}
In the first model, we seek to better understand the  factors associated with the prevalence of CIP and SA diversity as they evolve over time, and in particular their relation to the launching of the HB flagship programs. In order to model the article-level  factors (indicated by $p$) associated with  cross-domain research activity we define the binary indicator variable generically denoted as $I_{X,p}$. 

By way of example, if we are considering SA diversity, then the indicator variable $I_{X_{SA,p}}$ takes the value 1 if  $O_{SA}(\vec{F}_{p})=X_{SA}$ and 0 if $O_{SA}(\vec{F}_{p})=M$. We then  model the 2-state odds $Q\equiv \frac{P(O_{SA}(\vec{F}_{p})=X_{SA})}{P(O_{SA}(\vec{F}_{p})=M)} = \frac{P(X_{SA})}{P(M)}$, which represents the propensity for cross-domain research, where $P(X_{SA})+P(M)=1$. Likewise, in the case of CIP diversity we model the  odds as $Q\equiv \frac{P(O_{CIP}(\vec{F}_{p})=X_{CIP})}{P(O_{CIP}(\vec{F}_{p})=M)}$; and finally, we also consider the likelihood of research featuring both types of cross-domain activity, for SA \& CIP, represented as $Q\equiv \frac{P(O_{SA\&CIP}(\vec{F}_{p})=X_{SA\&CIP})}{P(O_{SA\&CIP}(\vec{F}_{p})=M)}$. 
Because all Scopus scholars map onto a single CIP, and since this model is primarily concerned with identifying factors associated with orientation towards cross-domain research, we exclude solo-authored research papers (i.e. those with $k_{p}=1$) from this analysis since the likelihood  for those articles is predetermined (i.e. $P(M)=1$); for the same reason, we also exclude articles with a single Major MeSH category (i.e. those with $w_{p}=1$).

For each article  we also include several covariates of $I_{X,p}$: the article publication year $y_{p}$; the mean journal citation impact, calculated as the average $z_{p}$ for articles from journal $j$, denoted by $\overline{z}_{j} = \langle z_{p} \vert \ \text{journal j} \rangle$;  the natural logarithm of the total number of coauthors,  $\ln k_{p}$; and the natural logarithm of the total number of Major MeSH terms, $\ln w_{p}$. As additional controls, we also include the total number of international regions associated with the authors' affiliations $N_{R,p}$ (with  min value 1 and max value 4), and also the total number of categories featured by the article,  $N_{SA,p}$ and $N_{CIP,p}$.

We then model the odds $Q$ by way of a  Logit regression  model, specified in the case of $X_{SA}$ as
\begin{eqnarray}
\left.\begin{aligned}
& \text{Logit}\Big(P(X_{SA})\Big) = \log\Big( \frac{P(X_{SA})}{P(M)}\Big) =   \\
& \beta_{0} +\beta_{y} y_{p}  +\beta_{\overline{z}} \overline{z}_{p}  + \beta_{k} \ln k_{p}  +  \beta_{w} \ln w_{p} +  \beta_{N_{R}} N_{R,p}  +  \beta_{N_{CIP}} N_{CIP,p}  + \epsilon \ ;
\end{aligned}\right.
\label{M1aSA}	 
\end{eqnarray}

in the case of $X_{CIP}$ as

\begin{eqnarray}
\left.\begin{aligned}
& \text{Logit}\Big(P(X_{CIP})\Big) = \log\Big( \frac{P(X_{CIP})}{P(M)}\Big) =   \\ 
& \beta_{0} +\beta_{y} y_{p}  +\beta_{\overline{z}} \overline{z}_{p}  + \beta_{k} \ln k_{p}  +  \beta_{w} \ln w_{p}  +  \beta_{N_{R}} N_{R,p}  +  \beta_{N_{SA}} N_{SA,p}  + \epsilon \ ;
\end{aligned}\right.
\label{M1aCIP}	 
\end{eqnarray}

and in the case of $X_{SA\&CIP}$ as 

\begin{eqnarray}
\left.\begin{aligned}
& \text{Logit}\Big(P(X_{SA\&CIP})\Big) = \log\Big( \frac{P(X_{SA\&CIP})}{P(M)}\Big) =   \\
&  \beta_{0} +\beta_{y} y_{p}  +\beta_{\overline{z}} \overline{z}_{p}  + \beta_{k} \ln k_{p}  +  \beta_{w} \ln w_{p}  +  \beta_{N_{R}} N_{R,p}  + \epsilon \ .
 \end{aligned}\right.
\label{M1aSACIP}	 
\end{eqnarray}
To account for errors that are geographically correlated  over time, we estimated the model using robust standard errors clustered on a regional categorical variable. 
The full set of parameter results are tabulated in models (1)-(3) in {\bf Tables \ref{TableS1.tab}-\ref{TableS3.tab}}, which report the exponentiated coefficients. To be specific, the exponentiated coefficient $\exp(\beta)$ is the odds ratio, representing the factor by which   $Q$ changes for each 1-unit increase in the corresponding independent variable, i.e. $Q_{+1}/Q=\exp(\beta)$. In real terms, $100 \beta \approx 100(\exp(\beta)-1)$ represents the percent change in  $Q$ corresponding to a 1-unit increase  in the corresponding independent variable (where the approximation holds for small $\beta$ values). As a result,  $\exp(\beta)$ values  that are less than (greater than) unity indicate variables that negatively (positively) correlate with the  likelihood $P(X)$.

 \vspace{-0.15in}
\subsubsection*{Quantifying shifts in  propensity for  CIP and SA diversity associated with the announcement of global Flagship HB projects circa 2013}
\vspace{-0.15in}
In order to identify shifts in the 5-year period after the 2013 ramp-up of HB projects worldwide, we incorporated an interaction between the pre-/post periods -- indicated by $I_{\text{2014+},p}$, which takes the value 1 for $y_{p} \geq 2014$ and 0 otherwise -- and  a categorical variable specifying the region, represented by $I_{R,p}$. We use the {\it Rest of World} region category (indicated by countries colored gray in {\bf Fig. \ref{Figure1.fig}}) as the baseline for regional comparison since these regions did not feature flagship HB programs on the scale of those announced in Australia, Canada, China, Japan, Europe, South Korea, and the United States.

By way of example, in the case of modeling the  likelihood $P(X_{SA})$, the interaction term is added in the second row, 
\begin{eqnarray}
\left.\begin{aligned}
&\text{Logit}\Big(P(X_{SA})\Big) = \log\Big( \frac{P(X_{SA})}{P(M)}\Big)  = \\
& \beta_{0} +\beta_{y} y_{p}  +\beta_{\overline{z}} \overline{z}_{p}  + \beta_{k} \ln k_{p} +  \beta_{w} \ln w_{p}  +  \beta_{N_{R}} N_{R,p}  +  \beta_{N_{CIP}} N_{CIP,p} +  \\
&  \gamma_{R} I_{R,p}  + \gamma_{\text{2014+}} I_{\text{2014+},p} +  \delta_{R+} (I_{R,p} \times I_{\text{2014+},p})   + \epsilon \ .
   \end{aligned}\right.
\label{M2aSA}	 
\end{eqnarray}
To differentiate different types of model variables,  $\beta$ is used to identify coefficients associated with continuous variables, $\gamma$ is used for indicator variables, and $\delta$ is used to indicate interactions between indicator variables. In particular, the coefficient $\delta_{R+}$ measures the Difference-in-Difference (DiD) estimate of the effect of HB projects on the propensity for research teams to pursue $X_{SA}$ approaches.  {\bf Figures \ref{FigureS6.fig}} and  {\bf  \ref{FigureS7.fig}}  demonstrate that historical trends in the prevalence of cross-domain diversity satisfy the parallel trend assumption for both CIP and SA, respectively. The full set of parameter results are tabulated in models (4)-(6) in {\bf Tables \ref{TableS1.tab}-\ref{TableS3.tab}}, and the point estimates for principal test variables are visually summarized in {\bf Fig. \ref{FigureS10.fig}}.

\vspace{-0.15in}
\subsection{Author-level Model B}
\vspace{-0.15in}
In the second model, we seek to measure the relation between the two different types of article diversity -- CIP and SA -- and the article's scientific impact,  proxied by $c_{p}$. Our approach leverages the hierarchical features of the article-level data grouped into  author-specific subgroups representing HB researcher publication portfolios. As a result, model coefficients represent estimates  net of author-specific time-independent factors. In other words, this fixed-effect specification yields parameter estimates that are  net of  the author-specific baseline $\alpha_{a}=\langle z_{a} \rangle $, where $a$ is an author index. This specification identifies a  clear counterfactual framework for identifying the different outcomes associated with $X$ and $M$ that are relevant to researcher problem identification and team-assembly strategies.

First, in order to measure relative differences in citation impact within and across publication cohorts,  we apply a logarithmic transform that facilitates removing the time-dependent trend in the location and scale of the underlying  log-normal citation distribution \citep{Radicchi:2008}. As such, the normalized citation impact defined  in Eq. (\ref{Zp}) is 
\begin{equation}
z_{p} \equiv \frac{\ln (1+{c}_{p,t}) - \mu_{t}}{\sigma_{t}} \ , \nonumber
\end{equation}
 where $\mu_{t} \equiv \overline{\ln (1+{c}_{t})}$ is the mean and $\sigma_{t} \equiv  \sigma[ \ln (1+{c}_{t})]$ is the standard deviation of the log-citation distribution for  articles grouped by publication year. We uniformly add 1 to each ${c}_{p,t}$ count to avoid the  divergence $\ln 0$ associated with uncited publications, a common method that does not alter the interpretation of our results. Importantly, the standard deviation $\sigma_{t} \approx  \langle \sigma  \rangle =  1.24$ is approximately constant over the focal period of our analysis. Consequently, we are able to transform the relation between $z_{p}$ and a given covariates into a percent change in ${c}_{p,t}$ associated with the same covariate. 
 
 More  specifically,  building on previous work \citep{Petersen:2018,petersen2018mobility} we define the  citation premium   as the percent change $100 \Delta c_{p}/c_{p}$ associated with shift in the independent  variable  $v$. For sake of simplicity, consider the basic linear model $Y(c)= z_{p}= \beta_{0}+\beta_{v} v$ with the decomposition of differentials,  $\partial Y(c)/ \partial {v} = (\partial Y/ \partial {c}) (\partial c/ \partial {v}) =\beta_{v}$; it  follows from the property of logarithms that $\partial Y/ \partial {c} = \frac{1}{\sigma_{t}(1+c)}$. 
 Calculating the percent change $100  \Delta c_{p}/c_{p}$ follows  from  rearranging the differential relations above, yielding $\frac{d c_{p}}{\sigma_{t}(1+c_{p}) dv} =   \beta_{v}$. Hence,  when the independent variable $\beta_{v}$  is a binary indicator variable, then the  shift   from value 0 to 1 corresponds to $dv =1$, and so the percent change  $100  \Delta c_{p}/c_{p} \approx 100  d c_{p}/c_{p} \approx 100 \times \sigma_{t} \times \beta_{v} \approx 100 \times \langle \sigma \rangle \times \beta_{v} $.  
By extension, when the independent variable  is a scalar quantity then the percent change in $c_{p}$ associated with a 1-unit increase  $dv$ is  also given by $100 \times \langle \sigma \rangle \times \beta_{v}$. And in the case that the scalar quantity enters in logarithm (e.g. $\ln k_{p}$), then  a 1\% increase in $v$ corresponds to a  $\langle \sigma \rangle \times \beta_{v} $  percent increase in $c_{p}$.

\vspace{-0.15in}
\subsubsection*{Quantifying the effect of cross-domain diversity on scientific impact}
\vspace{-0.15in}
While previous work  aimed to identify the role of $X_{CIP}$ in the ecosystem of biology and computing researchers that championed the genomics revolution \citep{Petersen:2018}, here we seek to simultaneously identify the relative impact of $X_{CIP}$ and $X_{SA}$ in the emerging ecosystem of HB science. In this way, we are able to compare research strategies that leverage combinations of diverse researcher expertise -- i.e. cross-disciplinary collaboration --  to those that do not, in the ultimate pursuit of interdisciplinary knowledge and research  \citep{nissani1995fruits}.
  
To this end, we   model the relation between  $z_{p}$ and $X_{CIP}$ \& $X_{SA}$ by applying ordinary least-squares (OLS) regression to estimate the coefficients of the  panel regression model implemented with researcher profile fixed effects:
\begin{eqnarray}
z_{a,p}=  \alpha_{a} + \beta_{k} \ln k_{p}  +  \beta_{w} \ln w_{p} +  \beta_{\tau}  \tau_{a,p}    +  \gamma_{X_{SA}} I_{X_{SA}}  + \gamma_{X_{CIP}} I_{X_{CIP}} +  \\ \nonumber  
  \gamma(y_{p,}\overrightarrow{SA}_{p},\overrightarrow{CIP}_{p},\overrightarrow{R}_{p})   + \epsilon_{a,p} \ ,   
\label{M2a}	 
\end{eqnarray}
where the model parameters are estimated using Huber-White robust standard errors, which account for heteroscedasticity and serial correlation within the publication set of each scholar, indexed by $a$. 

The  control  variables in Eq. (\ref{M2a}) include $\ln k_{p}$,  measuring the natural logarithm of the total number of coauthors;  $\ln w_{p}$ is  the natural logarithm of the total number of Major MeSH terms; the career age variable $\tau_{a,p}$, measuring the number of years since the researcher's  first publication, capturing variation attributable to the career life cycle; and we also include factor variables controlling for publication year and other article-level features measured by $\overrightarrow{SA}_{p}$, $\overrightarrow{CIP}_{p}$, $\overrightarrow{R}_{p}$.
We exclude solo-authored research papers (i.e. those with $k_{p}=1$) along with articles with a single Major MeSH category (i.e. those with $w_{p}=1$).

{\bf Table \ref{TableS4.tab}} shows the full parameter estimates for six similar models that differ primarily in the type of cross-domain diversity included as the principle test variable, represented generically by $I_{X}$. In  models (1)-(3) we vary the specification of the types of SA and CIP being tested. To be specific, in model (1) we include indicators $I_{X_{SA}}$ and $I_{X_{CIP}}$, where $I_{X_{CIP}}$ takes the value 1 if $O_{CIP}(\vec{F}_{p})=X_{CIP}$ and 0 if $O_{CIP}(\vec{F}_{p})=M$, and similarly for $I_{X_{SA}}$. These definitions of $X$ correspond to the {\it Broad} configuration,  calculated using all CIP and SA categories, as indicated in {\bf Figures \ref{Figure4.fig}}(A,D).  According to this definition, articles combining SA (CIP) from any two or more categories are classified as $X_{SA}$ ($X_{CIP}$). 

 In model (2) we use $X$ indicators defined according to the {\it Neighboring} configuration representing  shorter-distance cross-domain activity -- here capturing the neurobiological -vs- bioengineering interface. In our model specification, $X$ is represented by  the binary indicator variables $I_{X_{\text{Neighboring},SA}}$ and $I_{X_{\text{Neighboring},CIP}}$. In the case of $X_{\text{Neighboring},SA}$, this interface corresponds to articles combining   at least one MeSH mapping  onto SA 1 (Psychiatry \& Psychology) and at least one MeSH mapping onto SA  [2] (Anatomy \& Organisms), [3] (Phenomena \& Processes) or [4] (Health). 
 In the case of $X_{\text{Neighboring},CIP}$, this interface corresponds to articles combining at least one coauthor whose department maps onto CIP [1] (Neurosciences) or [3] (Psychology) and at least one coauthor whose department maps onto CIP  [2] (Biology), [4] (Biotechnology \& Genetics) or [5] (Medical Specialty) or [6] (Health Sciences) or [7] (Pathology \& Pharmacology). Note that all Scopus scholars map onto a single CIP, and so solo-authored research articles are by definition mono-disciplinary.

 In model (3) we use $X$ indicators defined according to the {Distant} configuration, representing longer-distance or ``Convergent''  cross-domain activity  -- here capturing the neuro-psycho-medical -vs- techno-computational interface. In our model specification, $X$ is represented by  the binary indicator variables $I_{X_{\text{Distant},SA}}$ and $I_{X_{\text{Distant},CIP}}$. In the case of $X_{\text{Distant},SA}$, this interface corresponds to articles combining SAs [1-4] (corresponding to Psychiatry \& Psychology (mind), Anatomy \& Organisms (structure), Phenomena \& Processes (function) and Health, respectively) and at least one MeSH mapping onto SAs [5,6] (Techniques \& Equipment and Technology \& Information Science, respectively). In the case of $X_{\text{Neighboring},CIP}$, this interface corresponds to articles combining at least one coauthor whose department maps onto CIPs [1,3,5] (Neurosciences, Psychology and Medical Specialty, respectively)  and at least one coauthor whose department maps onto CIPs  [4,8] (Biotechnology \& Genetics and Engineering \& Informatics, respectively). 
 
Likewise, Models (4-6)  instead focus on $X_{SA\&CIP}$ (represented by $I_{X_{SA\&CIP}}$);  each model corresponds to the either the {\it Broad},  {\it Neighboring} or {\it Distant} configurations defining $X_{SA}$ and $X_{CIP}$. As such, these models test the  citation premium associated with articles featuring cross-domain diversity in combination. 
Because we exclude the  confounding subsets of articles featuring $X_{SA}$ but not $X_{CIP}$, or vice versa, then the counterfactual to  $X_{SA\&CIP}$ in are articles that are mono-dimensional in both categories. Thus, since the counterfactual groups are similar, the the citation premium estimated by $\gamma_{X_{SA\&CIP}}$ are comparable with the $\gamma_{X_{SA}}$ and $\gamma_{X_{CIP}}$ estimated in models (1-3).  The full set of parameter results are reported in {\bf Table \ref{TableS5.tab}}, and the transformed point estimates measuring the percent increase in $c_{p}$ associated with each $X$ definition  are visually summarized in {\bf Fig. \ref{Figure5.fig}}(C).

\vspace{-0.15in}
\subsubsection*{Quantifying shifts in the effect of cross-domain diversity associated with the announcement of global Flagship HB projects circa 2013}
\vspace{-0.15in}
We test for shifts in the citation premium attributable to the advent of global Flagship HB projects by introducing an interaction between $I_{\text{2014+},p}$ and $I_{X_{SA\&CIP}}$, as indicated by the addition of two terms into the second row,

\begin{eqnarray}
z_{a,p}=  \alpha_{a} + \beta_{k} \ln k_{p}  +  \beta_{w} \ln w_{p} + \beta_{\tau}  \tau_{a,p}   +  \gamma_{X_{SA\&CIP}} I_{X_{SA\&CIP}}  +   \\ \nonumber  
+ \gamma_{\text{2014+}} I_{\text{2014+}}  +\delta_{X_{SA\&CIP}+} (I_{X_{SA\&CIP}} \times I_{\text{2014 +}} ) +  \\ \nonumber
 \gamma(y_{p,}\overrightarrow{SA}_{p},\overrightarrow{CIP}_{p},\overrightarrow{R}_{p})   + \epsilon_{a,p} \ ,   
\label{M2b}	 
\end{eqnarray}
As before, this Difference-in-Difference approach is based upon the counterfactual comparison of articles  featuring $X_{SA\&CIP}$ to those featuring $M$, integrating  an additional comparison between articles published after 2014  to those published before 2014. As in the previous citation model, the model parameter $\gamma_{X_{SA\&CIP}}$ represents the citation premium attributable to research endeavors simultaneously featuring cross-domain combinations of both SA and CIP. However, in this specification $\gamma_{X_{SA\&CIP}}$ applies to articles published before 2014. The analog estimate of the relative citation premium for articles published after 2014 is $\gamma_{X_{SA\&CIP}}+\delta_{X_{SA\&CIP}+}$. In other words, if all other covariates are held at the average values, then the  citation premium difference is given by $\delta_{X_{SA\&CIP}+}$, with positive (negative) values indicating an increase (decrease) in the citation premium after 2014. The principle test variables  $\gamma_{X_{SA\&CIP}}$, $\delta_{X_{SA\&CIP}+}$ and their sum are visually summarized in {\bf Fig. \ref{Figure5.fig}}(D).

\clearpage
\newpage

\begin{figure*}
\centering{\includegraphics[width=0.99\textwidth]{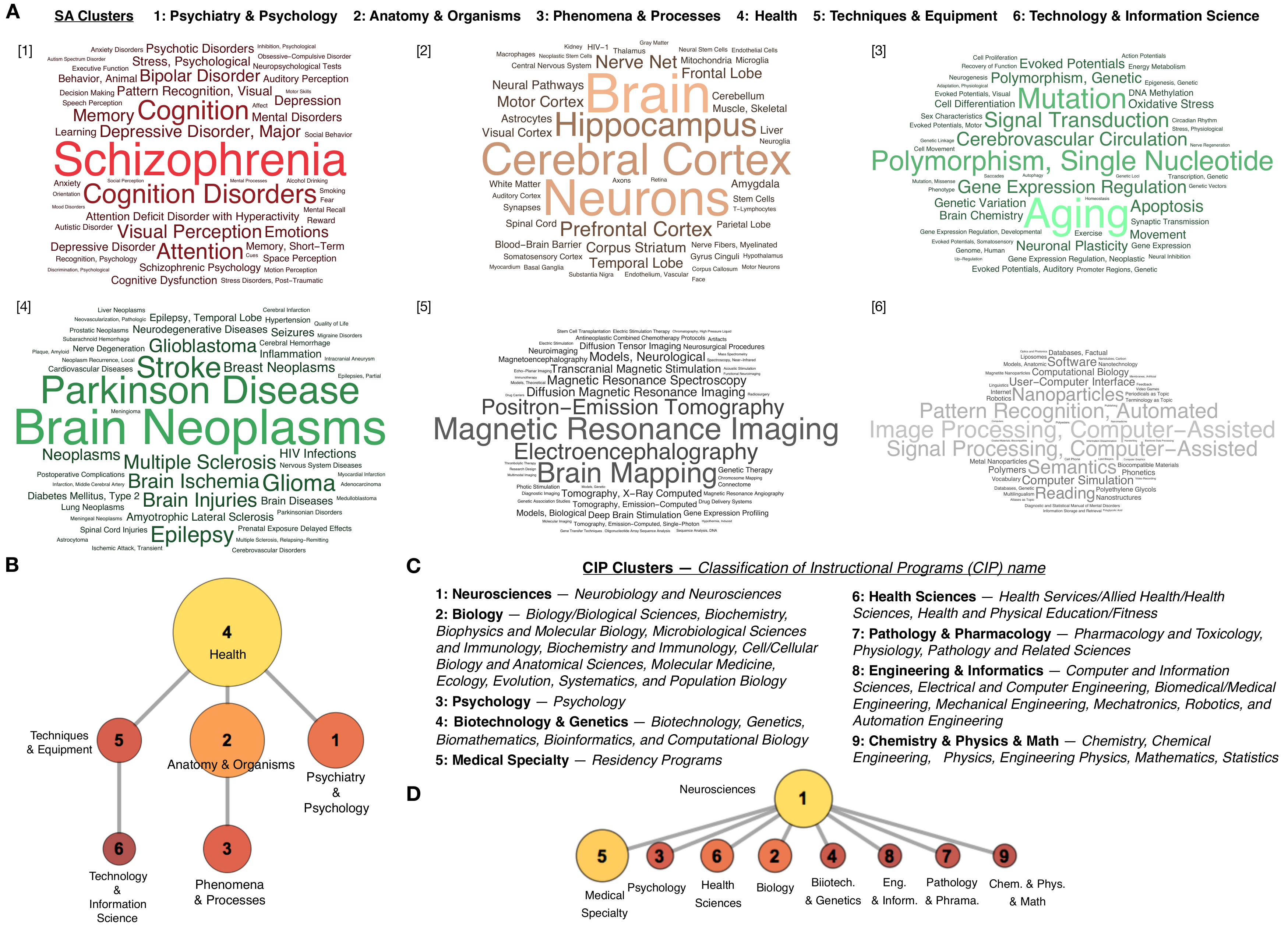}}
 \caption{  \small \label{FigureS2.fig} {\bf  Subject Area and Disciplinary clusters.} (A) Principal  MeSH terms comprising  6 Subject Area (SA) clusters. (B) Minimum spanning tree representation of topical hierarchy based upon SA co-occurrence within articles; node size proportional to total number of articles featuring a particular SA. (C) Department CIP codes comprising 9 disciplinary clusters. (D) Minimum spanning tree representation of disciplinary hierarchy based upon CIP co-occurrence within articles; node size proportional to total number of articles featuring a particular CIP. }
\end{figure*}

\begin{figure*}
\centering{\includegraphics[width=0.75\textwidth]{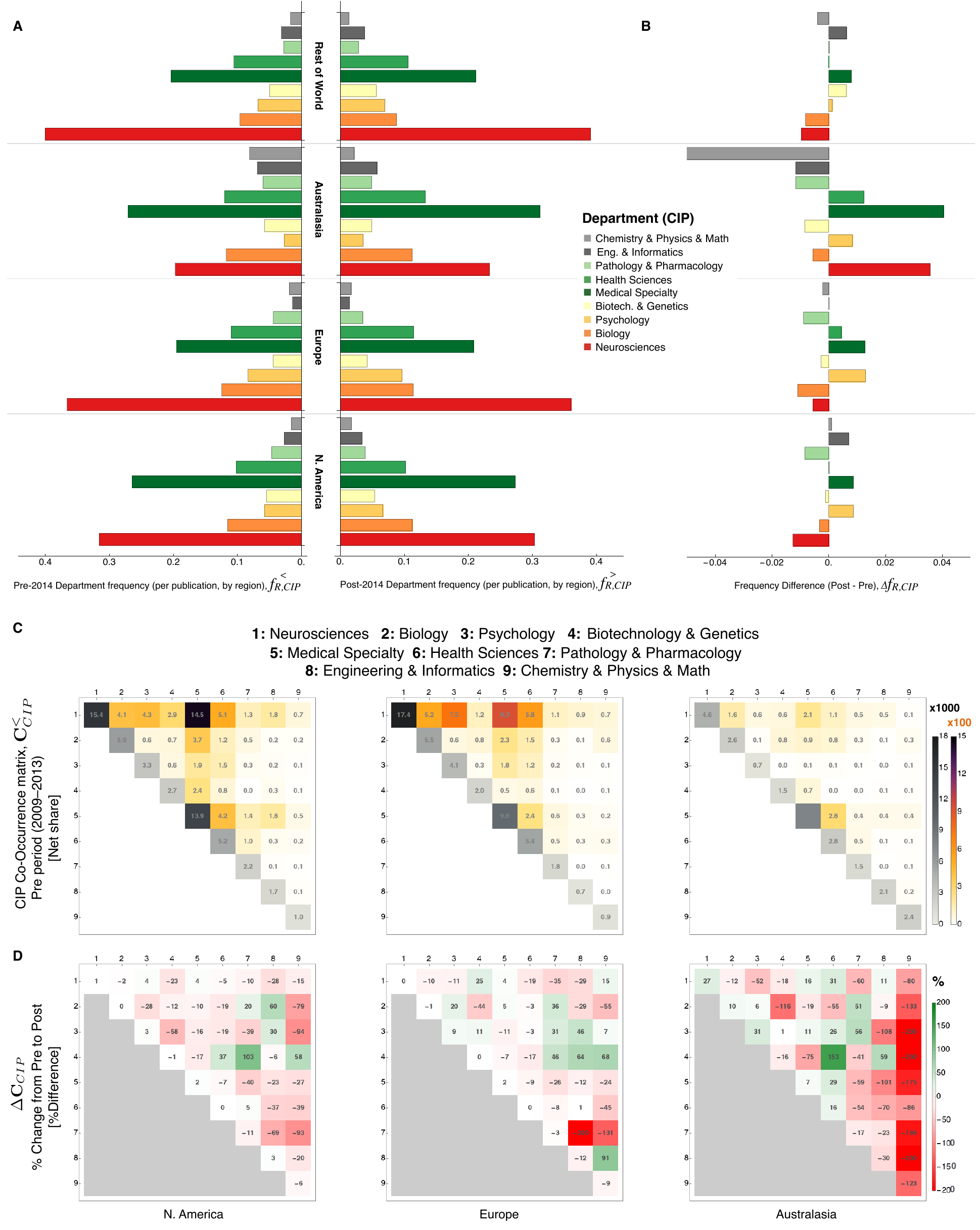}}
 \caption{  \footnotesize  \label{FigureS3.fig} {\bf Temporal and regional distributions of CIP-coded author departments in human brain research.} (A) Relative frequency of department CIP clusters in the 5-year period before 2014 ($f^{<}_{R,CIP}$) and after 2014 ($f^{>}_{R,CIP}$); $f$ values are normalized to unity within region. (B) Shift in CIP cluster frequencies given by the difference $\Delta f_{R,CIP} = f^{>}_{R,CIP}-f^{<}_{R,CIP}$. 
 (C) Disciplinary $\{CIP,CIP\}$ co-occurrence in human brain science -- by region. Each co-occurrence matrix ${\bf C}^{<}_{CIP}$ measures the frequency of a given $\{CIP,CIP\}$ pair over the 5-year pre-period 2009-2013  based upon publications associated with one of three broad geographic regions; see Eqn. (\ref{DTildematrixSI}) for its definition. By construction, matrix element values $C^{<}_{CIP,ij}$  are proportional  to the net share of publications featuring the indicated pair. Diagonal elements measure the frequency of publications  featuring only a single CIP category.  Note the use of two  legends, one for the mono-disciplinary diagonal elements (gray-scale legend reported in units of 1000 publications) and one for  off-diagonal elements (color-scale legend reported in units of 100 publications);  as indicated by the legend scales, mono-CIP publications occur with significantly higher frequency than multi-CIP publications. (D) Relative change (post - pre period) in the  co-occurrence matrix:  $\Delta C_{CIP,ij}$ measures the percent difference  in the frequency of publications characterized by each $\{CIP,CIP\}$ pair.
 }
\end{figure*}

\begin{figure*}
\centering{\includegraphics[width=0.90\textwidth]{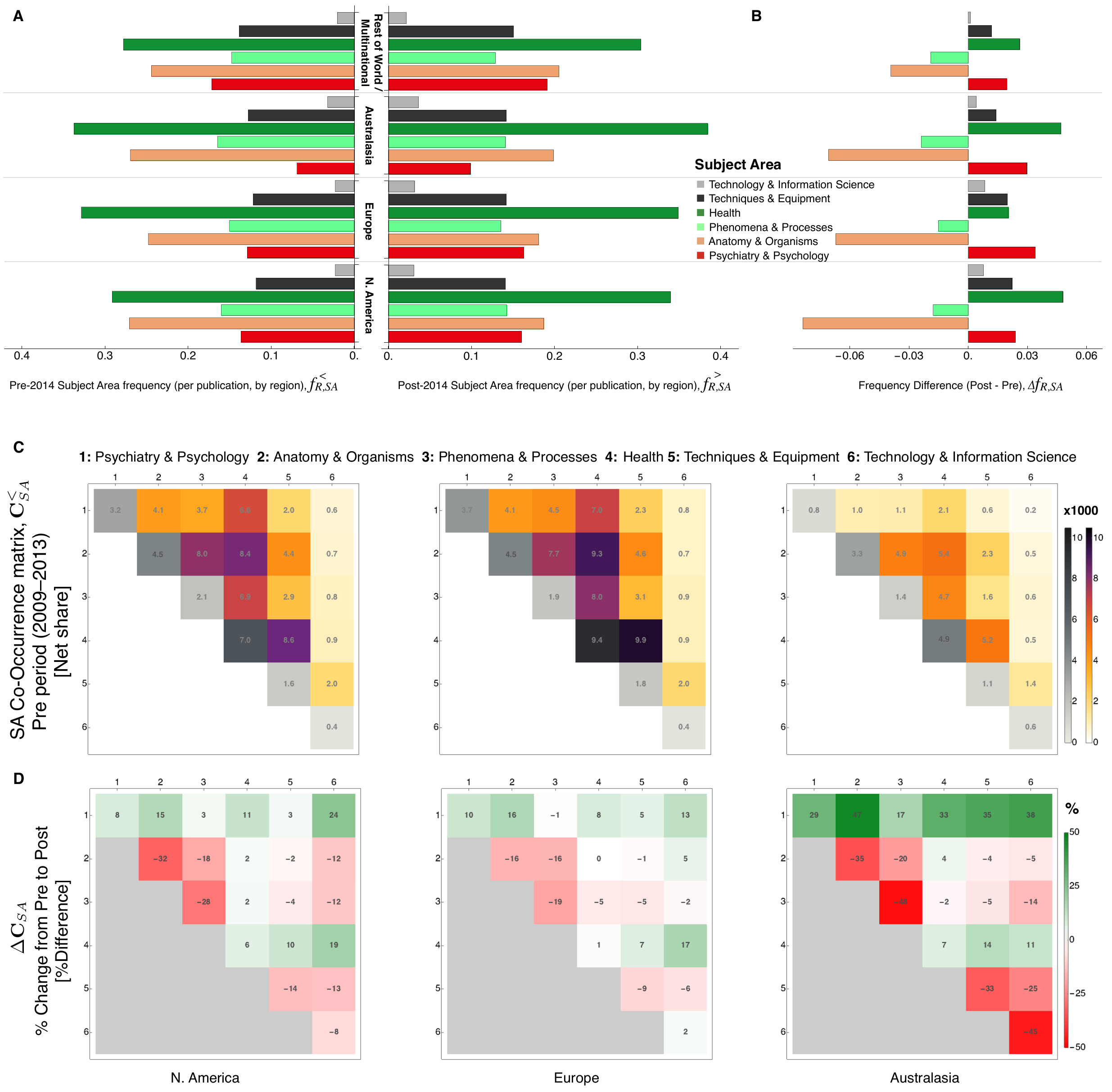}}
 \caption{ \small \label{FigureS4.fig} {\bf  Temporal and regional distributions of Subject Areas (SA) in human brain research.} (A) Relative frequency of topical SA clusters in the 5-year period before 2014 ($f^{<}_{R,SA}$) and after 2014 ($f^{>}_{R,SA}$); $f$ values are normalized to unity within region. (B) Shift in SA cluster frequencies given by the difference $\Delta f_{R,SA} = f^{>}_{R,SA}-f^{<}_{R,SA}$.
(C) Topical $\{SA,SA\}$ co-occurrence in human brain science -- by region. Each co-occurrence matrix ${\bf C}^{<}_{SA}$ measures the frequency of a given $\{SA,SA\}$ pair over the 5-year pre-period 2009-2013 based upon publications associated with one of three broad geographic regions; see Eqn. (\ref{DTildematrixSI}) for its definition. By construction, matrix element values $C^{<}_{SA,ij}$  are proportional  to the net share of publications featuring the indicated pair. Diagonal elements measure the frequency of publications  featuring only a single SA category.  Note the use of two  legends, one for the mono-dimensional diagonal elements (gray-scale legend) and one for off-diagonal elements (color-scale legend), both of which are reported in units of 1000 publications. (D) Dynamic co-occurrence matrix, $\Delta C_{SA,ij}$, measuring the percent difference  (post-pre) in the frequency of publications characterized by each $\{SA,SA\}$ pair.}
\end{figure*}

\begin{figure*}
\centering{\includegraphics[width=0.62\textwidth]{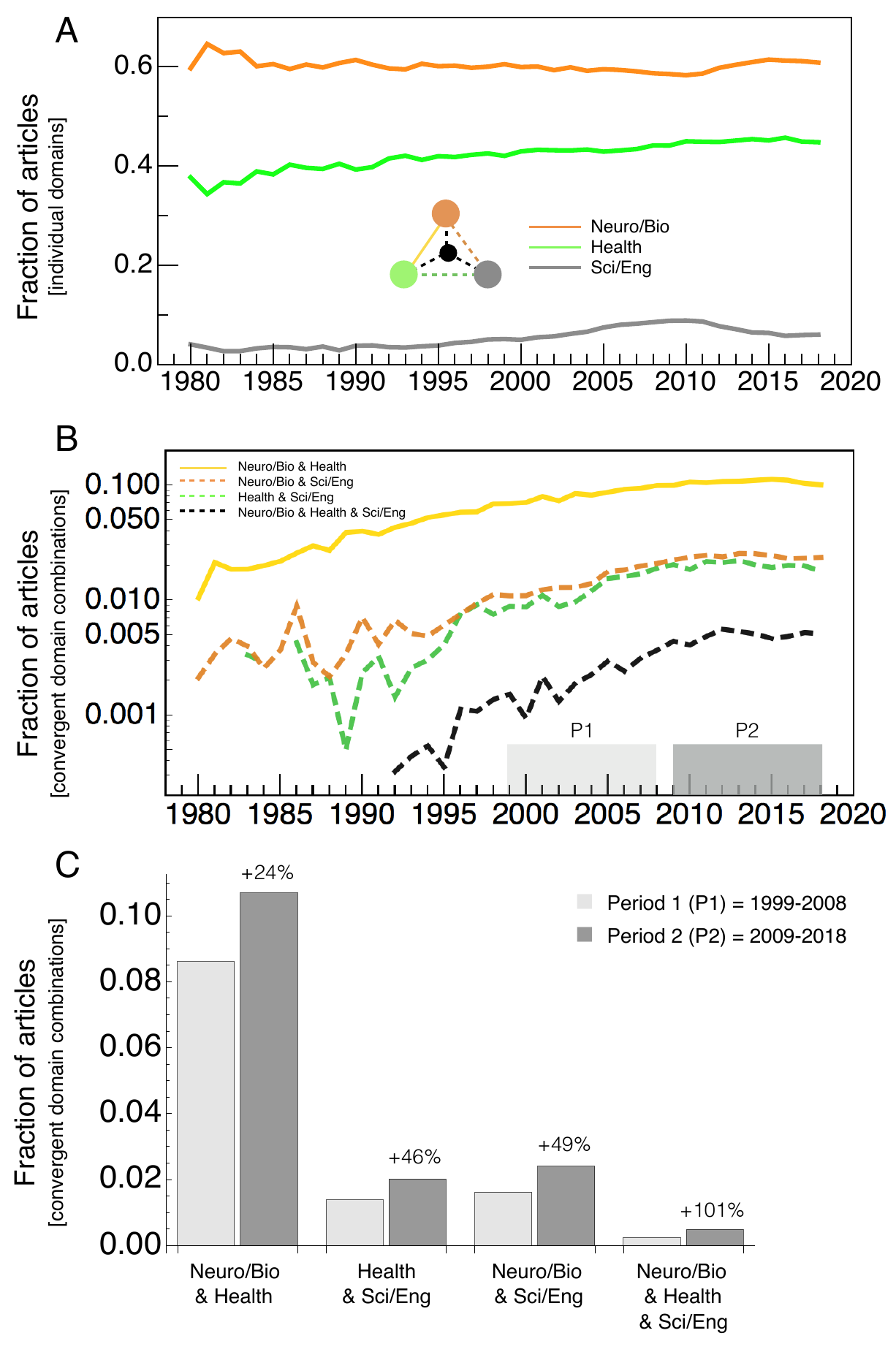}}
 \caption{  \label{FigureS11.fig} {\bf Increasing frequency of cross-disciplinary collaboration across convergent HBS interfaces.} (A) Fraction of articles featuring at least one coauthor belonging to a given CIP super-group (unconditioned on CIP of remaining coauthors): {\it Neuro/Bio} corresponds to CIP [1-4]; {\it Health} corresponds to CIP [5-7]; and {\it Sci/Eng} corresponds to CIP [8,9]  (see Fig. \ref{Figure1.fig}). The network inset indicates the 7 types of combinations across these 3 types. (B) Fraction of articles featuring combinations of scholars from two or more CIP super-groups, representing the densification of the convergent interfaces visualized in the networks shown in {\bf Fig. \ref{Figure2.fig}}(B), computed for two periods: $P1$ = [1999-2008] and $P2$ = [2009-2018].  (C) Increased frequency of convergent domain combinations between $P1$ and $P2$. For example, the most prominent convergent interface is between {\it Neuro/Bio} and {\it Health}, which was featured in 8.6\% of articles in $P1$ and 10.7\% in $P2$, corresponding to a +24\% growth in $P2$ relative to $P1$. All percent increases are significant at the $p<0.001$ level based on a two-sample two-tailed z-test comparing the proportions for $P1$ and $P2$.  }
\end{figure*}

\begin{figure*}
\centering{\includegraphics[width=0.9\textwidth]{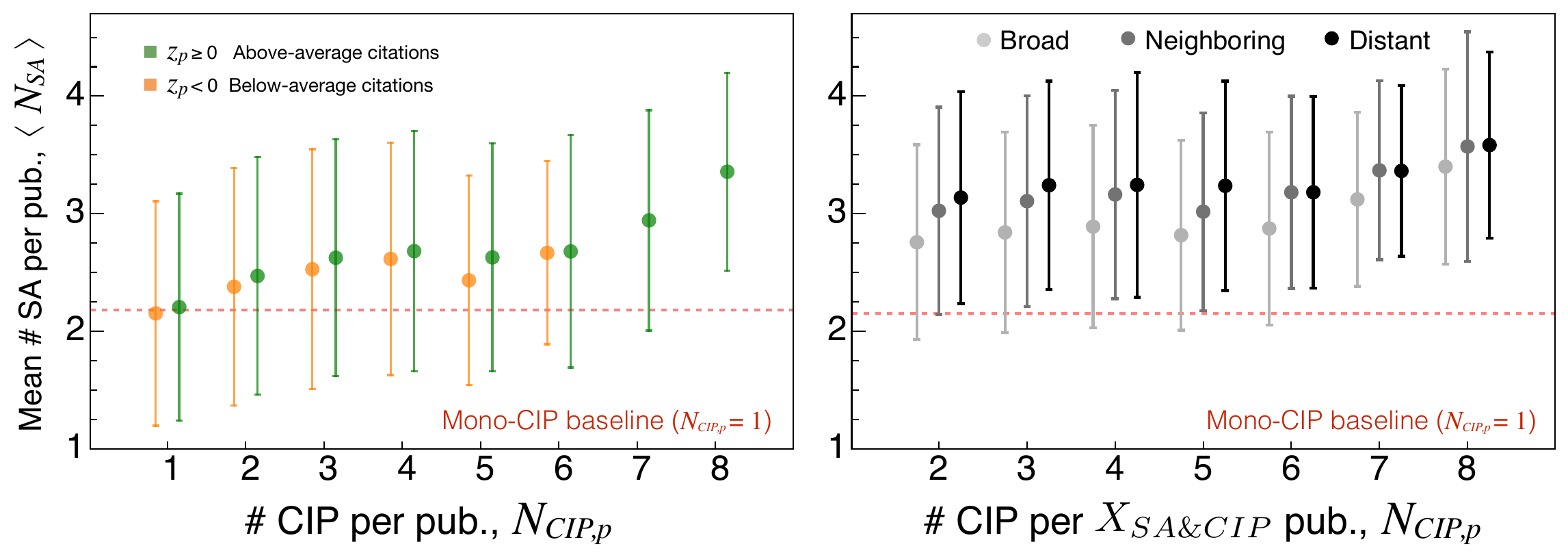}}
 \caption{  \label{FigureS12.fig} {\bf Expansive  topical integration facilitated by CIP diversity.} The number $N_{CIP,p}$ of distinct  CIP featured by a given article is a measure of disciplinary diversity. (A) Average number of SA per article, $\langle N_{SA} \rangle$, computed for articles with a given $N_{CIP,p}$ and conditioned on the normalized citation impact. 
 (B) Average number of SA per article, $\langle N_{SA}$, computed for articles  featuring $X_{SA\&CIP}$ according to a given configuration ({\it Broad}, {\it Neighboring} and {\it Distant}).  The {\it Distant} configuration consistently corresponds to the highest levels of SA diversity. Comparing panels (A) and (B),  $\langle N_{SA} \rangle $ are also consistently larger for the $N_{CIP}$ subsets in (B)  featuring $X_{SA\&CIP}$ .
 For both panels, the horizontal dashed red line represents the baseline for comparison, computed as the average number  of SA, $\langle N_{SA} \rangle$= 2.2, calculated for mono-disciplinary articles ($N_{CIP,p}=1$).   }
\end{figure*}

\begin{figure*}
\centering{\includegraphics[width=0.99\textwidth]{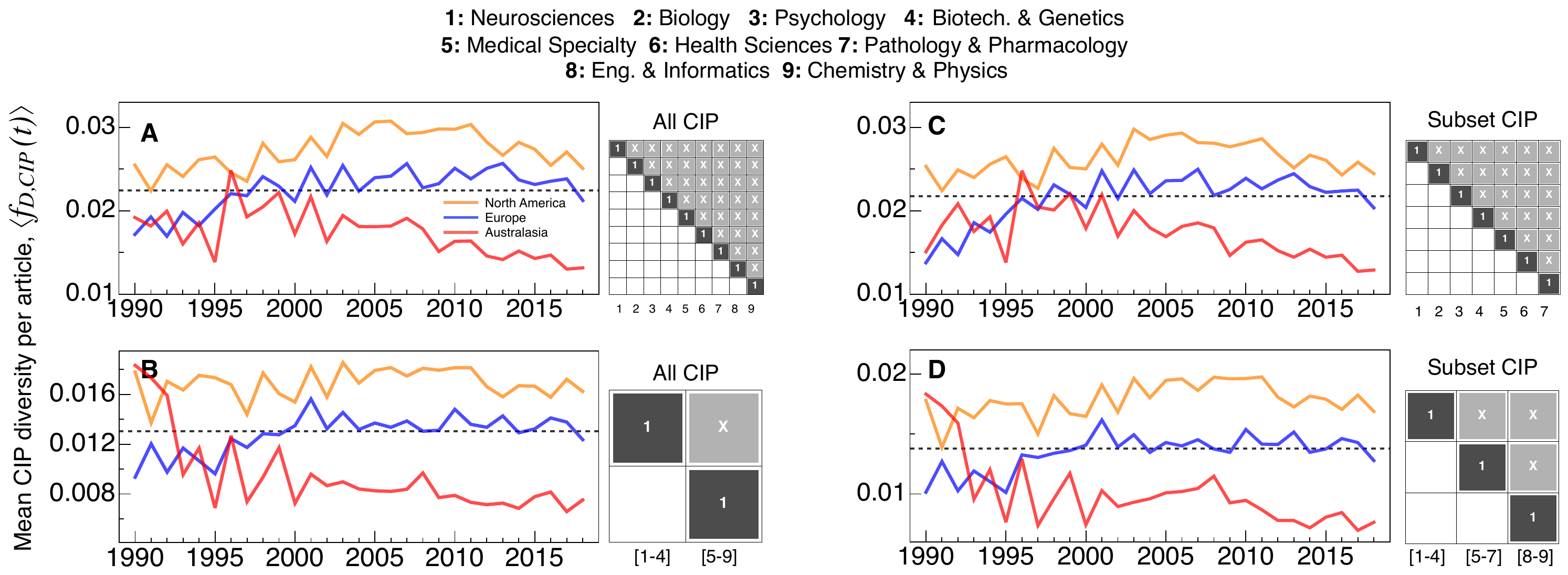}}
 \caption{   \label{FigureS6.fig} {\bf Trends in cross-disciplinary (CIP) scholarship in human brain science.}  Each curve corresponds to $\langle f_{D,CIP}(t) \rangle$, representing the average article diversity measured as categorical CIP co-occurrence in the off-diagonal matrix elements of ${\bf D}_{CIP,p}$, see Eq. (\ref{Dmatrix}); each curve is  calculated for articles belonging to a given geographic region, as determined by the  coauthors'   regional affiliations: Australasia (red), Europe (blue), and North America (orange). For each panel we provide a matrix motif indicating the set of focal CIP categories; counts for categories included in brackets are considered in union. For example, whereas panel (A) calculates $\langle f_{D,CIP}(t) \rangle$ across all 9 CIP categories (each category considered separately); instead, panel (B) calculates each ${\bf D}_{p}$ by considering just two super-groups, the first consisting of the union of CIP counts for categories [1-4], and the second comprised of categories [5-9].}
\end{figure*}

\begin{figure*}
\centering{\includegraphics[width=0.99\textwidth]{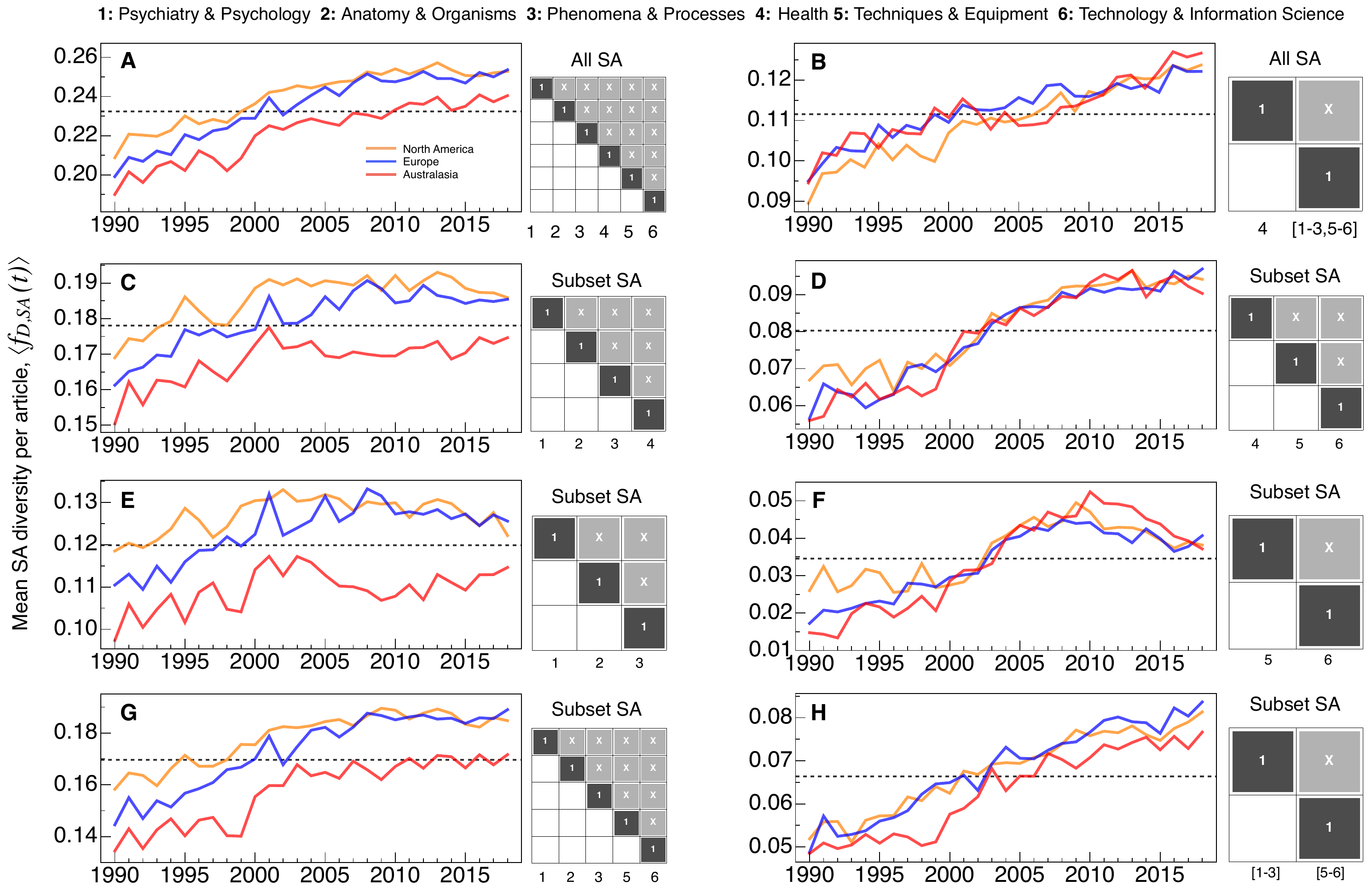}}
 \caption{   \label{FigureS7.fig} {\bf Trends in cross-topical (SA) scholarship in human brain science.}  Each curve corresponds to $\langle f_{D,SA}(t) \rangle$, representing the  average article diversity measured as categorical SA co-occurrence in the off-diagonal matrix elements of ${\bf D}_{SA,p}$, see Eq. (\ref{Dmatrix}); each curve is  calculated for articles belonging to a given geographic region, as determined by the  coauthors'   regional affiliations: Australasia (red), Europe (blue), and North America (orange). For each panel we provide a matrix motif indicating the set of focal SA categories; counts for categories included in brackets are considered in union. For example, whereas panel (A) calculates $\langle f_{D,SA}(t) \rangle$ across all 6 SA categories (each category considered separately); instead, panel (C) calculates each ${\bf D}_{SA,p}$ by considering a subset of four SA categories  1-4.}
\end{figure*}

\begin{figure*}
\centering{\includegraphics[width=0.79\textwidth]{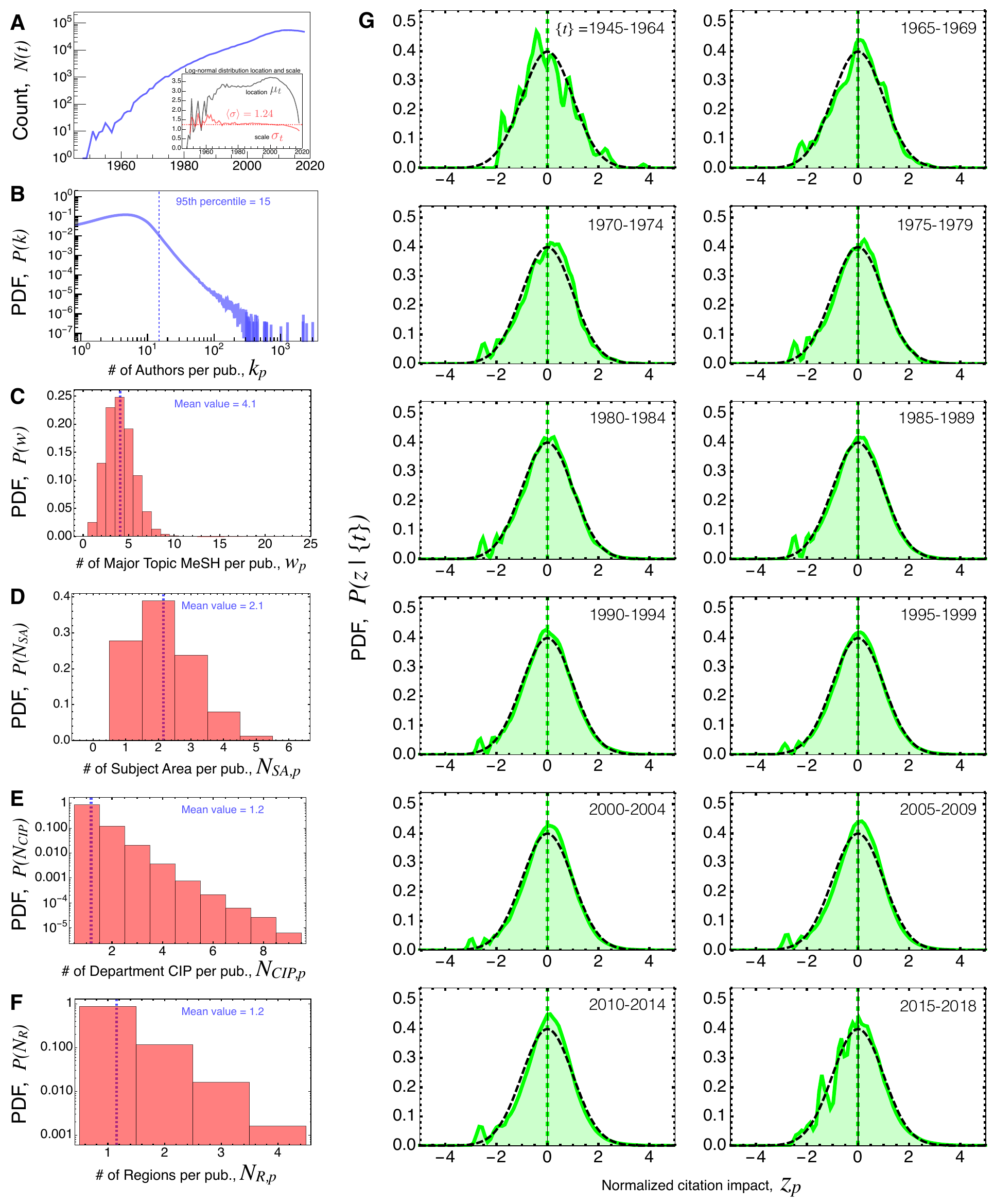}}
 \caption{ \footnotesize  \label{FigureS8.fig} {\bf Distributions of Article-level variables.} 
 (A) $N(t)$ is the number of HB articles by year. (inset) The log-citation distribution is well described by a log-normal distribution (see panel G). As such,  $\mu_{t}$ and $\sigma_{t}$ corresponding to log-transformed citation counts are appropriate measures of log-normal  location and scale; the  average and standard deviation are  $\langle \sigma \rangle \pm \text{SD} = 1.24 \pm 0.09$ over the 49-year period 1970-2018.
 (B) $P(k)$ is the probability distribution (PDF)  of the number of coauthors per article. 
 (C) $P(w)$ is the PDF  of the number of Major Topic MeSH ``keywords'' per publication, denoted by $w_{p}$. 
 (D) Each MeSH keyword maps onto one of the 6 SA clusters. Shown is the PDFof the number of  distinct SA categories per publication, $N_{SA,p}$.
 (E)  Each departmental affiliation maps onto one of the 9 CIP clusters. Shown is the PDF of the number of distinct CIP categories per publication, $N_{CIP,p}$.
 (F) Each Scopus Author's affiliation  maps onto one of 4 regions: Australasia, Europe, North America, and (rest of) World. Shown is the PDF of the number of region categories  per publication, $N_{R,p}$.
 (G) Probability distribution (PDF) of  $z_{p}$  disaggregated by publication cohort $\{t\}$; each green curve represents the smoothed kernel density estimate of the $P(z)$, calculated with kernel bandwith = 0.1.
 Data are split into 5-year periods from 1965-2018, with the first panel including data from 1945-1964. 
 Each PDF shows the baseline Normal distribution $N(0,1)$, demonstrating the stability of the distribution of  normalized citation impact values over time, thereby facilitating robust cross-temporal modeling. 
 }
\end{figure*}

\begin{figure*}
\centering{\includegraphics[width=0.99\textwidth]{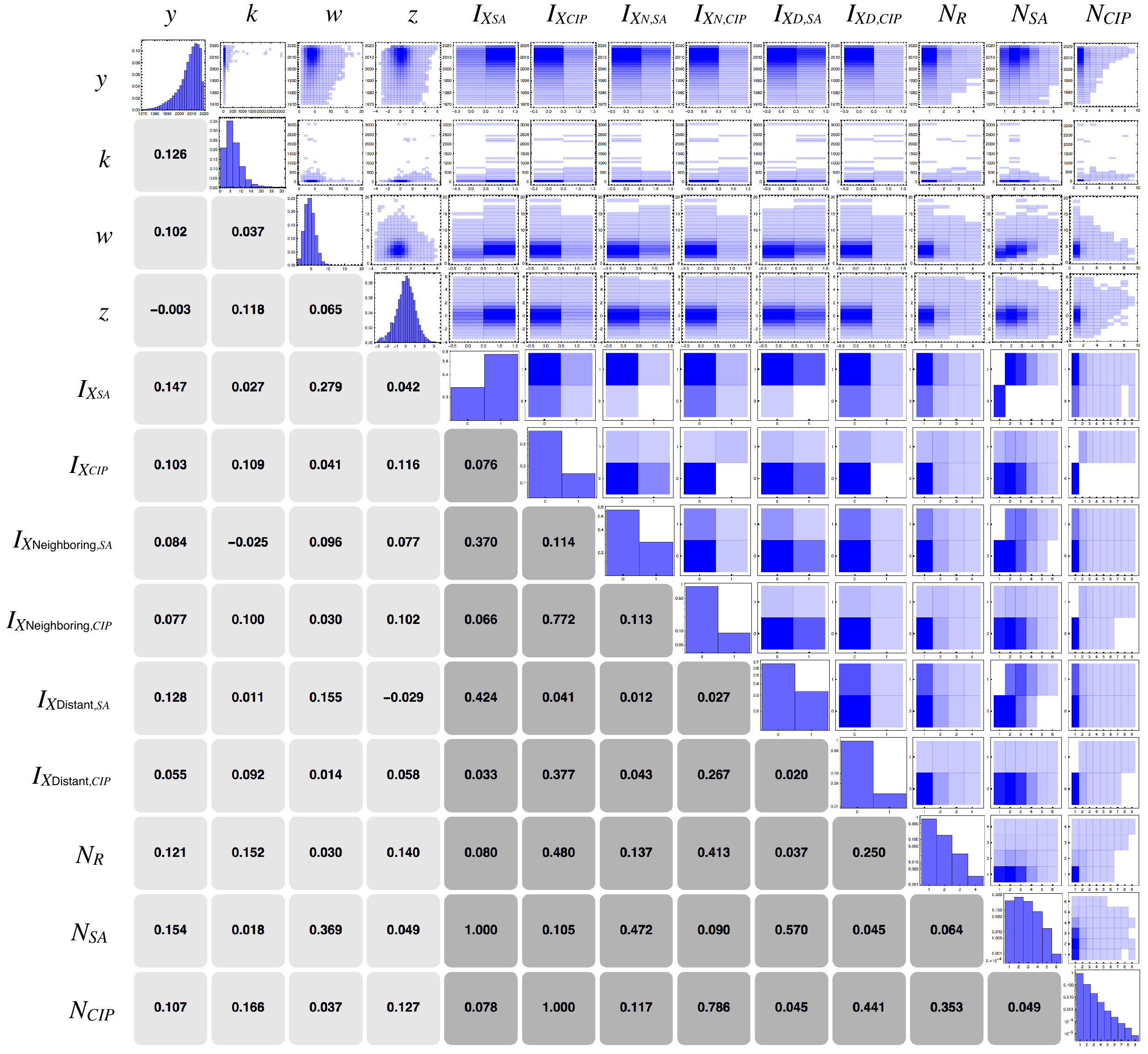}}
 \caption{  \label{FigureS9.fig} {\bf Cross-correlation and Descriptive statistics for regression model variables.} Upper-diagonal elements: bivariate histogram between row and column variables. Diagonal elements: histogram for variable indicated by the row/column labels. Lower-diagonal elements: bivariate cross-correlation coefficient: light-shaded squares indicate the Pearson's correlation coefficient between two variables that are both continuous measures; dark-shaded squares indicate the Cramer's V associate between two variables that are both nominal (categorical). 
 }
\end{figure*}

\begin{figure*}
\centering{\includegraphics[width=0.99\textwidth]{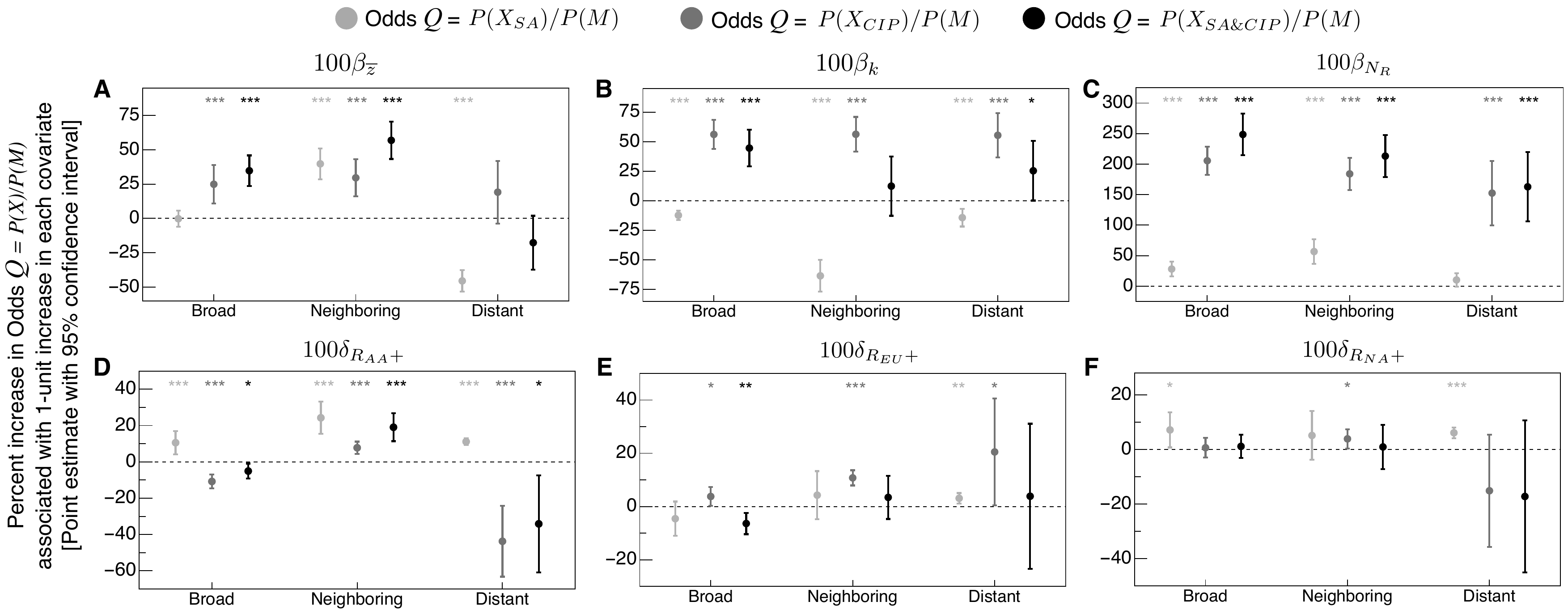}}
 \caption{  \label{FigureS10.fig} {\bf Summary of Logit model parameter estimates.} (A-C) Reported are $100 \beta$ for the main covariates of interest reported in Tables \ref{TableS1.tab}-\ref{TableS3.tab}, quantifying the percent increase in the odds $Q \equiv P(X)/P(M)$ associated with a one-unit increases in: (A) mean journal citation impact $\overline{z}_{j,p}$; (C) $\ln k$; (B) number of coauthors, $k_{p}$; (C) number of major MeSH terms (keywords), $w_{p}$; (D-F) difference-in-difference estimates ($100\delta_{R+}$) capturing the  effect of  Flagship project ramp-ups after 2013  on rates of cross-domain research -- at three levels of specificity regarding the diversity range captured by $X$.
The  {\it Broad} configuration correspond to unconstrained combinations of SA and CIP (represented by $X_{SA}$, $X_{CIP}$, $X_{SA\&CIP}$).  
The  {\it Neighboring}  configuration corresponds to specific set of category combinations capturing the neurobiological -vs- bioengineering interface, represented by SA [1] $\times$ [2-4] and CIP [1,3] $\times$  [2,4-7]  (and represented by $X_{\text{Neighboring},SA}$, $X_{\text{Neighboring},CIP}$, $X_{\text{Neighboring},SA\&CIP}$). 
 And  {\it Distant} also identifies a specific set of category combinations capturing  the neuro-psycho-medical -vs- techno- computational interface, represented by SA [1-4] $\times$ [5,6] and CIP [1,3,5] $\times$  [4,8] ($X_{\text{Distant},SA}$, $X_{\text{Distant},CIP}$, $X_{\text{Distant},SA\&CIP}$). Reported are percent increase in  $Q$, a ratio representing the propensity for cross-domain research relative to mono-domain research, directly associated with the ramp-up of Brain projects in: (D) Australasia; (E) Europe; (F) North America. Shown are point estimates with 95\% confidence interval. Standard errors clustered by region to account for residuals that are correlated within regions over time. Asterisks above each estimate indicate the associated $p-$value level:  $^{*} \ p<0.05$, $^{**} \ p<0.01$, $^{***} \ p<0.001$.}
\end{figure*}

\begin{figure*}
\centering{\includegraphics[width=0.99\textwidth]{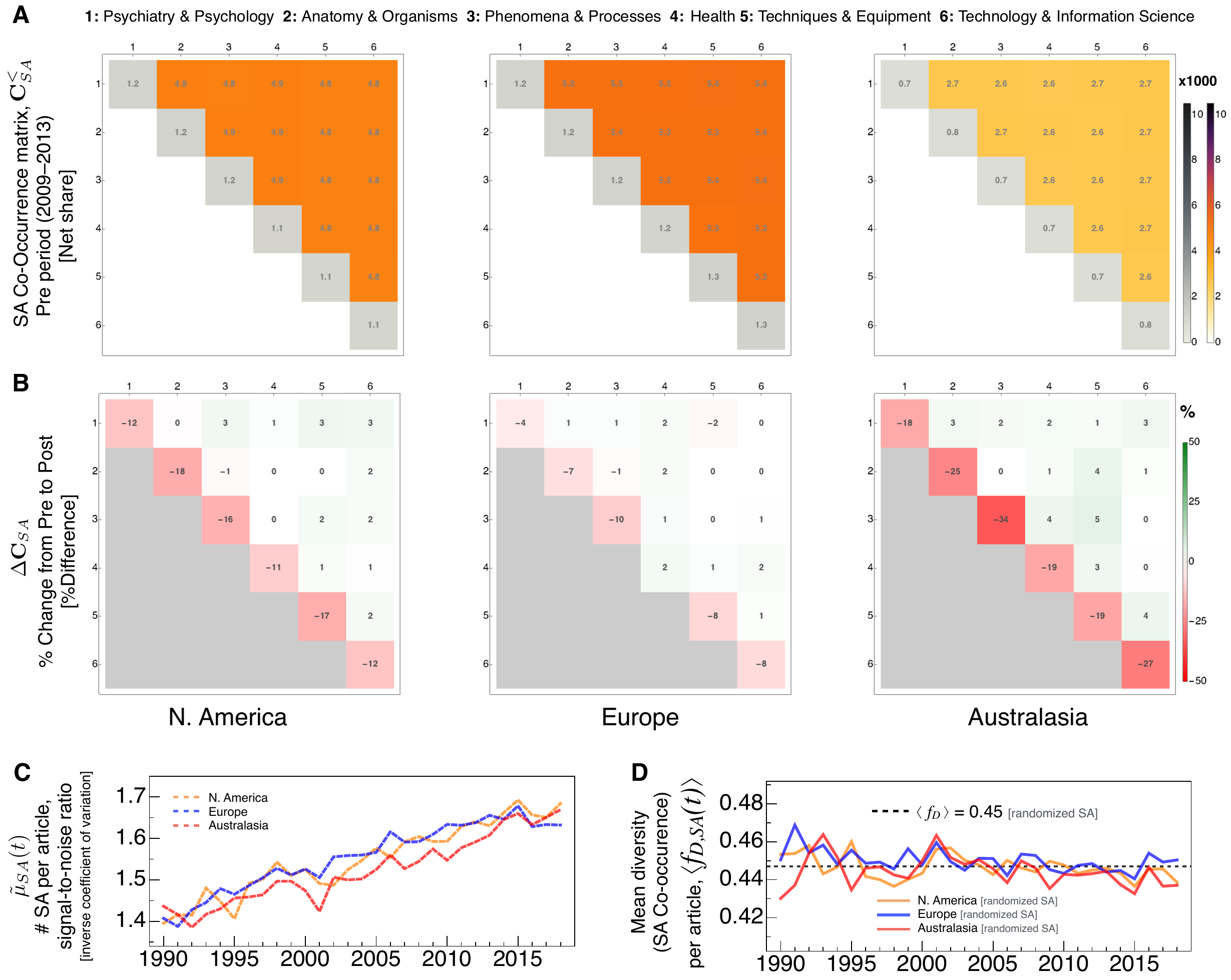}}
 \caption{ \label{FigureS5.fig} {\bf Co-occurrence metrics appropriately account for secular growth in research output and MeSH annotation.} Demonstration of consistent co-occurrence metrics calculated using randomized category vectors, $\vec{v}_{p,\text{rand.}}$, with vector elements shuffled such that the total counts $\vert \vec{v}_{p} \vert$ for each $p$ is conserved. (A) Elimination of variation among the diagonal and off-diagonal  elements of the co-occurrence matrix ${\bf C}^{<}_{SA,\text{rand.}}$ indicate that no other significant statistical biases underly this co-occurrence calculation. (B) Reduction of the relative change between two shuffled co-occurence matrices ${\bf C}^{<}_{SA,\text{rand.}}$ and ${\bf C}^{>}_{SA,\text{rand.}}$ to the level of noise; The largest off-diagonal value observed is 3\%, representing a  threshold for classifying significant shifts in the corresponding real data shown in Figs. \ref{FigureS3.fig}(D) and \ref{FigureS4.fig}(D). (C) Increase in the signal-to-noise ratio $\tilde{\mu}_{SA}(t) = \mu_{SA}(t)/\sigma_{SA}(t)$, measuring the average number of SA per article ($\mu_{SA}$), normalized by the standard deviation ($\sigma_{SA}$). Increase in the number of SA (and MeSH) per article is a source of secular growth that could introduce temporal bias challenging the interpretation of the results. (D) Accounting for this secular growth of $\tilde{\mu}_{SA}(t)$ yields a mean diversity per article $\langle f_{D,SA,\text{rand.}}$ which is approximately constant over time, consistent with the expected results for randomized category vectors, $\vec{v}_{p,\text{rand.}}$. }
\end{figure*}

\begin{figure*}
\centering{\includegraphics[width=0.99\textwidth]{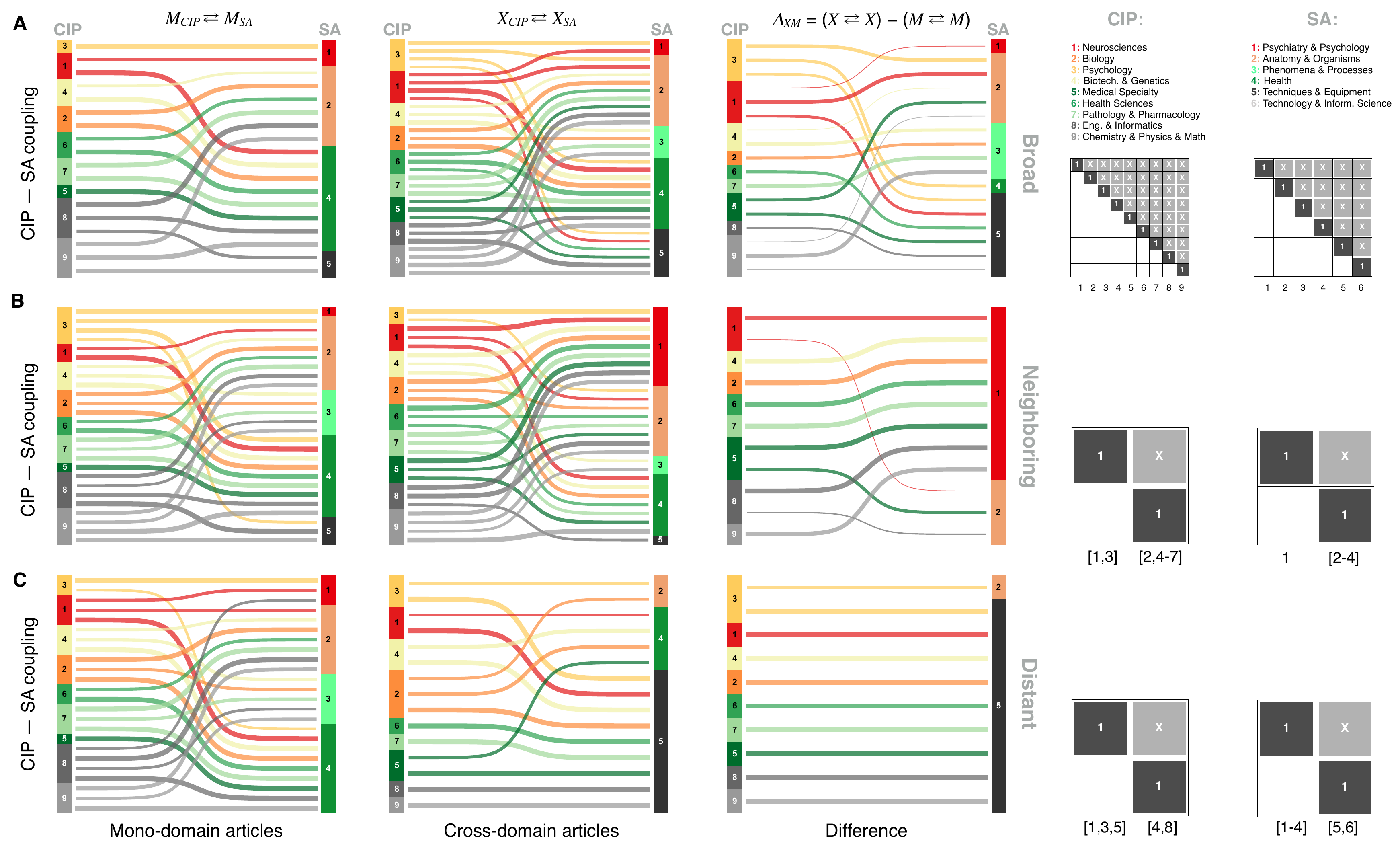}}
 \caption{   \label{FigureSX.fig} {\bf Cross-domain CIP-SA coupling.} (A) {\it Broad} configuration (as also shown in Fig. 3B, reproduced here to facilitate visual comparison). (B) {\it Neighboring}. (C)  {\it Distant}. 
 The  first column illustrates mono-mono coupling, calculated using only the mono-domain articles ($M$). For this case, the bi-partite CIP-SA networks are rather consistent across each configuration, indicating a common baseline for comparison across configurations.
 The second column shows the  CIP-SA coupling network calculated using only the cross-domain articles ($X_{SA\&CIP}$). 
The third column shows  the difference between the corresponding mono- and cross-domain networks in each row. As such, comparison across any two networks in the third column corresponds to a difference-in-difference.  
Comparison of $ \Delta_{XM}$ across configurations facilitates a qualitative difference-in-difference  for identifying emergent cross-domain coupling particular of a given configuration.
Regarding the {\it Neighboring}  configuration relative to the {\it Broad} configuration, we observe over-representation of   the links between CIP [2,4-9] and the Psychiatry \& Psychology domain SA [1], and to a lesser degree Engineering \& Informatics CIP [8] and Biology SA [2].
Likewise, in the case of the  {\it Distant} configuration,  the over-represented links  are CIP [2,4,7] with the Techniques \& Equipment SA [5]. 
}
\end{figure*}

\clearpage
\newpage

\begin{table}[h!]
\centering
\caption{Modeling the prevalence of cross-domain activity at the article level. Article-level analysis implemented using the logit model. The dependent variable is a binary indicator variable taking the value 1 if the article features cross-domain combinations (represented by $X_{SA,p}$ or $X_{CIP,p}$ or $X_{SA\&CIP,p}$) and 0 otherwise.
Publication data included:  articles published in period $ y_{p} \in [1970, 2018]$ with $k_{p} \geq 2$ and $w_{p} \geq 2$.
Robust standard errors are shown in parenthesis below each point estimate. 
Reported are odds ratios,  $\exp(\beta)$.}
\resizebox{0.99\columnwidth}{!}{ 
\def\sym#1{\ifmmode^{#1}\else\(^{#1}\)\fi}
\begin{tabular}{l*{6}{c}}
\hline\hline    &    \multicolumn{1}{c}{(1)}    &    \multicolumn{1}{c}{(2)}    &    \multicolumn{1}{c}{(3)}    &    \multicolumn{1}{c}{(4)}    &    \multicolumn{1}{c}{(5)}    &    \multicolumn{1}{c}{(6)}\\ 
    &    \multicolumn{1}{c}{$X_{SA} $}    &    \multicolumn{1}{c}{$X_{CIP}$}    &    \multicolumn{1}{c}{$X_{SA\&CIP} $}    &    \multicolumn{1}{c}{$X_{SA} $}    &    \multicolumn{1}{c}{$X_{CIP}$}    &    \multicolumn{1}{c}{$X_{SA\&CIP}$}\\ 
\hline
$y$    &    1.032\sym{***}    &    1.009\sym{***}    &    1.046\sym{***}    &    1.033\sym{***}    &    1.021\sym{***}    &    1.061\sym{***}\\ 
    &    (0.000866)    &    (0.00242)    &    (0.00339)    &    (0.00118)    &    (0.00315)    &    (0.00394)\\ 
$\overline{z}_{j}$    &    0.997    &    1.282\sym{***}    &    1.415\sym{***}    &    0.978    &    1.223\sym{*}    &    1.308\sym{**}\\ 
    &    (0.0298)    &    (0.0911)    &    (0.0805)    &    (0.0320)    &    (0.103)    &    (0.126)\\ 
$\ln k$    &    0.885\sym{***}    &    1.753\sym{***}    &    1.562\sym{***}    &    0.897\sym{***}    &    1.821\sym{***}    &    1.639\sym{***}\\ 
    &    (0.0184)    &    (0.109)    &    (0.124)    &    (0.0142)    &    (0.117)    &    (0.140)\\ 
$\ln w$    &    4.655\sym{***}    &    0.933\sym{*}    &    4.858\sym{***}    &    4.678\sym{***}    &    0.929\sym{*}    &    4.896\sym{***}\\ 
    &    (0.144)    &    (0.0285)    &    (0.217)    &    (0.150)    &    (0.0286)    &    (0.215)\\ 
$N_{R}$    &    1.324\sym{***}    &    7.810\sym{***}    &    12.00\sym{***}    &    1.211\sym{**}    &    3.028\sym{***}    &    4.569\sym{***}\\ 
    &    (0.0807)    &    (0.928)    &    (2.096)    &    (0.0755)    &    (0.789)    &    (0.898)\\ 
$N_{CIP}$    &    1.307\sym{***}    &        &        &    1.294\sym{***}    &        &    \\ 
    &    (0.0828)    &        &        &    (0.0818)    &        &    \\ 
$N_{SA}$    &        &    1.216\sym{***}    &        &        &    1.206\sym{***}    &    \\ 
    &        &    (0.0476)    &        &        &    (0.0487)    &    \\ 
\hline
$I_{\text{2014+}}$    &        &        &        &    0.949    &    0.754\sym{***}    &    0.716\sym{***}\\ 
    &        &        &        &    (0.0348)    &    (0.0420)    &    (0.0367)\\ 
$I_{R_{NA}}$    &        &        &        &    0.913    &    0.380\sym{***}    &    0.397\sym{***}\\ 
    &        &        &        &    (0.0680)    &    (0.103)    &    (0.0755)\\ 
$I_{R_{EU}}$    &        &        &        &    0.942    &    0.313\sym{***}    &    0.345\sym{***}\\ 
    &        &        &        &    (0.0712)    &    (0.0849)    &    (0.0669)\\ 
$I_{R_{AA}}$    &        &        &        &    0.746\sym{***}    &    0.229\sym{***}    &    0.191\sym{***}\\ 
    &        &        &        &    (0.0567)    &    (0.0651)    &    (0.0386)\\ 
$I_{R_{NA}} \times I_{\text{2014+}}$    &        &        &        &    1.074\sym{*}    &    1.007    &    1.012\\ 
    &        &        &        &    (0.0352)    &    (0.0186)    &    (0.0218)\\ 
$I_{R_{EU}} \times I_{\text{2014+}}$    &        &        &        &    0.955    &    1.038\sym{*}    &    0.938\sym{**}\\ 
    &        &        &        &    (0.0313)    &    (0.0186)    &    (0.0192)\\ 
$I_{R_{AA}} \times I_{\text{2014+}}$    &        &        &        &    1.111\sym{**}    &    0.898\sym{***}    &    0.950\sym{*}\\ 
\hline
\(N\)    &    602599    &    602599    &    207281    &    602599    &    602599    &    207281\\ 
Pseudo \(R^{2}\)  & 0.0725 & 0.1945 & 0.3081 & 0.0735 & 0.2045 & 0.3191 \\ 
\hline\hline
\multicolumn{7}{l}{\footnotesize Exponentiated coefficients; Standard errors in parentheses}\\ 
\multicolumn{7}{l}{\footnotesize \sym{*} \(p<0.05\), \sym{**} \(p<0.01\), \sym{***} \(p<0.001\)}\\ 
\end{tabular}}
\label{TableS1.tab} 
\end{table}

\begin{table}[htbp]\centering
\caption{Conditional definition of $X_{p}$ -- identifying ``Neighboring'' or shorter-distance cross-domain combinations. 
Article-level analysis implemented using the logit model. The dependent variable is a binary indicator variable taking the value 1 if the article features cross-domain combinations (represented by $X_{\text{Neighboring},SA,p}$ or $X_{\text{Neighboring},CIP,p}$ or $X_{\text{Neighboring},SA\&CIP,p}$) and 0 otherwise.
Publication data included:  articles published in period $ y_{p} \in [1970, 2018]$ with $k_{p} \geq 2$ and $w_{p} \geq 2$.
Robust standard errors are shown in parenthesis below each point estimate. 
Reported are odds ratios,  $\exp(\beta)$.}
\resizebox{0.99\columnwidth}{!}{ 
\def\sym#1{\ifmmode^{#1}\else\(^{#1}\)\fi}
\begin{tabular}{l*{6}{c}}
\hline\hline    &    \multicolumn{1}{c}{(1)}    &    \multicolumn{1}{c}{(2)}    &    \multicolumn{1}{c}{(3)}    &    \multicolumn{1}{c}{(4)}    &    \multicolumn{1}{c}{(5)}    &    \multicolumn{1}{c}{(6)}\\ 
    &    \multicolumn{1}{c}{$X_{\text{Neighboring},SA}$}    &    \multicolumn{1}{c}{$X_{\text{Neighboring},CIP}$}    &    \multicolumn{1}{c}{$X_{\text{Neighboring},SA\&CIP}$}    &    \multicolumn{1}{c}{$X_{\text{Neighboring},SA}$}    &    \multicolumn{1}{c}{$X_{\text{Neighboring},CIP}$}    &    \multicolumn{1}{c}{$X_{\text{Neighboring},SA\&CIP}$}\\ 
\hline
$y$    &    1.030\sym{***}    &    1.002    &    1.025\sym{***}    &    1.028\sym{***}    &    1.012\sym{**}    &    1.036\sym{***}\\ 
    &    (0.00243)    &    (0.00267)    &    (0.00284)    &    (0.00294)    &    (0.00378)    &    (0.00513)\\ 
$\overline{z}_{j}$    &    1.488\sym{***}    &    1.344\sym{***}    &    1.765\sym{***}    &    1.428\sym{***}    &    1.266\sym{***}    &    1.646\sym{***}\\ 
    &    (0.0855)    &    (0.0929)    &    (0.122)    &    (0.0746)    &    (0.0757)    &    (0.0954)\\ 
$\ln k$    &    0.530\sym{***}    &    1.755\sym{***}    &    1.132    &    0.543\sym{***}    &    1.832\sym{***}    &    1.188\\ 
    &    (0.0362)    &    (0.131)    &    (0.145)    &    (0.0346)    &    (0.133)    &    (0.154)\\ 
$\ln w$    &    1.756\sym{***}    &    0.889\sym{***}    &    1.816\sym{***}    &    1.788\sym{***}    &    0.889\sym{***}    &    1.799\sym{***}\\ 
    &    (0.0992)    &    (0.0304)    &    (0.106)    &    (0.0828)    &    (0.0237)    &    (0.0888)\\ 
$N_{R}$    &    1.763\sym{***}    &    6.297\sym{***}    &    8.424\sym{***}    &    1.853\sym{***}    &    2.617\sym{**}    &    4.071\sym{***}\\ 
    &    (0.181)    &    (0.845)    &    (1.485)    &    (0.268)    &    (0.867)    &    (1.679)\\ 
$N_{CIP}$    &    1.429\sym{**}    &        &        &    1.415\sym{**}    &        &    \\ 
    &    (0.172)    &        &        &    (0.171)    &        &    \\ 
$N_{SA}$    &        &    1.230\sym{***}    &        &        &    1.215\sym{***}    &    \\ 
    &        &    (0.0526)    &        &        &    (0.0529)    &    \\ 
\hline   
$I_{\text{2014+}}$    &        &        &        &    1.029    &    0.770\sym{***}    &    0.795\sym{**}\\ 
    &        &        &        &    (0.0560)    &    (0.0478)    &    (0.0653)\\ 
$I_{R_{NA}}$    &        &        &        &    1.122    &    0.405\sym{*}    &    0.511\\ 
    &        &        &        &    (0.182)    &    (0.151)    &    (0.225)\\ 
$I_{R_{EU}}$    &        &        &        &    1.192    &    0.327\sym{**}    &    0.404\sym{*}\\ 
    &        &        &        &    (0.202)    &    (0.121)    &    (0.179)\\ 
$I_{R_{AA}}$    &        &        &        &    0.626\sym{**}    &    0.172\sym{***}    &    0.156\sym{***}\\ 
    &        &        &        &    (0.110)    &    (0.0651)    &    (0.0700)\\ 
$I_{R_{NA}} \times I_{\text{2014+}}$    &        &        &        &    1.053    &    1.040\sym{*}    &    1.009\\ 
    &        &        &        &    (0.0481)    &    (0.0187)    &    (0.0418)\\ 
$I_{R_{EU}} \times I_{\text{2014+}}$    &        &        &        &    1.044    &    1.114\sym{***}    &    1.035\\ 
    &        &        &        &    (0.0481)    &    (0.0163)    &    (0.0429)\\ 
$I_{R_{AA}} \times I_{\text{2014+}}$    &        &        &        &    1.274\sym{***}    &    1.081\sym{***}    &    1.210\sym{***}\\ 
    &        &        &        &    (0.0578)    &    (0.0187)    &    (0.0475)\\ 
\hline
\(N\)    &    602599    &    602599    &    430801    &    602599    &    602599    &    430801\\ 
Pseudo \(R^{2}\)  & 0.0496 & 0.1716 & 0.1919 & 0.0554 &  0.1837 & 0.2041 \\
\hline\hline
\multicolumn{7}{l}{\footnotesize Exponentiated coefficients; Standard errors in parentheses}\\ 
\multicolumn{7}{l}{\footnotesize \sym{*} \(p<0.05\), \sym{**} \(p<0.01\), \sym{***} \(p<0.001\)}\\ 
\end{tabular}}
\label{TableS2.tab} 
\end{table}

\begin{table}[htbp]\centering
\caption{Conditional definition of $X_{p}$ -- identifying ``Distant'' or longer-distance cross-domain combinations. 
Article-level analysis implemented using the logit model. The dependent variable is a binary indicator variable taking the value 1 if the article features cross-domain combinations (represented by $X_{\text{Distant},SA,p}$ or $X_{\text{Distant},CIP,p}$ or $X_{\text{Distant},SA\&CIP,p}$) and 0 otherwise.
Publication data included:  articles published in period $ y_{p} \in [1970, 2018]$ with $k_{p} \geq 2$ and $w_{p} \geq 2$.
Robust standard errors are shown in parenthesis below each point estimate. 
Reported are odds ratios,  $\exp(\beta)$.}
\resizebox{0.99\columnwidth}{!}{ 
\def\sym#1{\ifmmode^{#1}\else\(^{#1}\)\fi}
\begin{tabular}{l*{6}{c}}
\hline\hline    &    \multicolumn{1}{c}{(1)}    &    \multicolumn{1}{c}{(2)}    &    \multicolumn{1}{c}{(3)}    &    \multicolumn{1}{c}{(4)}    &    \multicolumn{1}{c}{(5)}    &    \multicolumn{1}{c}{(6)}\\ 
    &    \multicolumn{1}{c}{$X_{\text{Distant},SA}$}    &    \multicolumn{1}{c}{$X_{\text{Distant},CIP}$}    &    \multicolumn{1}{c}{$X_{\text{Distant},SA\&CIP}$}    &    \multicolumn{1}{c}{$X_{\text{Distant},SA}$}    &    \multicolumn{1}{c}{$X_{\text{Distant},CIP}$}    &    \multicolumn{1}{c}{$X_{\text{Distant},SA\&CIP}$}\\ 
\hline
$y$    &    1.033\sym{***}    &    1.017\sym{***}    &    1.043\sym{***}    &    1.036\sym{***}    &    1.035\sym{***}    &    1.071\sym{***}\\ 
    &    (0.000769)    &    (0.00445)    &    (0.00451)    &    (0.00174)    &    (0.0101)    &    (0.00955)\\ 
$\overline{z}_{j}$    &    0.635\sym{***}    &    1.210    &    0.838    &    0.624\sym{***}    &    1.127    &    0.750\sym{**}\\ 
    &    (0.0253)    &    (0.141)    &    (0.0837)    &    (0.0202)    &    (0.119)    &    (0.0659)\\ 
$\ln k$    &    0.867\sym{***}    &    1.740\sym{***}    &    1.289\sym{*}    &    0.879\sym{***}    &    1.861\sym{***}    &    1.403\sym{*}\\ 
    &    (0.0331)    &    (0.166)    &    (0.166)    &    (0.0318)    &    (0.178)    &    (0.185)\\ 
$\ln w$    &    2.258\sym{***}    &    0.918    &    2.584\sym{***}    &    2.261\sym{***}    &    0.894    &    2.496\sym{***}\\ 
    &    (0.114)    &    (0.111)    &    (0.211)    &    (0.115)    &    (0.105)    &    (0.167)\\ 
$N_{R}$    &    1.107    &    4.594\sym{***}    &    5.094\sym{***}    &    0.986    &    1.652\sym{*}    &    1.924\sym{**}\\ 
    &    (0.0627)    &    (1.235)    &    (1.476)    &    (0.0549)    &    (0.340)    &    (0.432)\\ 
$N_{CIP}$    &    1.181\sym{***}    &        &        &    1.169\sym{***}    &        &    \\ 
    &    (0.0204)    &        &        &    (0.0204)    &        &    \\ 
$N_{SA}$    &        &    1.183    &        &        &    1.171    &    \\ 
    &        &    (0.114)    &        &        &    (0.115)    &    \\ 
 \hline
$I_{\text{2014+}}$    &        &        &        &    0.871\sym{***}    &    0.735\sym{*}    &    0.648\sym{**}\\ 
    &        &        &        &    (0.0130)    &    (0.105)    &    (0.0991)\\ 
$I_{R_{NA}}$    &        &        &        &    0.872\sym{*}    &    0.378\sym{***}    &    0.450\sym{***}\\ 
    &        &        &        &    (0.0494)    &    (0.0957)    &    (0.100)\\ 
$I_{R_{EU}}$    &        &        &        &    0.894    &    0.123\sym{***}    &    0.132\sym{***}\\ 
    &        &        &        &    (0.0537)    &    (0.0306)    &    (0.0298)\\ 
$I_{R_{AA}}$    &        &        &        &    0.725\sym{***}    &    0.188\sym{***}    &    0.130\sym{***}\\ 
    &        &        &        &    (0.0460)    &    (0.0505)    &    (0.0305)\\ 
$I_{R_{NA}} \times I_{\text{2014+}}$    &        &        &        &    1.063\sym{***}    &    0.860    &    0.842\\ 
    &        &        &        &    (0.0106)    &    (0.0900)    &    (0.120)\\ 
$I_{R_{EU}} \times I_{\text{2014+}}$    &        &        &        &    1.031\sym{**}    &    1.228\sym{*}    &    1.039\\ 
    &        &        &        &    (0.0104)    &    (0.125)    &    (0.144)\\ 
$I_{R_{AA}} \times I_{\text{2014+}}$    &        &        &        &    1.118\sym{***}    &    0.646\sym{***}    &    0.711\sym{*}\\ 
    &        &        &        &    (0.0106)    &    (0.0643)    &    (0.0968)\\  
\hline
\(N\)    &    602599    &    602599    &    396471    &    602599    &    602599    &    396471\\  
Pseudo \(R^{2}\)  & 0.0375 & 0.1492 & 0.1474 & 0.0496 &  0.1716 & 0.1919 \\
\hline\hline
\multicolumn{7}{l}{\footnotesize Exponentiated coefficients; Standard errors in parentheses}\\ 
\multicolumn{7}{l}{\footnotesize \sym{*} \(p<0.05\), \sym{**} \(p<0.01\), \sym{***} \(p<0.001\)}\\ 
\end{tabular}}\label{TableS3.tab}
\end{table}

\begin{table}[htbp]\centering
\caption{Career-level analysis using panel model with individual researcher fixed effects.
Publication data included:  articles published in period $ y_{p} \in [1970, 2018]$ with  $k_{p} \geq 2$ and $w_{p} \geq 2$; only includes researchers with $N_{a} \geq 10$ articles satisfying these criteria.
Robust standard errors are shown in parenthesis below each point estimate. Y indicates additional fixed effects included in the regression model.}
\resizebox{0.99\columnwidth}{!}{ 
\def\sym#1{\ifmmode^{#1}\else\(^{#1}\)\fi}
\begin{tabular}{l*{6}{c}}
\hline\hline    &    \multicolumn{1}{c}{(1)}    &    \multicolumn{1}{c}{(2)}    &    \multicolumn{1}{c}{(3)}    &    \multicolumn{1}{c}{(4)}    &    \multicolumn{1}{c}{(5)}    &    \multicolumn{1}{c}{(6)}\\ 
    &    \multicolumn{1}{c}{$z_{p}$}    &    \multicolumn{1}{c}{$z_{p}$}    &    \multicolumn{1}{c}{$z_{p}$}    &    \multicolumn{1}{c}{$z_{p}$}    &    \multicolumn{1}{c}{$z_{p}$}    &    \multicolumn{1}{c}{$z_{p}$}\\ 
\hline
$\ln k$    &    0.415\sym{***}    &    0.416\sym{***}    &    0.421\sym{***}    &    0.438\sym{***}    &    0.424\sym{***}    &    0.406\sym{***}\\ 
    &    (0.00470)    &    (0.00468)    &    (0.00469)    &    (0.00551)    &    (0.00547)    &    (0.00527)\\ 
$\ln w$    &    0.0320\sym{***}    &    0.0376\sym{***}    &    0.0379\sym{***}    &    0.0244\sym{***}    &    0.0544\sym{***}    &    0.0385\sym{***}\\ 
    &    (0.00468)    &    (0.00465)    &    (0.00466)    &    (0.00615)    &    (0.00528)    &    (0.00536)\\ 
$\tau$    &    -0.0107\sym{***}    &    -0.0106\sym{***}    &    -0.0105\sym{***}    &    -0.0115\sym{***}    &    -0.00987\sym{***}    &    -0.0101\sym{***}\\ 
    &    (0.00148)    &    (0.00148)    &    (0.00149)    &    (0.00229)    &    (0.00170)    &    (0.00157)\\ 
\hline
$I_{X_{SA}}$    &    0.0480\sym{***}    &        &        &        &        &    \\ 
    &    (0.00367)    &        &        &        &        &    \\ 
$I_{X_{CIP}}$    &    0.0691\sym{***}    &        &        &        &        &    \\ 
    &    (0.00466)    &        &        &        &        &    \\ 
\hline
$I_{X_{\text{Neighboring},SA}}$    &        &    0.0878\sym{***}    &        &        &        &    \\ 
    &        &    (0.00461)    &        &        &        &    \\ 
$I_{X_{\text{Neighboring},CIP}}$    &        &    0.0675\sym{***}    &        &        &        &    \\ 
    &        &    (0.00496)    &        &        &        &    \\ 
\hline
$I_{X_{\text{Distant},SA}}$    &        &        &    -0.00993\sym{**}    &        &        &    \\ 
    &        &        &    (0.00376)    &        &        &    \\ 
$I_{X_{\text{Distant},CIP}}$    &        &        &    0.0205\sym{*}    &        &        &    \\ 
    &        &        &    (0.0102)    &        &        &    \\
\hline
$I_{X_{SA\&CIP}}$    &        &        &        &    0.132\sym{***}    &        &    \\ 
    &        &        &        &    (0.00728)    &        &    \\ 
$I_{X_{\text{Neighboring},SA\&CIP}}$    &        &        &        &        &    0.132\sym{***}    &    \\ 
    &        &        &        &        &    (0.00886)    &    \\ 
$I_{X_{\text{Distant},SA\&CIP}}$    &        &        &        &        &        &    0.0424\sym{**}\\ 
    &        &        &        &        &        &    (0.0158)\\  
\hline
constant    &    -0.719\sym{***}    &    -0.719\sym{***}    &    -0.736\sym{***}    &    -0.677\sym{***}    &    -0.781\sym{***}    &    -0.711\sym{***}\\
    &    (0.0541)    &    (0.0540)    &    (0.0543)    &    (0.0770)    &    (0.0547)    &    (0.0583)\\ 
Year  ($y$)  dummy    &        Y       &     Y     &        Y  &        Y         &      Y     &      Y                \\
Topic category ($\overrightarrow{SA}$) dummy     &        Y       &     Y     &        Y  &        Y         &      Y       &      Y             \\
Department  category ($\overrightarrow{CIP}$)  dummy      &        Y       &     Y     &        Y  &        Y         &      Y     &      Y                \\
 Region ($\overrightarrow{R}$) dummy     &        Y       &     Y     &        Y  &        Y         &      Y        &      Y             \\
\hline
\(N\)    &    825147    &    825147    &    825147    &    358237    &    552254    &    527347\\ 
adj. \(R^{2}\)   &    0.102    &    0.102    &    0.101    &    0.131    &    0.090    &    0.093\\ 
F    &    265.7    &    262.0    &    251.4    &    231.3    &    198.5    &    193.5\\
\# researcher profiles ($df_{r}+1$)     &    8448    &    8448    &    8448    &    8422    &    8435    &    8441\\
\hline\hline
\multicolumn{7}{l}{\footnotesize Standard errors in parentheses}\\ 
\multicolumn{7}{l}{\footnotesize \sym{*} \(p<0.05\), \sym{**} \(p<0.01\), \sym{***} \(p<0.001\)}\\
\end{tabular}}\label{TableS4.tab} 
\end{table}

\begin{table}[htbp]\centering
\caption{Flagship Project effect: Career-level analysis using panel model with researcher fixed effects.
Publication data included:  articles published in period $ y_{p} \in [1970, 2018]$ with $k_{p} \geq 2$ and $w_{p} \geq 2$; only includes researchers with $N_{a} \geq 10$ articles satisfying these criteria.
Robust standard errors are shown in parenthesis below each point estimate. Y indicates additional fixed effects included in the regression model.}
\resizebox{0.89\columnwidth}{!}{ 
\def\sym#1{\ifmmode^{#1}\else\(^{#1}\)\fi}
\begin{tabular}{l*{3}{c}}
\hline\hline    &    \multicolumn{1}{c}{(1)}    &    \multicolumn{1}{c}{(2)}    &    \multicolumn{1}{c}{(3)}\\ 
    &     \multicolumn{1}{c}{$z_{p}$}    &    \multicolumn{1}{c}{$z_{p}$}    &    \multicolumn{1}{c}{$z_{p}$} \\ 
\hline
$\ln k$    &    0.438\sym{***}    &    0.425\sym{***}    &    0.406\sym{***}\\ 
    &    (0.00551)    &    (0.00547)    &    (0.00528)\\ 
$\ln w$    &    0.0250\sym{***}    &    0.0546\sym{***}    &    0.0385\sym{***}\\ 
    &    (0.00616)    &    (0.00528)    &    (0.00536)\\ 
$\tau$    &    -0.0108\sym{***}    &    -0.00965\sym{***}    &    -0.00878\sym{***}\\ 
    &    (0.00255)    &    (0.00189)    &    (0.00175)\\ 
$I_{\text{2014+}}$    &    0.0137    &    0.00818    &    -0.0571\sym{***}\\ 
    &    (0.0160)    &    (0.0133)    &    (0.0111)\\ 
\hline
$I_{X_{SA\&CIP}}$    &    0.155\sym{***}    &        &    \\ 
    &    (0.00796)    &        &    \\ 
$I_{X_{SA\&CIP}} \times I_{\text{2014+}}$    &    -0.0884\sym{***}    &        &    \\ 
    &    (0.00985)    &        &    \\  
\hline
$I_{X_{\text{Neighboring},SA\&CIP}}$    &        &    0.182\sym{***}    &    \\ 
    &        &    (0.00973)    &    \\ 
$I_{X_{\text{Neighboring},SA\&CIP}} \times I_{\text{2014+}}$    &        &    -0.160\sym{***}    &    \\ 
    &        &    (0.0103)    &    \\ 
\hline
$I_{X_{\text{Distant},SA\&CIP}}$    &        &        &    0.0276\\ 
    &        &        &    (0.0180)\\ 
$I_{X_{\text{Distant},SA\&CIP}} \times I_{\text{2014+}}$    &        &        &    0.0432\sym{*}\\ 
    &        &        &    (0.0180)\\
    \hline
constant      &    -0.666\sym{***}    &    -0.777\sym{***}    &    -0.690\sym{***}\\ 
   &    (0.0729)    &    (0.0515)    &    (0.0559)\\ 
Year  ($y$)  dummy   &        Y       &     Y     &        Y   \\
Topic category ($\overrightarrow{SA}$) dummy     &        Y       &     Y     &        Y   \\
Department  category ($\overrightarrow{CIP}$)  dummy     &        Y       &     Y     &        Y   \\
 Region ($\overrightarrow{R}$) dummy       &        Y       &     Y     &        Y   \\
\hline
\(N\)   &    358237    &    552254    &    527347\\ 
adj. \(R^{2}\)    &    0.131    &    0.091    &    0.093\\
F    &    229.0    &    198.6    &    191.3\\  
\# researcher profiles ($df_{r}+1$)    &    8422    &    8435    &    8441\\ 
\hline\hline
\multicolumn{4}{l}{\footnotesize Standard errors in parentheses}\\ 
\multicolumn{4}{l}{\footnotesize \sym{*} \(p<0.05\), \sym{**} \(p<0.01\), \sym{***} \(p<0.001\)}\\
\end{tabular}}\label{TableS5.tab} 
\end{table}

\end{widetext}

\end{document}